\let\mypdfximage\pdfximage
\def\pdfximage{\immediate\mypdfximage}

\documentclass[journal,onecolumn,10pt]{IEEEtran1}

\makeatletter
\def\markboth#1#2{\def\leftmark{\@IEEEcompsoconly{\sffamily}\MakeUppercase{\protect#1}}%
\def\rightmark{\@IEEEcompsoconly{\sffamily}\MakeUppercase{\protect#2}}}
\makeatother

\let\labelindent\relax

\usepackage[T1]{fontenc}
\usepackage{tgtermes}
\usepackage{marvosym}
\usepackage{longtable}
\usepackage{graphics}
\usepackage{graphicx}
\usepackage{caption}
\usepackage[bgreek,english]{babel}
\usepackage{subfig}
\usepackage{epsfig}
\usepackage{color}
\usepackage{multirow}
\usepackage{mathrsfs}
\usepackage{textcomp}
\usepackage{amsfonts}
\usepackage{amsmath}
\usepackage{amssymb}
\usepackage{nccmath}
\usepackage[numbers,square,sort&compress,comma]{natbib}
\usepackage{chapterbib}
\usepackage[ruled,vlined,linesnumbered]{algorithm2e}
\usepackage{setspace}
\usepackage{verbatim}
\usepackage{wrapfig}
\usepackage{dblfloatfix}
\usepackage{comment}
\usepackage[strict]{changepage}
\usepackage{ragged2e}
\usepackage{microtype}
\usepackage{enumitem}
\usepackage{nccmath}
\usepackage{hyperref}
\usepackage{doi}
\usepackage{afterpage}
\usepackage{multirow,bigdelim}
\usepackage{booktabs}
\usepackage{color}
\usepackage[outline]{contour}
\usepackage{xcolor}
\usepackage{cancel}
\usepackage{tikz}
\usetikzlibrary{shapes,snakes,arrows,automata,positioning,matrix,calc,fadings,arrows.meta,decorations.markings,decorations.pathmorphing}
\usepackage{hf-tikz}

\setlist{parsep=0pt,listparindent=\parindent}

\DeclareCaptionFont{singlespacing}{\setstretch{1}}
\numberwithin{figure}{section}
\renewcommand \thesection{\arabic{section}}
\numberwithin{equation}{section}

\hypersetup{bookmarks=false,bookmarksopen=false,pdfpagemode=empty,pdfstartview=}

\title{\singlespacing\sf\huge Reduction of Markov Chains using a Value-of-Information-Based Approach}
\author{Isaac J. Sledge and Jos\'{e} C. Pr\'{i}ncipe%
\thanks{\fontdimen2\font=1.55pt Isaac J. Sledge a Research Engineer with the Advanced Signal Processing and Automated Target Recognition Branch of the US Naval Surface Warfare Center (NSWC-PCD), Panama City, FL 32407, USA (email: isaac.j.sledge@navy.mil).}
\thanks{\fontdimen2\font=1.55pt Jos\'{e} C. Pr\'{i}ncipe is the Don D. and Ruth S. Eckis Chair and Distinguished Professor with both the Department of Electrical and Computer Engineering and the Department of Biomedical Engineering, University of Florida, Gainesville, FL 32611, USA (email: principe@cnel.ufl.edu).  He is the director of the Computational NeuroEngineering Laboratory (CNEL) at the University of Florida.\vspace{0.1cm}}
\thanks{The work of the authors was funded by grants N00014-15-1-2013 (Jason Stack), N00014-14-1-0542 (Marc Steinberg), and N00014-19-WX-00636 (Marc Steinberg) from the US Office of Naval Research.  The first author was additionally supported by in-house laboratory independent research (ILIR) grant N00014-19-WX-00687 (Frank Crosby) from the US Office of Naval Research, a University of Florida Research Fellowship, a Robert C. Pittman Research Fellowship, and an ASEE Naval Research Enterprise Fellowship while at the University of Florida.  This authors would like to thank Sean P. Meyn for his suggestion to use the value of information for aggregating Markov chains.}%
}
\begin{document}
\bstctlcite{IEEEexample:BSTcontrol}

\maketitle
\RaggedRight\parindent=1.5em
\fontdimen2\font=2.1pt
\vspace{-1.55cm}\begin{abstract}\normalsize\singlespacing
\vspace{-0.25cm}{\small{\sf{\textbf{Abstract}}}}---In this paper, we propose an approach to obtain reduced-order models of Markov chains.  Our approach is composed of two information-theoretic processes.  The first is a means of comparing pairs of stationary chains on different state spaces, which is done via the negative Kullback-Leibler divergence defined on a model joint space.  Model reduction is achieved by solving a value-of-information criterion with respect to this divergence.  Optimizing the criterion leads to a probabilistic partitioning of the states in the high-order Markov chain.  A single free parameter that emerges through the optimization process dictates both the partition uncertainty and the number of state groups.  We provide a data-driven means of choosing the `optimal' value of this free parameter, which sidesteps needing to a priori know the number of state groups in an arbitrary chain.

\end{abstract}%
\begin{IEEEkeywords}\normalsize\singlespacing
\vspace{-1.25cm}{{\small{\sf{\textbf{Index Terms}}}}---Markov chains, value of information, aggregation, model reduction, dynamics reduction, information theory}
\end{IEEEkeywords}
\IEEEpeerreviewmaketitle
\allowdisplaybreaks
\singlespacing

\vspace{-0.4cm}\subsection*{\small{\sf{\textbf{1$\;\;\;$Introduction}}}}\addtocounter{section}{1}

Markov models have seen a widespread adoption in a variety of disciplines.  Part of their appeal is that the application and simulation of such models is rather efficient, provided that the corresponding state-space has a small to moderate size.  Dealing with large state spaces is often troublesome, in comparison, as it may not be possible to adequately simulate the underlying models.  Such large-scale spaces are frequently encountered in reinforcement learning, for instance \cite{ArrudaEF-conf2009a,AldhaheriRW-jour1991a,RenZ-jour2005a,SunT-jour2007a,JiaQS-jour2011a}.  

A means of rendering the simulation of large-scale models tractable is crucial for many applications.  One way for doing this is to reduce the overall size of the Markov-chain state space by aggregation \cite{AokiM-jour1978a}.  Aggregation entails either explicitly or implicitly defining and utilizing a function to partition nodes in the probability transition graph associated with the large-scale chain.  Groups of nodes, which are related by the their inter-state transition probabilities and have strong interactions, are combined and treated as a single aggregated node in a new graph.  This results in a lower-order chain with a reduced state space.  A stochastic matrix for the lower-order chain is then specified, which describes the transitions from one super-state to another.  This stochastic matrix should roughly mimic the dynamics of the original chain despite the potential loss in information incurred from the state combination.


There are a variety of methods for aggregating Markov chains, as we review in the next section.  In this paper, we develop and analyze a novel approach for aggregating Markov chains, which is composed of two information-theoretic processes \cite{PrincipeJC-book2010a}.  The first process entails quantifying the dissimilarity of nodes in the original and reduced-order probability transition graphs, despite the difference in state space sizes.  The second process involves iteratively partitioning similar nodes without explicit knowledge of the number of groups. 

For the first process, we adopt the reasonable view that nodes in a pair of chains are dissimilar if their associated rows of the stochastic matrix are sufficiently distinct.  We employ an information-theoretic measure, the negative Kullback-Leibler divergence \cite{RachedZ-jour2004a}, to gauge the distinctiveness and hence identify candidate nodes in the original chain for aggregation.  This divergence assesses the overlap between probability distributions.  It coincides with the Donsker-Varadhan rate function appearing in the large-deviations theory of Markov chains \cite{DonskerMD-jour1975a,DonskerMD-jour1975b} and measures the distance between two Markov chains defined on the same discrete state space.  

In the aggregation process that we consider, a reduced-order model is constructed on a discrete state space of a different size than the original model.  To facilitate assessing the negative Kullback-Leibler divergence between the original and reduced-order models, we construct a so-called joint model that incorporates details from both the original and reduced-order models.  This model encodes the salient properties of the lower-order transition matrix and is of the proper dimensionality to compare against rows of the original transition matrix.  A byproduct of using this joint model is that we can sidestep considering all possible liftings of the reduced-order models to the original space by averaging their dynamics according to a given distribution \cite{DengK-jour2011a,GeigerBC-jour2015a}.  Our approach therefore avoids having to solve an additional optimization problem, which is a boon when aggregating large-state-space chains.

The problem of finding an aggregated Markov chain that captures much of the dynamics in the original chain can be posed as a cost function that uses the above divergence.  For the second process, we consider the use of an information-theoretic criterion known as the value of information \cite{SledgeIJ-jour2017a,SledgeIJ-jour2017b,SledgeIJ-jour2017c} to efficiently partition the probability transition graph.  The value of information is a constrained, modified-free-energy-difference criterion that describes the maximum benefit associated with a given quantity of information in order to minimize average losses \cite{StratonovichRL-jour1965a,StratonovichRL-jour1966a}.  It is an optimal, non-linear conversion between information, in the Shannon sense \cite{CoverTM-book2006a}, and either costs or utilities, in the von-Neumann-Morgenstern sense \cite{vonNeumannJ-book1953a}.

In the context of aggregating Markov chains, the value of information describes the change in the distortion between the high- and low-order transition models that occurs from potentially modifying the number of state groups and elements of those groups.  The number of groups is implicitly determined by the bounded information that a given row of the original chain's transition matrix shares with a corresponding row of the reduced-order chain's transition matrix.  Low information bound amounts lead to small numbers of groups with many states per group.  A potentially good qualitative partitioning is often observed in such cases, as the reduced-order chain is parsimonious.  Higher information bound amounts can lead to large numbers of groups with fewer states per group.  The partitioning of the original chain can be over-complete, as related states may be unnecessarily split to yield a lower free energy.

Optimizing the value of information in a grouped-coordinate-descent manner yields a single free parameter that represents the effect of the information bound.  Increasing this parameter from some base value yields a hierarchy of partitions.  Each element of this hierarchy corresponds to a partition with an increasing information bound amount and hence a potentially increasing number of state groups.  Finer-scale group structure in the transition matrix is captured as the parameter value rises.  After some value, however, there are diminishing returns on the quality of the aggregation results.  Determining the `optimal' value, in a completely data-driven fashion, is hence crucial.  To find such values for arbitrary Markov chains, we apply perturbation theory.  In particular, we calculate the underestimation error of the information constraint in the value of information that occurs when considering finite-state chains.  We then augment the value-of-information criterion by subtracting out this overestimation.  Finally, we determine a lower bound for the free parameter that minimizes the underestimation error.  The corresponding aggregation process empirically avoids fitting more to the noise than the structure in the stochastic matrix of the high-order model.

As a part of our treatment of the value of information, we furnish convergence and convergence-rate proofs to demonstrate the optimality of the criterion for the aggregation problem.

The remainder of this short paper is organized as follows.  We begin, in section 2, with a survey of aggregation techniques for Markov chains.  Our approach is given in section 3.  In section 3.1, we introduce our notation and some fundamental concepts for binary-partition-based aggregation.  In section 3.1.1, we introduce the concept of a joint model so that the differently-sized transition matrices of the original and reduced-order chains can be compared.  We outline, in section 3.1.2, how this joint model facilitates the formulation of an minimum-dissimilarity aggregation optimization problem.  Properties of this problem are analyzed for general divergence measures.  At the end of section 3.1.2, we discuss practical issues associated with this initial optimization problem, which motivates the use of the value of information.  We show, in sections 3.2.1 and 3.2.2, how this information-theoretic criterion can be applied to probabilistically partition transition matrices.  We also cover how the criterion can be efficiently solved, how to construct the reduced-order transition matrices after partitioning, and how to find an upper bound on the free parameter that emerges from optimizing this information-theoretic criterion.  Lastly, in section 3.2.3, we furnish a bound on the expected performance of this criterion.

In section 4, we assess the empirical capabilities of the value of information for Markov chain aggregation.  We begin by covering our experimental protocols in section 4.1.  In section 4.2, we present our simulation results for series of synthetic datasets.  We first assess the performance of our value-of-information-based reduction for manually-selected and perturbation-theory-derived multiplier values.  We also comment on the convergence properties.  The appropriateness of the Shannon information constraint over an entropy constraint is additionally investigated in these sections.  Discussions of these results are given at the end of this section.  We summarize these findings in the broader context of our theoretical results in section 5.  Additionally, we outline directions for future research.



\subsection*{\small{\sf{\textbf{2$\;\;\;$Literature Review}}}} \addtocounter{section}{1}

A variety of Markov model aggregation techniques have been proposed over the years.  Some of the earliest work exploited the strong-weak interaction structure of nearly completely decomposable Markov chains to obtain reduced-order approximations \cite{SimonHA-jour1961a,CourtoisPJ-jour1975a}.  Both uncontrolled \cite{PervozvanskiiAA-jour1974a,GaitsgoriVG-jour1975a} and controlled Markov chains \cite{TeneketzisD-conf1980a,DelebecqueF-jour1981a,ZhangQ-jour1997a} have been extensively studied in the literature.  

The aggregation of nearly completely decomposable Markov chains have been investigated by Courtois \cite{CourtoisPJ-book1977a} and other researchers \cite{AldhaheriRW-conf1989a,KotsalisG-conf2003a,DeyS-jour2000a}.  Courtois developed an aggregation procedure that yields an approximation of the steady-state probability distribution with respect to a parameter that represents the weak interaction between state groups.  This process was later augmented to provide more accurate approximations \cite{VantilborghH-jour1985a}.  It was also combined with various iterative schemes, like the Gauss-Seidel method, to improve the speed of convergence \cite{CaoWL-jour1985a,KouryJR-jour1984a,BarkerGP-jour1986a,DayarT-jour1996a}.  Years later, Phillips and Kokotovic presented a singular perturbation interpretation of Courtois' aggregation approach \cite{PhillipsR-jour1981a}.  They developed a similarity transformation that converts the system into a singularly perturbed form, whose slow model coincides with the aggregated matrix found by Courtois' approach.  The use of singular perturbation has also been considered by other researchers \cite{PeponidesG-jour1983a,ChowJ-jour1985a,FilarJA-jour2001a}.

There are additional approaches that have been developed.  A few are worth noting here, as they resemble our contributions in various ways \cite{DengK-conf2009a,DengK-conf2012a,VidyasagarM-conf2010a,DengK-jour2011a,VidyasagarM-jour2012a,GeigerBC-jour2015a}.  For example, Deng and Huang \cite{DengK-conf2012a} used the Kullback-Leibler divergence as a cost function to obtain a low-rank approximation of the original transition matrix via nuclear-norm regularization.  This preserved the cardinality of the state space.  Here, we employ the negative Kullback-Leibler divergence as a means of measuring the change in the original and modified chains.  We, however, consider a modified chain that is of a reduced order, not the same order.  This change should provide more tangible benefits for the simulation of large-scale systems.

Another scheme that is related to ours is that of Vidyasagar \cite{VidyasagarM-jour2012a}.  Vidyasagar investigated an information-theoretic metric, the variation of information, between distributions on sets with different cardinalities.  Actually computing the metric that he proposed turns out to be computationally intractable for large-scale systems, however.  He therefore considered an efficient greedy approximation for finding an upper bound of the distance and studied its use for optimal order reduction.  He demonstrated that the optimal reduced-order distribution a set of of a particular cardinality is obtained by projecting the original distribution.  That is, the reduced-order distribution should have maximal entropy.  This condition is equivalent to to requiring that the partition function induces the minimum information loss.  In our work, the metric that we consider is tractable for different-cardinality sets.  The partitioning process is not, however, which motivates the use of the value of information for efficiently finding approximate partitions.  An advantage of using the value of information is that it directly minimizes the information loss, as it relies on a Shannon information constraint that quantifies the mutual dependence of the high-order and low-order chain states.

In \cite{DengK-conf2009a,DengK-jour2011a,GeigerBC-jour2015a}, Deng et al. and Geiger et al. developed two-step, information-theoretic approaches for Markov chain aggregation.  In the first step, the optimal model reduction problem is solved on the reduced space defined by a fixed partition function.  In the second step, Deng et al. \cite{DengK-conf2009a,DengK-jour2011a} select an optimal partition function according to a non-convex relaxation of the bi-partition problem, while Geiger et al. \cite{GeigerBC-jour2015a} find an approximate partition using the information-bottleneck method.  In both works, the distortion between the original and reduced-order models was assessed via Kullback-Leibler divergence.  The authors defined an optimization-based lifting procedure so that both chains would have the same cardinality.  The lifting employed by Geiger et al. incorporates one-step transition probabilities of the original chain, which minimizes information loss.  They obtained a tight bound for lumpable chains.  Deng et al. lift based only on the stationary distribution of the original chain, which maximizes the redundancy of the aggregated Markov chain.  Here, we consider the formation of an joint model based on a similar approach to Deng et al.: we form a probabilistically weighted average of the entries from the original stochastic matrix.  However, the formulation of this joint model occurs naturally versus being defined as an optimization problem.  

There are other topical differences between these approaches.  For example, Geiger et al. \cite{GeigerBC-jour2015a}, through the use of the information bottleneck, attempt to compress the original-model states into reduced-model states, in a lossy way, while keeping as much information about the original transition probabilities as possible.  Optimizing the value of information achieves a similar effect, albeit in a different manner.  It limits the information lost during quantization by both bounding the divergence between the original and reduced-order models and simultaneously maximizing the mutual dependence between the states in both models.  Despite this similar effect, the value of information has practical advantages.  We prove that the dynamical system underlying the partitioning process undergoes phase changes, for certain values of the criterion's single free-parameter, where a new state-group emerges in the reduced-order model.  Between critical values of the free parameter, no phase changes occur, which implies that only a finite number of distinct values must be considered.  For the information bottleneck, investigators would have to sweep over many parameter values, often far more than we consider, and repeatedly solve the aggregation problem.  Using an information-bottleneck scheme can hence be computationally prohibitive for large-scale Markov chains.  

We further enhance the practicality of the value of information by deriving an expression for the `optimal' free-parameter value.  This value performs a second-order minimization of the estimation error associated with the Shannon-mutual-information term in the value of information.  Empirically, using this value causes the partitioning process to fit more to the structure of well-defined state groups in the original model than outlier states.  It also tends to yield parsimonious partitions that neither over- nor under-quantize the state space.


Our motivation for considering a value-of-information-based methodology arose from our use of this criterion in reinforcement learning.  We have previously applied this criterion, in \cite{SledgeIJ-jour2017a,SledgeIJ-jour2017b,SledgeIJ-jour2017c}, for resolving the exploration-exploitation dilemma in Markov-decision-process-based reinforcement learning.  In our experiments on a variety of complicated application domains, we found that the value of information would consistently outperform existing search heuristics.  We originally attributed this improved learning rate solely to a systematic partitioning of the state space.  That is, groups of states would be partitioned, according to their action-value function magnitude, and assigned the same action.  The problem of determining an action that works well for an entire group of related states is easier than doing the same for each state individually.  However, it is our hypothesis that there is an aggregation of the Markov chains underlying the Markov decision processes.  The aggregation theory developed in this paper represents a necessary first step to showing that the criterion can perform reinforcement learning on a simpler Markov decision process whose dynamics roughly mirror those of the original problem.

We are not the first to consider the aggregation of Markov chains that appear in Markov-decision-process-based reinforcement learning, though \cite{ArrudaEF-conf2009a,AldhaheriRW-jour1991a,RenZ-jour2005a,SunT-jour2007a,JiaQS-jour2011a}.  Aldhaheri and Khalil \cite{AldhaheriRW-jour1991a} focused on the optimal control of nearly completely decomposable Markov chains.  They adapted Howard's policy-iteration algorithm to work on an aggregated model.  They showed that they could provide optimal control that minimizes the average cost over an infinite horizon.  Sun et al. \cite{SunT-jour2007a} employed time aggregation to reduce the state space for complicated Markov decision processes.  They divided the original process into segments, by certain states, to form an embedded Markov decision process.  Value iteration is then executed on this lower-order model.   In \cite{JiaQS-jour2011a}, Jia provided a polynomial-time means of aggregating states of a Markov decision process when the optimal value function is known.  For approximate value functions, he showed how to apply ordinal optimization to uncover a good state reduction with a high probability of being the correct aggregation.  A commonality of these works is that they are model-based: they assume that the transition probabilities are explicitly known.  Our previous work \cite{SledgeIJ-jour2017b,SledgeIJ-jour2017c}, however, focused on model-free learning, where these probabilities are not available a priori.  Model reduction according to a value-of-information-based should therefore occur implicitly during the exploration process if the concepts we develop as part of our aggregation theory extend to Markov decision processes.


\subsection*{\small{\sf{\textbf{3$\;\;\;$Methodology}}}} \addtocounter{section}{1}

Our approach for aggregating Markov chains can be described as follows.  Given a stochastic matrix of transition probabilities between states, we seek to partition this matrix to produce a reduced-size matrix which we refer to as an aggregated stochastic matrix.  The aggregated stochastic matrix has an equivalent graph-based interpretation, as it characterizes the edge weights of an undirected graph.  The vertices in this matrix correspond to states of a reduced-order chain.  There is a one-to-many mapping of a vertex from the aggregated stochastic matrix to the vertices of the original transition-matrix graph for the high-order chain.  Edges of the aggregated stochastic matrix codify the transition probability between pairs of states in the low-order model.

There are many possible aggregated stochastic matrices that can be formed for a given Markov chain.  We would like to find a matrix that yields the least total distortion for some measure, particularly the negative Kullback-Leibler divergence.  Due to the different sizes of the original transition matrix and the aggregated stochastic matrix, though, directly applying this divergence is not possible.  While we could re-define the Kullback-Leibler divergence for probability vectors with different cardinalities, we have opted to instead transform the aggregated stochastic matrices so that they are of the same size as the original transition matrix.  We specify how to construct a so-called joint model that encodes all of the dynamics of the reduced-order chain.  We provide an straightforward objective function for constructing a binary partition of the original transition matrix to uncover the optimal aggregated stochastic matrix.

The objective function that we specify leads to another issue: finding the optimal aggregated stochastic matrix is not trivial due to the binary-valuedness of the one-to-many mappings.  It can quickly become computationally intractable as the size of the state space rises.  To make our aggregation approach more computationally efficient, we relax the binary assumption by considering an alternate objective function, which is based on the value of information.  Optimization of the value of information yields a probabilistic partitioning process for finding part of the aggregated stochastic matrix.  A single parameter associated with this function dictates both the uniformity of the probabilistic partitions and the number of state groups that emerge.  In the limits of the parameter value, the solution of the value of information approaches the global solutions of the original objective function.  A hierarchy of possible partitions, each with a different number of groups, are produced for other parameter values; these are approximate solutions of the binary-partition-based objective function.

\subsection*{\small{\sf{\textbf{3.1$\;\;\;$Aggregating Markov Chains}}}}

\subsection*{\small{\sf{\textbf{3.1.1$\;\;\;$Preliminaries}}}}


For our approach, we consider a first-order, homogeneous Markov chain defined on a finite state space.  Our analyses of such chains focus on graph-based transition abstractions.\vspace{0.05cm}
\begin{itemize}
\item[] \-\hspace{0.0cm}{\small{\sf{\textbf{Definition 3.1.}}}} The transition model of a first-order, homogeneous Markov chain is a weighted, directed graph $R_\pi$ given by the three-tuple $(V_\pi,E_\pi,\Pi)$ with the following elements
\begin{itemize}
\item[] \-\hspace{0.5cm}(i) A set of $n$ vertices $V_\pi \!=\! v_\pi^1 \cup \ldots \cup v_\pi^n$ representing the states of the Markov chain.\\
\item[] \-\hspace{0.5cm}(ii) A set of $n \!\times\! n$ edge connections $E_\pi \subset V_\pi \!\times\! V_\pi$ between reachable states in the Markov chain.\\
\item[] \-\hspace{0.5cm}(iii) A stochastic transition matrix $\Pi \!\in\! \mathbb{R}_+^{n \times n}$.  Here, $[\Pi]_{i,j} \!=\! \pi_{i,j}$ represents the non-negative transition probability between states $i$ and $j$.  We impose the constraint that the probability of experiencing a state transition is independent of time.
\end{itemize}\vspace{0.05cm}

\noindent The subscripts on the vertices and edges represent the dependence on the matrix $\Pi$.\vspace{0.025cm}
\end{itemize}


\noindent Throughout, we assume that all Markov chains are irreducible and aperiodic.  As a consequence, there is a unique invariant probability distribution $\gamma$ associated with the chain such that $\gamma^\top \Pi \!=\! \gamma^\top$.  We will sometimes write this\\ \noindent distribution as $\gamma(\Pi)$ to denote to which matrix the distribution is associated.

We are interested in comparing pairs of Markov chains.  A means to do this is by considering given rows of the stochastic transition matrix.  We represent the $i$th row of $\Pi$ by $\pi_{i,1:n} \!=\! [\pi_{i,1},\ldots,\pi_{i,n}]$.  $\pi_{i,1:n}$ is a probability vector describing the chance of transitioning from state $v_\pi^i$ to any possible next states.  We assume that $\pi_{i,j} \!=\! 0$ if and only if there is no directed edge from state $v_\pi^i$ to state $v_\pi^j$ and hence no chance of transitioning between these states.  

If a pair of transition models for different Markov chains, $R_\pi$ and $R_\varphi$, have the same number of states, then they can be compared according to a measure $g : \mathbb{R}_+^{n} \times \mathbb{R}_+^{n}\! \to \mathbb{R}_+$ acting on $\pi_{i,1:n}$ and $\varphi_{i,1:n}$ $\forall i$.  Here, we take this measure to be the negative Kullback-Leibler divergence; the theory that follows is applicable to many general measures, though.\vspace{0.05cm}
\begin{itemize}
\item[] \-\hspace{0.0cm}{\small{\sf{\textbf{Definition 3.2.}}}} Let $R_\pi \!=\! (V_\pi,E_\pi,\Pi)$ and $R_\varphi \!=\! (V_\varphi,E_\varphi,\Phi)$ be transition models of two Markov chains\\ \noindent over $n$ states.  The negative relative entropy, or negative Kullback-Leibler divergence, between a given set of states between these two chains is a function given by $g(\pi_{i,1:n},\varphi_{i,1:n}) \!=\! \sum_{j=1}^n \gamma_i \pi_{i,j} \textnormal{log}(\pi_{i,j}/\varphi_{i,j})$, where $\gamma$ is the invariant probability distribution associated with $R_\pi$.  The divergence rate is finite provided that $\Pi$ is absolutely continuous with respect to $\Phi$.\vspace{0.05cm}
\end{itemize}

\begin{figure*}
\centering
   \includegraphics[width=0.825\linewidth]{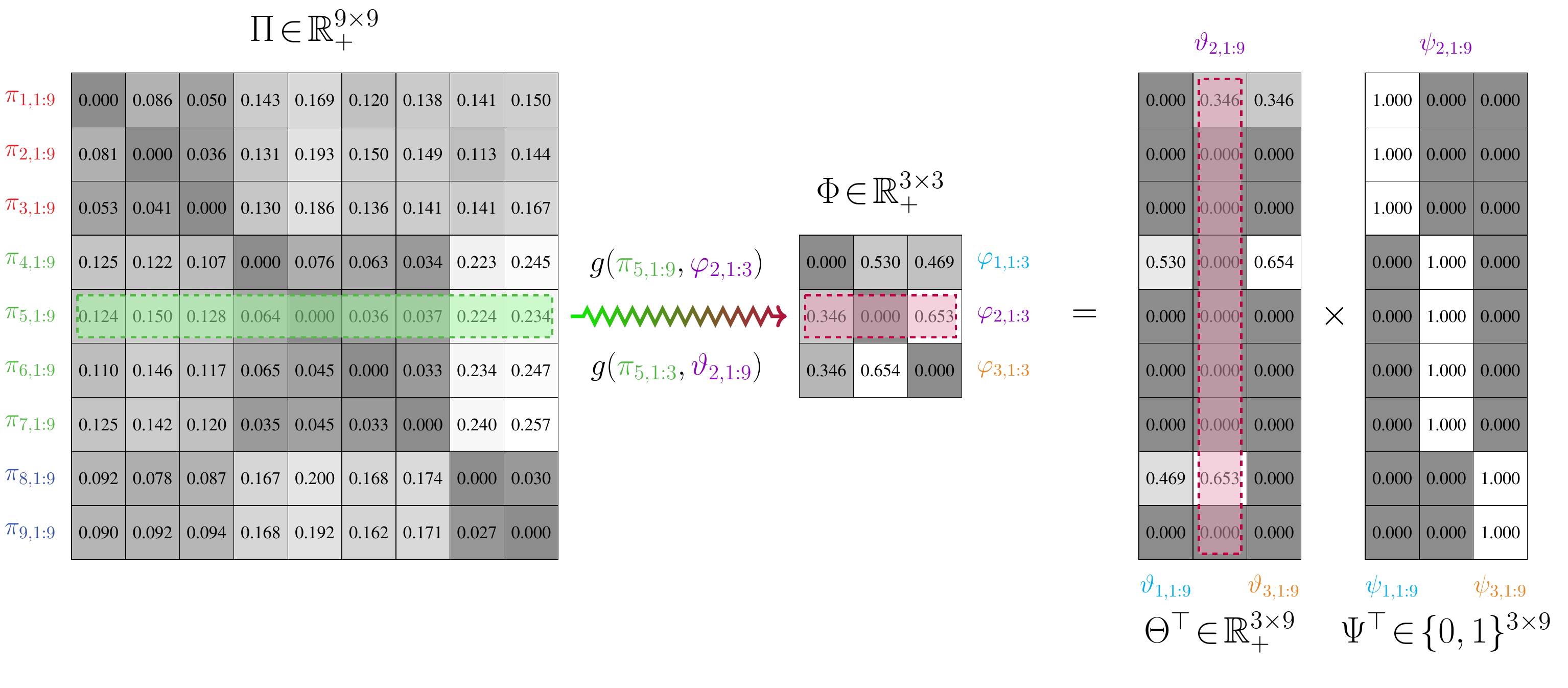}\vspace{-0.3cm}
\caption[]{Depiction of the comparison process for exact, binary aggregation of a nine-state Markov chain.  The transition matrix $\Phi$ associated with the low-order, three-state Markov chain cannot be directly compared to the transition matrix $\Pi$ of the high-order, nine-state chain for general measures $g$.  For example, we may want to compare the fifth row of $\Pi$, $\pi_{5,1:9}$, which is highlighted in green, with the second row of $\Phi$, $\varphi_{2,1:3}$: $g(\pi_{5,1:9},\varphi_{2,1:3})$, which is highlighted in purple.  To facilitate this comparison, we consider a joint model whose accumulation matrix $\Theta$ is of the proper size for comparison against $\Pi$.  $\Theta$ when multiplied with the binary partition matrix $\Psi$ equals the low-order transition matrix $\Phi$.\vspace{0.15cm}

It can be seen that the accumulation matrix $\Theta$ of the joint model encodes all of the dynamics of $\Phi$.  This relationship ensures that $g(\pi_{5,1:9},\vartheta_{2,1:9})$ is actually comparing the dynamics of $\Pi$ and $\Phi$.  $\Theta$ has been automatically padded with zero entries, by way of the exact aggregation process developed in this section, to ensure that it is of the same size as $\Pi$.  For this example, only the first, fourth, and eighth entries of any row in $\Pi$ are relevant for the comparison of the entries in $\Phi$.  Such entries lead to the maximal preservation of the information between $\Pi$ and $\Phi$ when using the negative Kullback-Leibler divergence for the measure $g$.\vspace{-0.4cm}}
\end{figure*}

\noindent Since we are considering the problem of chain aggregation, the state spaces will be different.  One chain $R_\pi$ will have $n$ states while another $R_\varphi$ will have $m$ states, with $m \!<\! n$.  The dimensionalities of given rows in the corresponding transition matrices will hence not be equivalent, which precludes a direct comparison using conventional measures.  

To facilitate the application of measures to $R_\pi$ and $R_\varphi$ when they have different discrete state spaces, we consider construction of a joint model $R_\vartheta$.  This joint model defines a joint state space composed of $V_\pi$ and $V_\varphi$.  It consequently possesses a weighting matrix $\Theta$ with the same number of columns as $\Pi$, which is outlined in definition 3.4 and illustrated in figure 3.1.

The joint model relies on the specification of a binary partition function $\psi : \mathbb{Z}_+ \!\to \mathbb{Z}_+$, which is given in definition 3.3.  This function provides a one-to-many mapping between states in $V_\pi$ and $V_\varphi$ and hence can be seen as a means of delineating which states of the original chain should be combined.\vspace{0.05cm} 
\begin{itemize}
\item[] \-\hspace{0.0cm}{\small{\sf{\textbf{Definition 3.3.}}}} Let $R_\pi \!=\! (V_\pi,E_\pi,\Pi)$ and $R_\varphi \!=\! (V_\varphi,E_\varphi,\Phi)$ be transition models of two Markov chains\\ \noindent over $n$ and $m$ states, respectively.  A binary partition function $\psi$ is a surjective mapping between two state index sets, $\mathbb{Z}_{1:n}$ and $\mathbb{Z}_{1:m}$, such that $\psi^{-1}(\mathbb{Z}_{1:m})$ is a partition of $\mathbb{Z}_{1:n}$.  That is, $\psi^{-1}(j) \!\subset\! \mathbb{Z}_{1:n}$ is not empty, $\psi^{-1}(1) \cup \ldots \cup \psi^{-1}(m) \!=\! \mathbb{Z}_{1:n}$, and $\psi^{-1}(j) \cap \psi^{-1}(k) \!=\! \emptyset$, for $j \!\neq\! k$.

It can be seen that a partition of a state index set induces a binary partition matrix $[\Psi]_{i,j} \!=\! \psi_{i,j}$, where\\ \noindent $\psi_{i,j} \!=\! 1$ if $i \!\in\! \psi^{-1}(j)$ and $\psi_{i,j} \!=\! 0$ if $i \!\notin\! \psi^{-1}(j)$.  Thus, $[\Psi]_{1:n,k} \!=\! \sum_{i \in \psi^{-1}(k)} e_i$, where $e_i$ is the $i$th unit vector.  The set of all binary partition matrices is given by $\{\Psi \!\in\! \mathbb{R}^{n \times m}_+|[\Psi]_{i,j} \!=\! \psi_{i,j} \!\in\! \{0,1\},\; \sum_{j=1}^m \psi_{i,j} \!=\! 1\}$.\vspace{0.075cm}
\end{itemize}
\begin{itemize}
\item[] \-\hspace{0.0cm}{\small{\sf{\textbf{Definition 3.4.}}}}  Let $R_\pi \!=\! (V_\pi,E_\pi,\Pi)$ and $R_\varphi \!=\! (V_\varphi,E_\varphi,\Phi)$ be transition models of two Markov chains over\\ \noindent $n$ and $m$ states, respectively, where $m \!<\! n$. $R_\vartheta \!=\! (V_\vartheta,E_\vartheta,\Theta)$ is a joint model, with $m \!+\! n$ states, that is defined by
\begin{itemize}
\item[] \-\hspace{0.5cm}(i) A vertex set $V_\vartheta \!=\! V_\pi \cup V_\varphi$, which is the union of all state vertices in $R_\pi$ and $R_\varphi$.  For simplicity, we\\ \noindent assume that the vertex set for the intermediate transition model is indexed such that the first $m$ nodes are from $R_\varphi$ and the remaining $n$ nodes are from $R_\pi$.
\item[] \-\hspace{0.5cm}(ii) An edge set $E_\vartheta \subset V_\varphi \!\times\! V_\pi$, which are one-to-many mappings from the states in the original transition model $R_\pi$ to the reduced-order transition model $R_\varphi$.  
\item[] \-\hspace{0.5cm}(iii) A weighting matrix $\Theta$, which is such that $\Theta \!\in\! \mathbb{R}_+^{m \times n}$, $\Theta \!=\! [\vartheta_{1,1:n}^\top,\vartheta_{2,1:n}^\top,\ldots,\vartheta_{m,1:n}^\top]^\top$.  The partition function $\psi$ provides a relationship between the stochastic matrices $\Phi$ and $\Theta$ of $R_\varphi$ and $R_\vartheta$, respectively.  This is given by $\varphi_{j,k} \!=\! \sum_{i \in \psi^{-1}(k)} \vartheta_{j,k}$ $\forall j,k$, or, rather, $\Theta\Psi \!=\! \Phi$, where $[\Psi]_{1:n,k} \!=\! \sum_{i \in \psi^{-1}(k)} e_i$.


\end{itemize}\vspace{0.05cm}
\end{itemize}



\begin{figure*}
\hspace{-0.7cm}\begin{tabular}{c c}
   \includegraphics[width=3.55in]{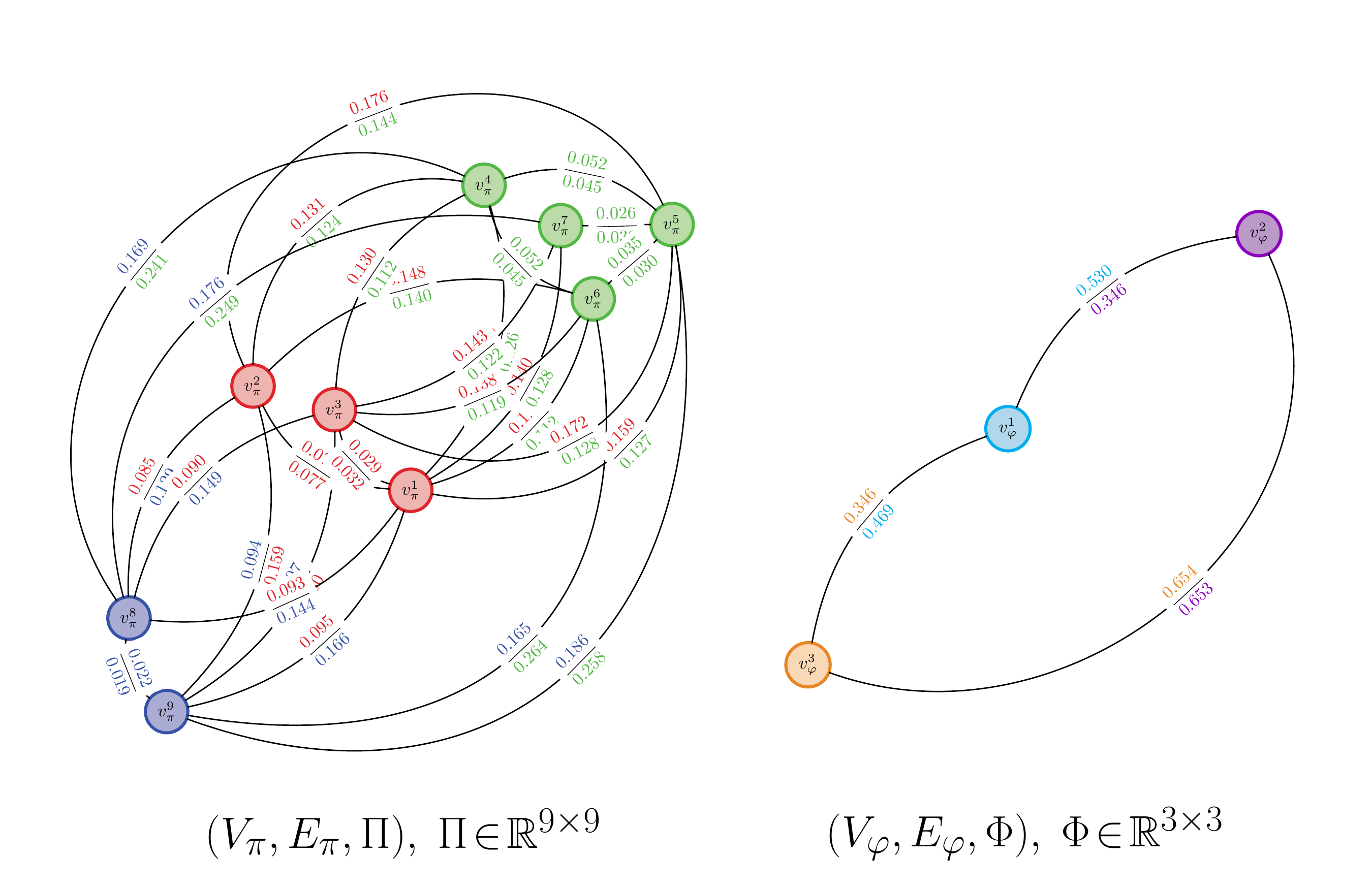} & \hspace{-0.6cm}\includegraphics[width=3.55in]{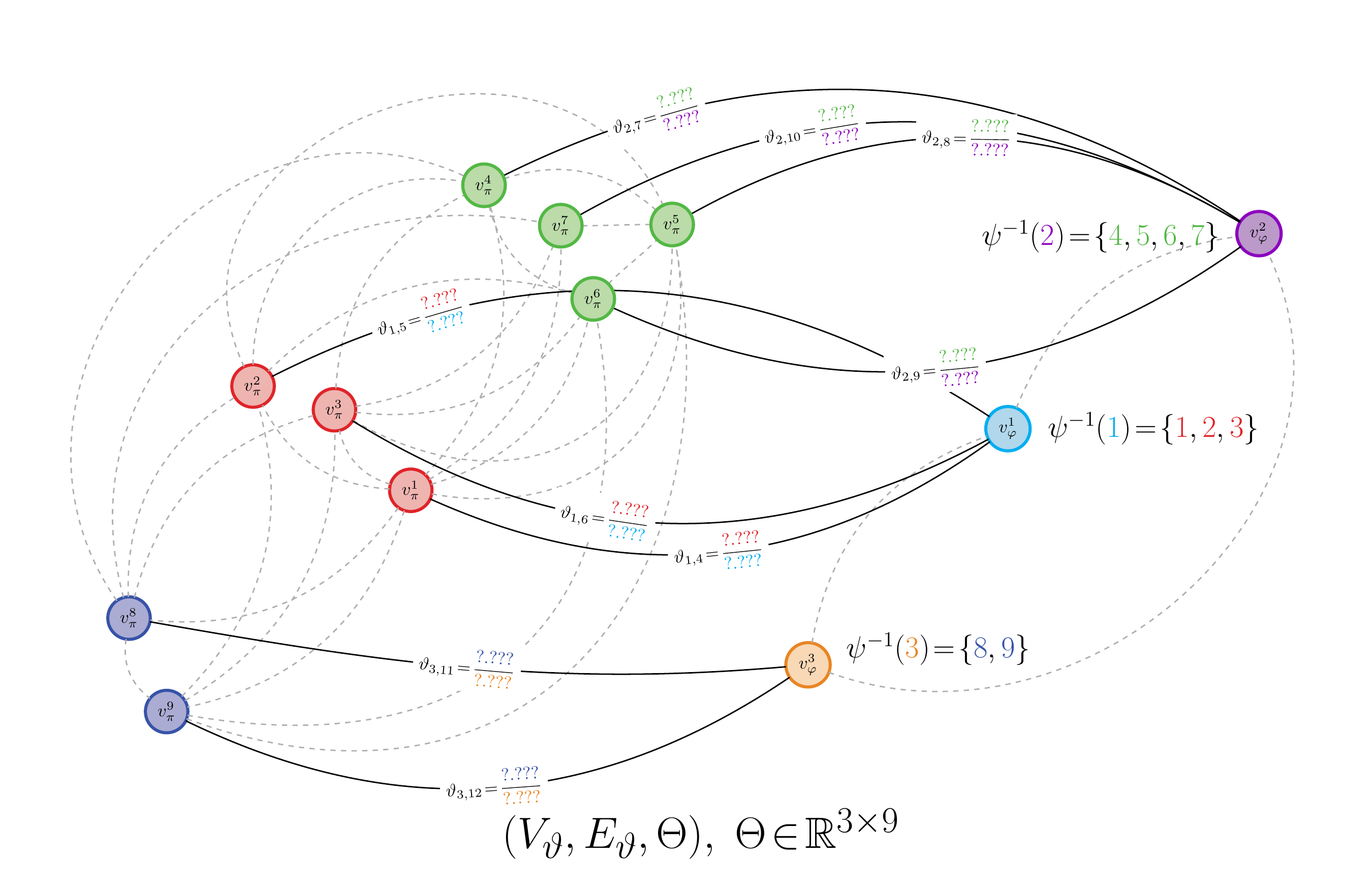}\vspace{-0.25cm}\\ \footnotesize{(a) Original (left) and reduced-order (right) transition model} & \hspace{-0.6cm}\footnotesize{(b) Joint model}
\end{tabular}
\caption[]{Depictions of the various models when using binary-valued partitions for the transition matrices in figure 3.1.  In (a), we show the transition model for a high-order, nine-state Markov chain (right) and its low-order, three-state transition model representation (left) after the aggregation process.  The numbers along the edges represent the probabilities of transitioning to and from pairs of states.  In (b), we show the joint model.  The vertices of the joint model represent states in both the high-order and low-order chains.  The edges between state pairs in both the high- and low-order chains, which are depicted using dashed lines, are removed.  In the joint model, edges are inserted to connect states in high-order chain with those in the low-order chain, thereby providing the state aggregation.  Note that the edge weights in the joint model are unknown a priori and must be uncovered.\vspace{-0.4cm}}
\end{figure*}

\noindent An illustration of the joint model relative to the other models is provided in figure 3.2.

\subsection*{\small{\sf{\textbf{3.1.2$\;\;\;$Partitioning Process and State Aggregation}}}}

For any given transition model $R_\pi$, we would like to find, by way of the joint model $R_\vartheta$, another transition model $R_\varphi$ with fewer states that resembles the dynamics encoded by $R_\pi$.  We therefore seek a $R_\vartheta$ with a weighting matrix $\Theta$ that has the least total distortion with respect to the transition matrix $\Pi$ of $R_\pi$ for some partition.  In this section, we specify how to find $R_\varphi$.

Before we can define the notion of least total distortion, we must first specify the concept of the total distortion of $\Theta$ with respect to $\Pi$.\vspace{0.05cm}
\begin{itemize}
\item[] \-\hspace{0.0cm}{\small{\sf{\textbf{Definition 3.5.}}}} Let $R_\pi \!=\! (V_\pi,E_\pi,\Pi)$, $R_\varphi \!=\! (V_\varphi,E_\varphi,\Phi)$, and $R_\vartheta \!=\! (V_\vartheta,E_\vartheta,\Theta)$ be transition models of two\\ \noindent Markov chains over $n$ and $m$ states and the joint model over $n \!+\! m$ states, respectively.  The total distortion between $\Pi$ and $\Theta$ and hence $\Pi$ and $\Phi$ is 
\begin{equation*}
q(R_\pi,R_\varphi) = \textnormal{min}_{\Theta \in \mathbb{R}_+^{m \times n}}\Bigg(\sum_{i=1}^n p(i)g(\pi_{i,1:n},\vartheta_{\psi(i),1:n})\,\Bigg|\,R_\vartheta \!\in\! \mathcal{R}_{\pi\varphi}\Bigg)
\end{equation*}
for some unit-sum weights $p(i)$.  We can take these weights to be the invariant distribution of the original Markov chain, i.e., $p(i) \!=\! \gamma_i$.

For this objective function we have the constraint that $\mathcal{R}_{\pi\varphi}$ must be a member of the set of all joint models for $R_\pi$ and $R_\varphi$.  \vspace{0.05cm}
\end{itemize}

  
\noindent It can be seen from definition 3.5 that the total distortion is over the set of all possible binary partitions.  We, however, seek the best binary partition.  Best, in this context, means that it would yield an $R_\varphi$ with the least total distortion to $R_\pi$.  It hence would lead to a lower-order model $R_\varphi$ that most resembles $R_\pi$ according to the chosen measure $g$.\vspace{0.05cm}
\begin{itemize}
\item[] \-\hspace{0.0cm}{\small{\sf{\textbf{Definition 3.6.}}}} Let $R_\pi \!=\! (V_\pi,E_\pi,\Pi)$ and $R_\varphi \!=\! (V_\varphi,E_\varphi,\Phi)$ be transition models of two Markov chains\\ \noindent over $n$ and $m$ states, respectively.  The accumulation matrix $\Phi \!=\! [\varphi_{1,1:m}^\top,\varphi_{2,1:m}^\top,\ldots,\varphi_{m,1:m}^\top]^\top$ for $R_\varphi$ that\\ \noindent achieves the least total distortion to $\Pi$ of $R_\pi$, according to $g : \mathbb{R}_+^{n} \times \mathbb{R}_+^{n}\! \to \mathbb{R}_+$, is given by
\begin{equation*}
\textnormal{arg min}_{\Psi \in \mathbb{R}_+^{n \times m},\,\Phi \in \mathbb{R}_+^{m \times m}} \Bigg(q(R_\pi,R_\varphi)\Bigg|\, [\Psi]_{1:n,k} \!=\! \sum_{i \in \psi^{-1}(k)} e_i\Bigg).
\end{equation*}
Here, $q : \mathbb{R}_+^{n \times n} \times \mathbb{R}_+^{m \times m} \to \mathbb{R}_+$ is the total distortion.
\vspace{0.05cm}
\end{itemize}
At least one minimizer exists for both assessing total distortion and least total distortion.  This is because both are continuous functions operating on closed and bounded sets and hence, according to the Weierstrass extreme value theorem, obtain both a maximum and minimum on those sets.


From definition 3.6, we can now specify the optimization problem of aggregating a Markov chain described.  This problem can be solved in a two-step process.  The first step entails finding the optimal partition that leads to the least total distortion between the original chain $\Pi$ and $\Phi$, as described by $\Theta$.  The second step involves constructing the corresponding low-order transition matrix $\Phi$ from $\Theta$ and $\Psi$.
\begin{itemize}
\item[] \-\hspace{0.0cm}{\small{\sf{\textbf{Definition 3.7.}}}} Let $R_\pi \!=\! (V_\pi,E_\pi,\Pi)$ and $R_\varphi \!=\! (V_\varphi,E_\varphi,\Phi)$ be transition models of two Markov chains\\ \noindent over $n$ and $m$ states, respectively.  The optimal reduced-order transition model $R_\varphi$ with respect to the original model $R_\pi$ can be found as follows

\begin{itemize}
\item[] \-\hspace{0.5cm}(i) Optimal partitioning: Find a binary partition matrix $\Psi$ that leads to the least total distortion between the models $R_\pi$ and $R_\varphi$.  As well, find the corresponding weighting matrix $\Theta$ that satisfies
\begin{equation*}
\textnormal{arg min}_{\Psi \in \mathbb{R}_+^{n \times m},\, \Theta \in \mathbb{R}_+^{m \times n}}\Bigg(\sum_{i=1}^n p(i)g(\pi_{i,1:n},\vartheta_{\psi(i),1:n})\,\Bigg|\,R_\vartheta \!\in\! R_{\pi\varphi},\, [\Psi]_{1:n,k} \!=\! \sum_{i \in \psi^{-1}(k)} e_i\Bigg).
\end{equation*}
Solving this problem has the effect of partitioning the $n$ vertices of the relational matrix $R_\pi$ into $m$ groups.
\item[] \-\hspace{0.5cm}(ii) Transition matrix construction: Obtain the transition matrix for $R_\varphi$ from the following expression: $\varphi_{j,k} \!=\! \sum_{i \in \psi^{-1}(k)} \vartheta_{j,k}$ using the optimal weights $\Theta$ and the binary partition matrix $\Psi$ from step (i).
\end{itemize}
\end{itemize}
It is important to notice for the first step in definition 3.7 that there is no efficient way to find a $R_\varphi$ with least total distortion to $R_\pi$.  This is due to the binary nature of the partitions, which leads to a problem with an NP-hard computational complexity.  For practical problems, which may contain thousands or even millions of states, this aggregation procedure will not be tractable.  A more efficient alternative is therefore required.


\subsection*{\small{\sf{\textbf{3.2$\;\;\;$Approximately Aggregating Markov Chains}}}}

\subsection*{\small{\sf{\textbf{3.2.1$\;\;\;$Preliminaries}}}}

A straightforward way to make the aggregation problem more efficient is by approximating the least total distortion optimization given in definition 3.6.  This can be effectuated by relaxing the constraint that the state-state assignments specified by the partition matrix are binary.  Instead, each state from the high-order chain can have a chance to map to states in the low-order chain.  Such changes lead to the notion of a probabilistic partition matrix.\vspace{0.05cm} 


\begin{itemize}
\item[] \-\hspace{0.0cm}{\small{\sf{\textbf{Definition 3.8.}}}} Let $R_\pi \!=\! (V_\pi,E_\pi,\Pi)$ and $R_\varphi \!=\! (V_\varphi,E_\varphi,\Phi)$ be transition models of two Markov chains\\ \noindent over $n$ and $m$ states, respectively.  A probabilistic partition function $\psi$ is a surjective mapping between two state index sets, $\mathbb{Z}_{1:n}$ and $\mathbb{Z}_{1:m}$, such that $\psi^{-1}(\mathbb{Z}_{1:m})$ is a partition of $\mathbb{Z}_{1:n}$, which has a given probabilistic chance of occurring.  That is, $\psi^{-1}(j) \!\subset\! \mathbb{Z}_{1:n} \times \mathbb{R}_+^n$ is not empty and where $\psi^{-1}(1) \cup \ldots \cup \psi^{-1}(m) \!=\!  \mathbb{Z}_{1:n}^m \times \mathbb{R}_+^{m \times n}$, with the real-valued responses being non-negative and summing to one.  

The probabilistic partition of a state index set induces a probabilistic partition matrix $[\Psi]_{i,j} \!=\! \psi_{i,j}$, where\\ \noindent $\psi_{i,j} \!=\! \zeta$ if $i \!\in\! \psi^{-1}(j)$ occurs with probability $\zeta$.  The set of all probabilistic partition matrices for the two chains specified above is given by \noindent $\{\Psi \!\in\! \mathbb{R}_+^{n \times m}|[\Psi]_{i,j} \!=\! \psi_{i,j} \!\in\! [0,1],\; \sum_{j=1}^m \psi_{i,j} \!=\! 1\}$.


\end{itemize}

\noindent An example of a probabilistic partitioning is given in figure 3.3.

\begin{figure*}
\centering
   \includegraphics[width=0.9\linewidth]{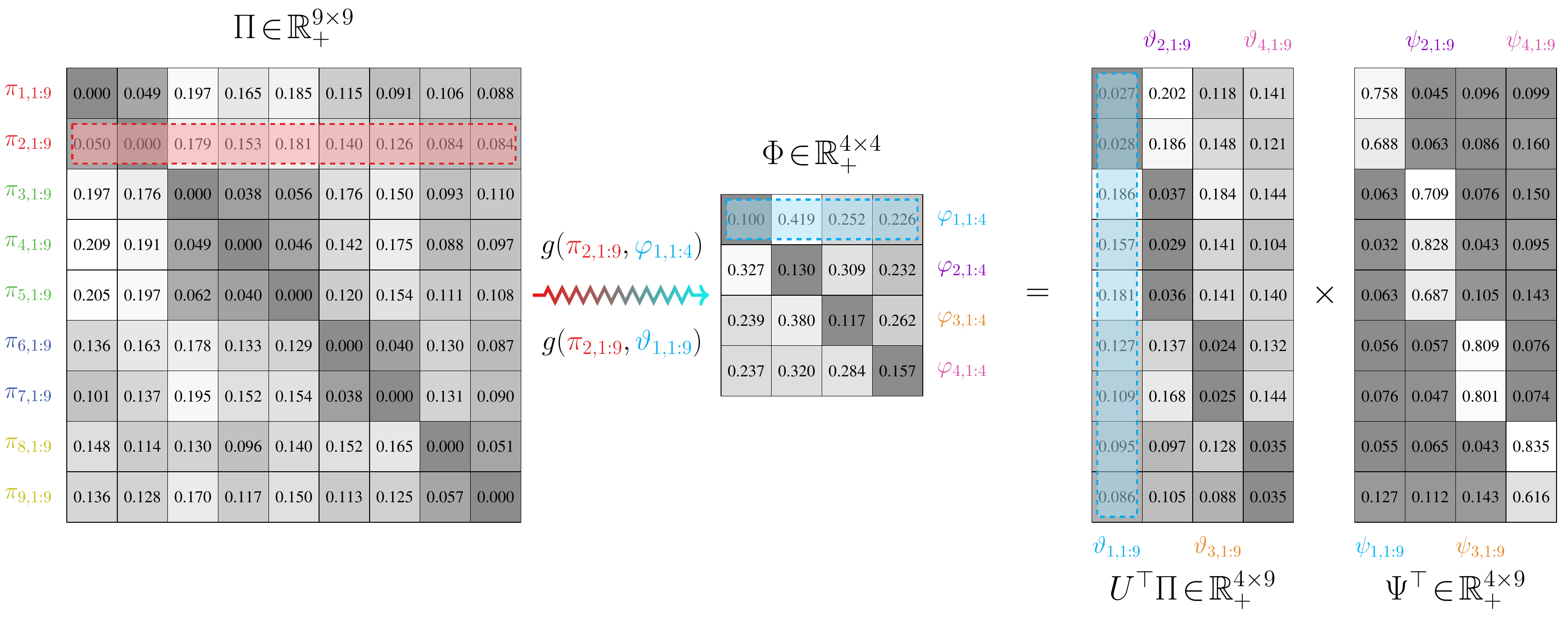}\vspace{-0.3cm}
\caption[]{Depiction of the comparison process for approximate, probabilistic aggregation of a nine-state Markov chain.  Here, we want to compare the second row of the high-order transition matrix $\Pi$ with the first row of a potential low-order stochastic matrix $\Phi$.  We show the transition matrix $\Pi$ associated with a nine-state Markov chain on the left; four state clusters are visible along the main diagonal.  The corresponding low-order transition matrix $\Phi$ for a four-state chain is given in the right.  As before, comparisons between $\Pi$ and $\Phi$ occur by comparing rows of $\Pi$ with rows of $\Theta\Psi$.  $\Phi$ can be found via the joint model weight matrix $\Theta \!=\! U^\top\Pi$ and the probabilistic partition $\Psi$: $\Phi \!=\! U^\top\Pi\Psi$.  The least expected distortion between the high-order $\Pi$ and low-order $\Phi$ transition matrices is determined by way of $\Theta$ and $\Psi$. \vspace{0.15cm}

When performing exact aggregation, the dynamics of $\Phi$ are directly encoded in $\Theta$.  $\Psi$ is only used to determine which columns of $\Theta$ can be ignored.  For approximate aggregation, the dynamics of $\Phi$ are split between $\Theta$ and $\Psi$.  This is because each state in the high-order model can have have the chance to map to multiple states in the low-order model.\vspace{-0.4cm}}
\end{figure*}

As before, we will partition and compare the dynamics for pairs of chains according to rows of the corresponding stochastic transition matrices.  We will, therefore, still encounter issues when trying to compare transition models $R_\pi$ and $R_\varphi$ with differing state spaces.  We again consider the construction of a joint model $R_\vartheta$ to avoid this issue.  The only difference between this joint model and the one defined for binary-valued partitions is that the weighting matrix has a different form.  The connectivity of the joint-space graph can hence be different.
\begin{itemize}
\item[] \-\hspace{0.0cm}{\small{\sf{\textbf{Definition 3.9.}}}}  Let $R_\pi \!=\! (V_\pi,E_\pi,\Pi)$ and $R_\varphi \!=\! (V_\varphi,E_\varphi,\Phi)$ be transition models of two Markov chains over\\ \noindent $n$ and $m$ states, respectively, where $m \!<\! n$. $R_\vartheta \!=\! (V_\vartheta,E_\vartheta,\Theta)$ is a joint model, with $m \!+\! n$ states, that is defined by
\begin{itemize}
\item[] \-\hspace{0.5cm}(i) A vertex set $V_\vartheta \!=\! V_\pi \cup V_\varphi$, which is the union of all state vertices in $R_\pi$ and $R_\varphi$.
\item[] \-\hspace{0.5cm}(ii) An edge set $E_\vartheta \subset V_\varphi \!\times\! V_\pi$, which are one-to-many mappings from the states in the original transition model $R_\pi$ to the reduced-order transition model $R_\varphi$.  
\item[] \-\hspace{0.5cm}(iii) A weighting matrix $\Theta \!\in\! \mathbb{R}_+^{m \times n}$.  The partition function $\psi$ provides a relationship between the stochastic matrices $\Phi$ and $\Theta$ of $R_\varphi$ and $R_\vartheta$, respectively.  This is given by $\varphi_{i,j} \!=\! \sum_{k=1}^n \vartheta_{k,j}\psi_{k,i}$ $\forall i,j$, or, rather,\\ \noindent $\Phi \!=\! \Theta \Psi$, where $\Psi \!\in\! \mathbb{R}_+^{n \times m}$ is the probabilistic partition matrix.
\end{itemize}\vspace{0.05cm}
\end{itemize}

\noindent An illustration of this joint model is given in figure 3.4 for the stochastic matrix presented in figure 3.3.  Unlike the joint model for binary-valued partitions, using probabilistic partitions allows for each state in the high-order chain to map to multiple states in the low-order chain.

\subsection*{\small{\sf{\textbf{3.2.2$\;\;\;$Partitioning Process and State Aggregation}}}}

For any transition model $R_\pi$ we, again, would like to find a joint model $R_\vartheta$ that facilitates the construction of another transition model $R_\varphi$.  $R_\varphi$ should have fewer states than $R_\pi$ while still possessing similar intra-group transition dynamics.  Since we are now considering probabilistic partitions, we instead seek a $\Theta$ with the least expected distortion to $\Pi$ to ensure that the dynamics of $\Phi$ largely match those of $\Pi$.  Definition 3.5 is hence modified as follows.\vspace{0.05cm}
\begin{itemize}
\item[] \-\hspace{0.0cm}{\small{\sf{\textbf{Definition 3.10.}}}} Let $R_\pi \!=\! (V_\pi,E_\pi,\Pi)$, $R_\varphi \!=\! (V_\varphi,E_\varphi,\Phi)$, and $R_\vartheta \!=\! (V_\vartheta,E_\vartheta,\Theta)$ be transition models of two\\ \noindent Markov chains over $n$ and $m$ states and the joint model over $n \!+\! m$ states, respectively.  The least expected\\ \noindent distortion between $\Pi$ and $\Theta$ and hence $\Pi$ and $\Phi$ is
\begin{equation*}
q(R_\pi,R_\varphi) = \textnormal{min}_{\Psi \in \mathbb{R}_+^{n \times m},\,\Theta \in \mathbb{R}_+^{m \times n}}\Bigg(\sum_{i=1}^n\sum_{j=1}^m \gamma_i \psi_{i,j}g(\pi_{i,1:n},\vartheta_{j,1:n})\,\Bigg|\!\begin{array}{c}0 \!\leq\! \vartheta_{i,k}, \psi_{i,k} \!\leq\! 1,\; \sum_{k=1}^m \vartheta_{i,k} \!=\! 1,\vspace{0.01cm}\\ \sum_{k=1}^m \psi_{i,k} \!=\! 1\end{array}\!\!\Bigg).
\end{equation*}
\vspace{0.05cm}
\end{itemize}

\begin{figure*}
\hspace{-0.8cm}\begin{tabular}{c c}
   \includegraphics[width=3.55in]{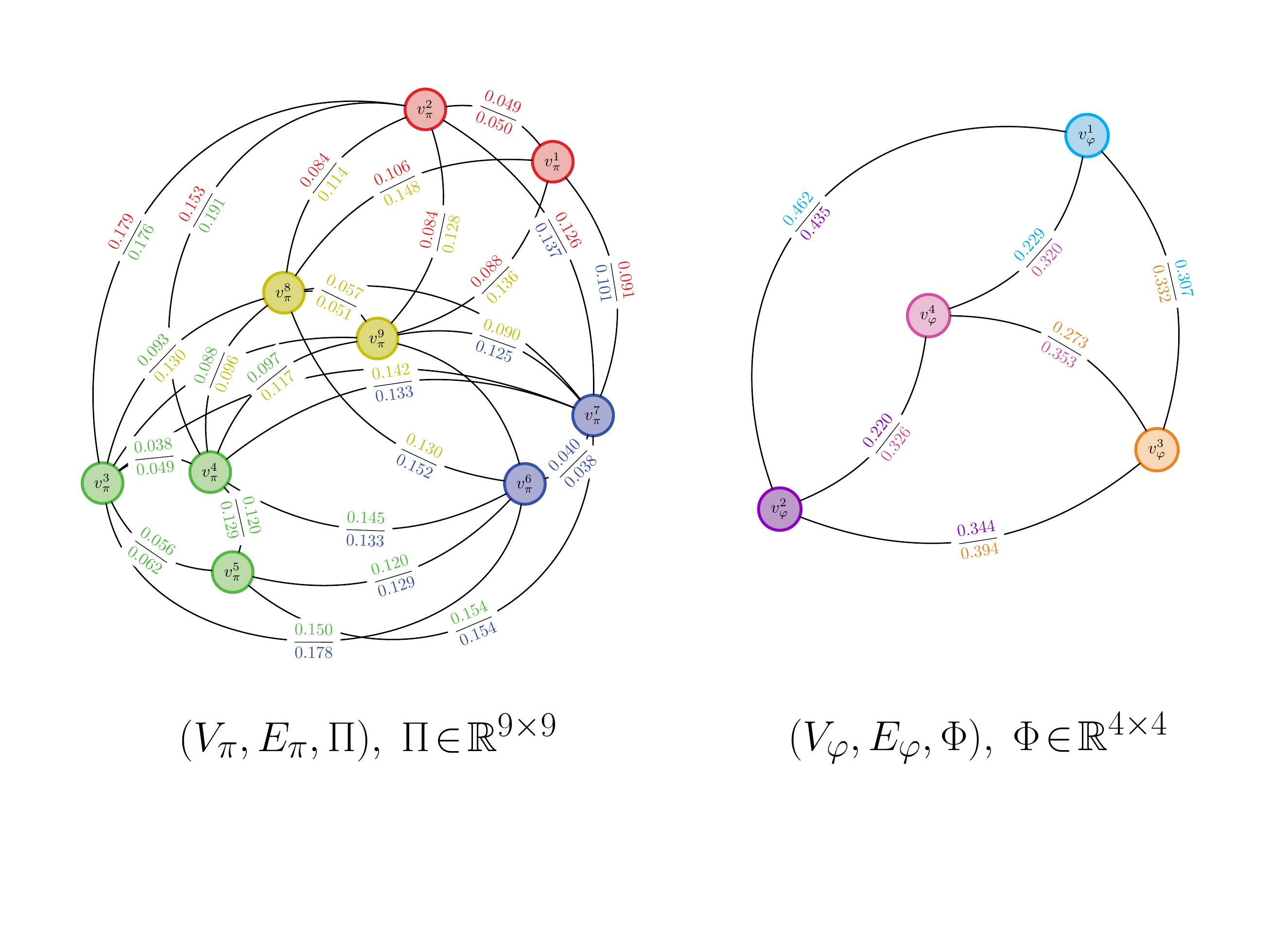} & \hspace{-0.6cm}\includegraphics[width=3.55in]{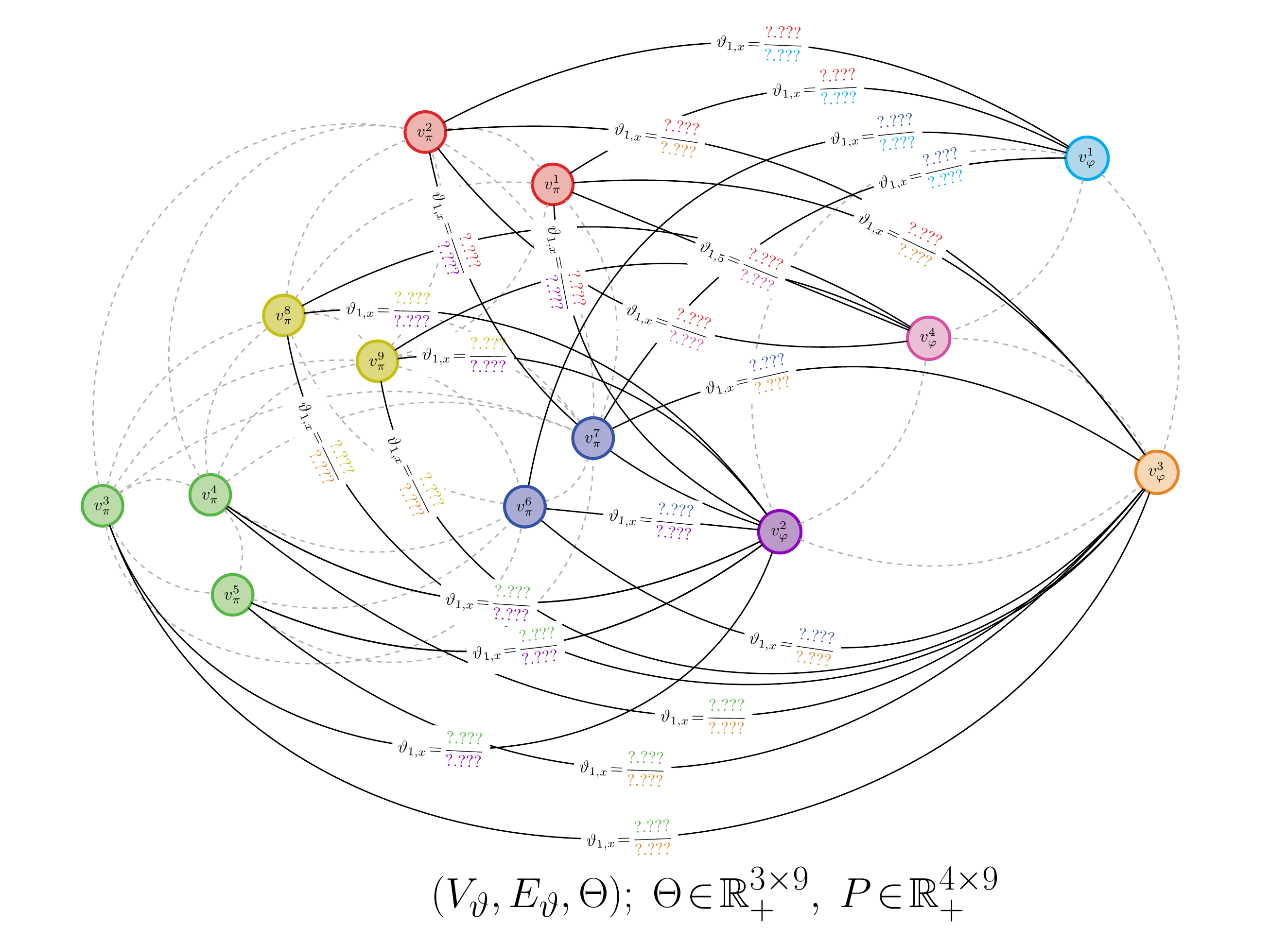}\vspace{-0.25cm}\\ \footnotesize{(a) Original (left) and reduced-order (right) transition model} & \hspace{-0.6cm}\footnotesize{(b) Joint model}
\end{tabular}
\caption[]{Depictions of the various models for the transition matrices in figure 3.3 when using probabilistic partitions.  In (a), we show the transition model for a high-order, nine-state Markov chain (right) and its low-order, four-state transition model representation (left) after the approximate aggregation process.  In (b), we show the joint model defined by $\Theta \!=\! U^\top \Pi$ and a relatively low value of $\beta$ for this example.  As before, each edge in both chains is removed and mappings between states in the two chains are established.  For probabilistic partitions, each state in the high-order chain have the chance to map to more than one state in the low-order chain.  This contrasts with the binary-valued partition case, where each state in the high-order chain could only be associated with a single state in the low-order chain.\vspace{-0.4cm}}
\end{figure*}

There are few constraints on the probabilistic partitions in definition 3.10, which can make finding viable solutions difficult.  To address this issue, we impose that the partitions should minimize the information loss associated with the state quantization process.  That is, the mutual dependence between states in the high-order and low-order chains should be maximized with respect to a supplied upper bound.  Simultaneously, the least expected distortion, for this supplied bound, should be achieved.

Aggregating Markov chains in this fashion can be done via a two-step process similar to that in definition 3.7.\vspace{0.05cm}
\begin{itemize}
\item[] \-\hspace{0.0cm}{\small{\sf{\textbf{Definition 3.11.}}}} Let $R_\pi \!=\! (V_\pi,E_\pi,\Pi)$ and $R_\varphi \!=\! (V_\varphi,E_\varphi,\Phi)$ be transition models of two Markov chains\\ \noindent over $n$ and $m$ states, respectively.  The optimal reduced-order transition model $R_\varphi$ with respect to the original model $R_\pi$ can be found as follows:

\begin{itemize}
\item[] \-\hspace{0.5cm}(i) Optimal partitioning: Find a probabilistic partition matrix $\Psi$ that leads to the least expected distortion between the models $R_\pi$ and $R_\varphi$.  As well, find the corresponding weighting matrix $\Theta$ that satisfies
\begin{equation*}
\textnormal{arg min}_{\Psi \in \mathbb{R}_+^{n \times m},\,\Theta \in \mathbb{R}_+^{m \times n}}\Bigg(\sum_{i=1}^n\sum_{j=1}^m \gamma_i \psi_{i,j}g(\pi_{i,1:n},\vartheta_{j,1:n})\,\Bigg|\!\!\begin{array}{c}\sum_{j=1}^m \alpha_j \sum_{i=1}^n \psi_{i,j}\textnormal{log}(\psi_{i,j}/\gamma_i) \!\leq\! r\vspace{0.1cm},\\ \; 0 \!\leq\! \vartheta_{i,k},\psi_{i,k} \!\leq\! 1,\; \sum_{k=1}^m \vartheta_{i,k} \!=\! 1,\; \sum_{k=1}^m \psi_{i,k} \!=\! 1\end{array}\!\!\Bigg)
\end{equation*}
for some positive value of $r$; $r$ has an upper bound of $-\sum_{i=1}^n \gamma_i\textnormal{log}(\gamma_i)$.  The variables $\alpha$, $\gamma$, and $\psi$ all have probabilistic interpretations: $\alpha_j \!=\! p(v_\varphi^j)$ and $\gamma_i \!=\! p(v_\pi^i)$ correspond to marginal probabilities of states $v_\varphi^j$ and $v_\pi^i$, while $\psi_{i,j} \!=\! p(v_\varphi^j|v_\pi^i)$ is the conditional probability of state $v_\pi^i$ mapping to state $v_\varphi^j$.
\item[] \-\hspace{0.5cm}(ii) Transition matrix construction: Obtain the transition matrix for $R_\varphi$ from the following expression: $\varphi_{i,j} \!=\! \sum_{k=1}^n \vartheta_{k,j}\psi_{k,i}$ using the optimal weights $\Theta$ and the probabilistic partition matrix $\Psi$ from step (i).\vspace{0.05cm}
\end{itemize}
\end{itemize}

\noindent The optimization problem presented in definition 3.11 trades off between the minimum expected distortion and the information  contained by the states in the low-order chain $R_\phi$ about those in the original, high-order chain $R_\pi$ after partitioning.  It is hence describing the value of quantizing the high-order model by a certain amount \cite{StratonovichRL-jour1965a,StratonovichRL-jour1966a}; this is the value of information formulated for Markov chains, which is, itself, an analogue of rate-distortion theory \cite{CoverTM-book2006a}.  Coarsely quantizing $\Pi$, as dictated by the parameter $r$, leads to a parsimonious low-order stochastic matrix $\Phi$ that may not greatly resemble the dynamics of $\Pi$.  Finely quantizing $\Pi$, again determined by $r$, yields a $\Phi$ that is similar to the high-order model's transition matrix $\Pi$ yet may contain many redundant details.

In the value of information, the role of the Shannon mutual information term $\sum_{j=1}^m \alpha_j \sum_{i=1}^n \psi_{i,j}\textnormal{log}(\psi_{i,j}/\gamma_i)$ is to impose a certain level of randomness, or uncertainty, in the partition matrix to ensure that the entries can be non-binary.  A similar effect could be achieved by considering a Shannon entropy constraint on the partition matrix $-\sum_{i=1}^n\sum_{j=1}^m \psi_{i,j}\textnormal{log}(\psi_{i,j})$.  However, a Shannon entropy constraint is rather non-restrictive on the entries of the partition matrix: there is the potential that a given row $\psi_{1:n,j} \!\in\! \mathbb{R}_+^n$ could be a duplicate of another, thereby over-\\ \noindent inflating the number of states in the reduced-order chain and leading to a poor aggregation.  We have found, empirically, that Shannon mutual information does not share this defect, except when all states have a uniform chance of being grouped together in every group.  This is because we are bounding the informational overlap between the original and aggregated states.  Coincident partitions often violate this bound.  In the Shannon-entropy case, however, we are only bounding the uncertainty on the entries of the partition, so there is no direct constraint between the original and aggregated states.


Definitions 3.8, 3.9, and 3.11 provide a means of approximating the computationally intractable aggregation process outlined in the previous section.  Actually solving the constrained optimization problem in definition 3.11 can be efficiently performed in a few ways.  Here, we opt to form the Lagrangian and differentiate it.  This provides an expectation-maximization-like procedure for specifying the probabilistic partitions $\Psi$.\vspace{0.05cm}
\begin{itemize}
\item[] \-\hspace{0.0cm}{\small{\sf{\textbf{Proposition 3.1.}}}} For a transition model $R_\pi \!=\! (V_\pi,E_\pi,\Pi)$ over $n$ states and a joint model $R_\vartheta \!=\! (V_\vartheta,E_\vartheta,\Theta)$\\ \noindent and $m \!+\! n$ states, the Lagrangian of the relevant terms for the minimization problem given in definition 3.11 is $F(\Psi,\alpha;\Pi,\Theta,\gamma) = \mathbb{E}[\mathbb{E}[g(\Pi,\Theta)|\Psi]|\gamma] \!-\! \mathbb{E}[D_\textnormal{KL}(\gamma\|\Psi)]/\beta$, or, rather,
\begin{equation*}
F(\Psi,\alpha;\Pi,\Theta,\gamma) = \Bigg(\sum_{i=1}^n\sum_{j=1}^m \gamma_i \psi_{i,j}g(\pi_{i,1:n},\vartheta_{j,1:n})\Bigg) - \frac{1}{\beta}\Bigg(\sum_{j=1}^m \alpha_j \sum_{i=1}^n \psi_{i,j}\textnormal{log}(\psi_{i,j}/\gamma_i)\Bigg).
\end{equation*}
Here, $\beta \!\geq\! 0$ is a Lagrange multiplier that emerges from the Shannon mutual information constraint in the value of information.

Probabilistic partitions $[\Psi]_{i,j} \!=\! \psi_{i,j}$, which are local solutions of $\nabla F(\Psi,\alpha;\Pi,\Theta,\gamma) \!=\! 0$, can be found by the following expectation-maximization-based alternating updates
\begin{equation*}
\alpha_j \leftarrow \sum_{i=1}^n \gamma_i\psi_{i,j},\;\;\;\;\; \psi_{i,j} \leftarrow \!\Bigg(\alpha_j e^{-\beta g(\pi_{i,1:n},\vartheta_{j,1:n})}\!\Bigg)\!\Bigg/\!\Bigg(\sum_{p=1}^m \alpha_p e^{-\beta g(\pi_{i,1:n},\vartheta_{p,1:n})}\Bigg),
\end{equation*}
which are iterated until convergence.\vspace{0.05cm}
\end{itemize}

\noindent Proposition 3.2 shows that the alternating optimization updates in proposition 3.1 yield monotonic decreases in the modified free-energy associated with the value of information.  Global convergence to solutions can therefore be obtained.  Proposition 3.3 bounds the approximation error as a function of the number of alternating-optimization iterations.  Linear-speed convergence to solutions is hence obtained, which coincides with the interpretation of the updates as an expectation-maximization-type algorithm.

\begin{itemize}
\item[] \-\hspace{0.0cm}{\small{\sf{\textbf{Proposition 3.2.}}}} Let $R_\pi \!=\! (V_\pi,E_\pi,\Pi)$ and $R_\varphi \!=\! (V_\varphi,E_\varphi,\Phi)$ be transition models of two Markov chains\\ \noindent over $n$ and $m$ states, respectively, where $m \!<\! n$.  If $[\Psi^*]_{i,j} \!=\! \psi_{i,j}^*$ is an optimal probabilistic partition and\\ \noindent $[\alpha^*]_j \!=\! \alpha^*_j$ an optimal marginal probability vector, then, for the updates in proposition 3.1, we have that: 
\begin{itemize}
\item[] \-\hspace{0.5cm}(i) The approximation error is non-negative
\begin{equation*}
\Bigg(F(\Psi^{(k)},\alpha^{(k)};\Pi,\Theta,\gamma) - F(\Psi^*,\alpha^*;\Pi,\Theta,\gamma)\Bigg) = \sum_{i=1}^n \gamma_i \textnormal{log}\Bigg(\frac{\sum_{j=1}^m \alpha^*_je^{-\beta g(\pi_{i,1:n},\vartheta_{j,1:n})}}{\sum_{j=1}^m \alpha^{(k)}_je^{-\beta g(\pi_{i,1:n},\vartheta_{j,1:n})}}\Bigg) \geq 0.
\end{equation*}
\item[] \-\hspace{0.5cm}(ii) The modified free energy monotonically decreases $F(\Psi^{(k)},\alpha^{(k)};\Pi,\Theta,\gamma) \!\geq\! F(\Psi^{(k+1)},\alpha^{(k+1)};\Pi,\Theta,\gamma)$ across all iterations $k$.
\item[] \-\hspace{0.5cm}(iii) For any $K \!\geq\! 1$, we have the following bound for the sum of approximation errors
\begin{equation*}
\Bigg(\sum_{k=1}^K F(\Psi^{(k)},\alpha^{(k)};\Pi,\Theta,\gamma) - F(\Psi^*,\alpha^*;\Pi,\Theta,\gamma)\Bigg) \leq \sum_{i=1}^n \sum_{j=1}^m \gamma_i \psi_{i,j}^* \textnormal{log}\Bigg(\frac{\psi_{i,j}^*}{\psi_{i,j}^{(1)}}\Bigg).
\end{equation*}
\end{itemize}
Here, $F(\Psi,\alpha;\Pi,\Theta,\gamma) \!=\! \mathbb{E}[\mathbb{E}[g(\Pi,\Theta)|\Psi]|\gamma] \!-\! \mathbb{E}[D_\textnormal{KL}(\gamma\|\Psi)]/\beta$ is the Lagrangian.\vspace{0.05cm}
\end{itemize}

\begin{itemize}
\item[] \-\hspace{0.0cm}{\small{\sf{\textbf{Proposition 3.3.}}}} Let $R_\pi \!=\! (V_\pi,E_\pi,\Pi)$ and $R_\varphi \!=\! (V_\varphi,E_\varphi,\Phi)$ be transition models of two Markov chains\\ \noindent over $n$ and $m$ states, respectively, where $m \!<\! n$.  If $[\Psi^*]_{i,j} \!=\! \psi_{i,j}^*$ is an optimal probabilistic partition and\\ \noindent $[\alpha^*]_j \!=\! \alpha^*_j$ an optimal marginal probability vector, then, the approximation error
\begin{equation*}
\Bigg(F(\Psi^*,\alpha^*;\Pi,\Theta,\gamma) - F(\Psi^{(k)},\alpha^{(k)};\Pi,\Theta,\gamma)\Bigg) \leq \frac{1}{k}\sum_{i=1}^n \sum_{j=1}^m \gamma_i \psi_{i,j}^* \textnormal{log}\Bigg(\frac{\psi_{i,j}^*}{\psi_{i,j}^{(1)}}\Bigg).
\end{equation*}
falls off as a function of the inverse of the iteration count $k$.  Here, the constant factor of the error bound is a Kullback-Leibler divergence between the initial partition matrix $\Psi^{(1)}$ and the global-best partition matrix $\Psi^*$.
\end{itemize}


As shown in proposition 3.1, a Lagrange multiplier $\beta$ is introduced to account for the mutual information constraint.  The effects of $\beta$ are as follows.  As $\beta$ tends to zero, minimizing the Lagrangian is approximately same as minimizing the negative Shannon information.  The information loss associated with the quantization process takes precedence, albeit at the expense of a potentially poor reconstruction.  In this case, there are few state clusters defined by the partition; that is, there are few rows in $\Theta$.  Every state in the high-order transition model $R_\pi$ has an almost uniform chance to map to each state in the low-order model $R_\varphi$.  The alternating updates from proposition 3.1 yield a global minimizer of the value of information, which follows from the convexity of the dual criterion and the Picard-iteration theory of Zangwill \cite{ZangwillWI-book1969a}.

As $\beta$ is increased, the probabilistic partitions become more binary.  Higher probabilities are therefore assigned for a state in $R_\pi$ to map to either a small set of states or a single state in $R_\varphi$.  This is because the effects of the Shannon information term are increasingly ignored in favor of achieving the minimum expected distortion.  The value of information problem given in definition 3.11 therefore approaches the binary aggregation problem from definition 3.7.  An increasing number of clusters are formed by the partition matrix, which increases the number of rows in the weighting matrix $\Theta$ and hence $\Phi$.  When $\beta$ tends to infinity, we obtain a completely binary partition matrix.  We hence recover the least total distortion function given in definition 3.6.  This binary partition can contain as many clusters as states in the high-order model $R_\pi$; that is, no aggregation may be performed, so $R_\varphi$ is typically equal to $R_\pi$.

\begin{figure*}
\centering
   \includegraphics[width=5.95in]{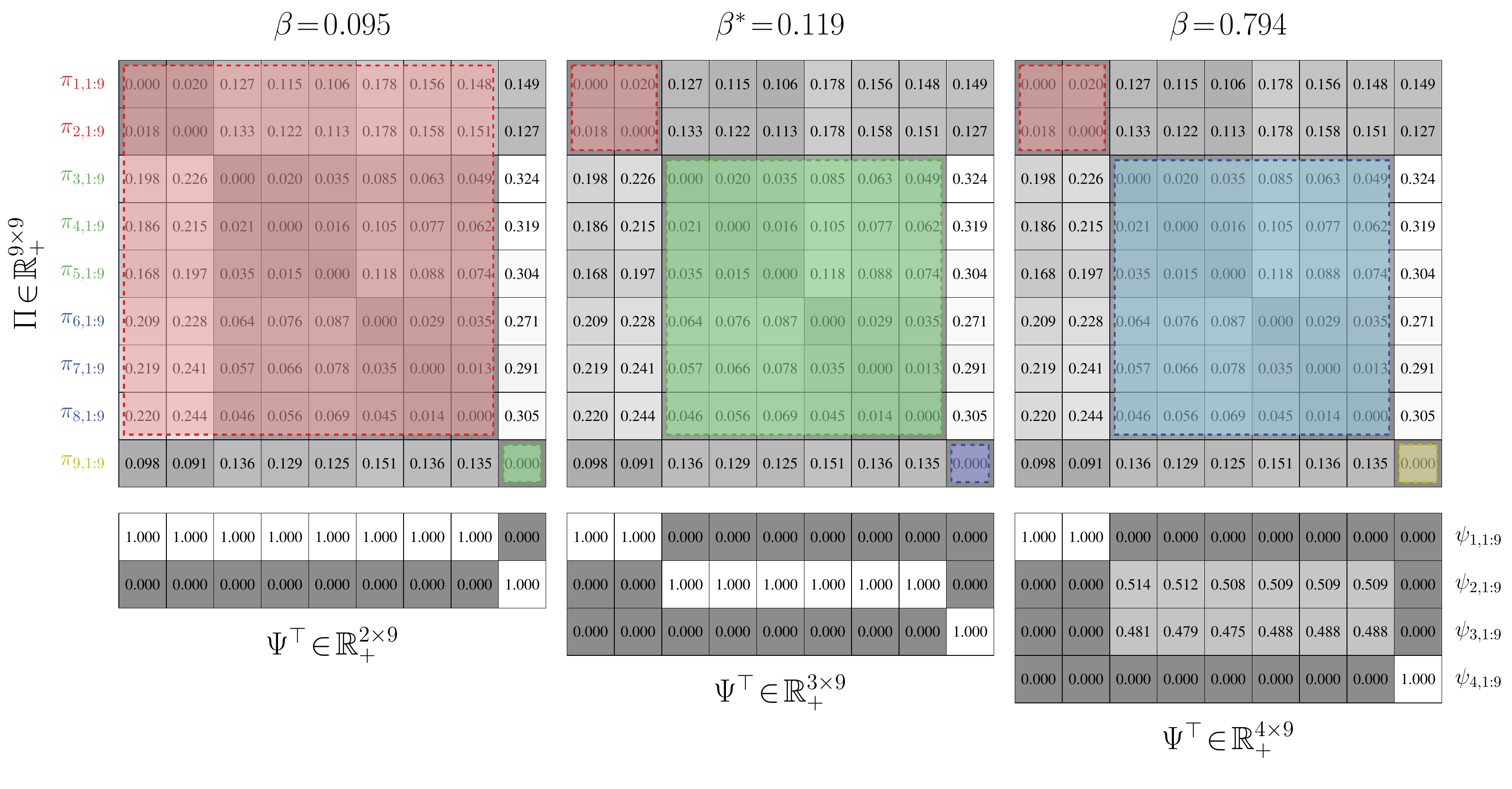}\vspace{-0.225cm}
\caption[]{An illustration of the phase change property when the Lagrange multiplier $\beta$ is increased above three critical values.  For $0 \!\leq\! \beta \!<\! 0.095$, all of the states in the original chain are grouped together.  As $\beta$ is slightly increased beyond this upper threshold, a new state group emerges, as we highlight on the left-hand side of the figure.  For any $0.095 \!\leq\! \beta \!<\! 0.119$ only two state groups are formed.  As $\beta$ is increased to $\beta \!\geq\! 0.119$ and $\beta \!\geq\! 0.794$ three and four state groups are formed, respectively; these results are shown in the middle and right-hand side of the figure.   The `optimal' value of $\beta$, predicted by our perturbation-theory results, is close to $\beta \!=\! 0.119$.  This yields a parsimonious aggregation where the state-groups are compact and well separated.  For $\beta \!\geq\! 0.794$, the original chain is over-partitioned: near-coincident clusters are defined in $\Psi$.  The value of information hence starts to fit more to the noise in the state transitions than to the well-defined state groupings as $\beta$ is increased beyond the next critical point after the `optimal' value. \vspace{-0.4cm}}
\end{figure*}

The number of state clusters in the high-order chain, or, rather, the number of distinct rows of the weight matrix $\Theta$, does not increase continuously as a function of $\beta$.  Instead, it increases only for certain critical values of $\beta$ where a bifurcation occurs in the underlying gradient flow of the Lagrangian.  Critical values of $\beta$ can be explicitly determined when using the negative Kullback-Leibler divergence by looking at the second derivative of the Lagrangian at $\Theta$.\vspace{0.05cm}
\begin{itemize}
\item[] \-\hspace{0.0cm}{\small{\sf{\textbf{Proposition 3.4.}}}} Let $R_\pi \!=\! (V_\pi,E_\pi,\Pi)$ and $R_\varphi \!=\! (V_\varphi,E_\varphi,\Phi)$ be transition models of two Markov chains\\ \noindent over $n$ and $m$ states, respectively, where $m \!<\! n$.  Let $g(\pi_{i,1:n},\vartheta_{i,1:n}) \!=\! \sum_{j=1}^n \gamma_i \pi_{i,j} \textnormal{log}(\pi_{i,j}/\vartheta_{i,j})$, where\\ \noindent $R_\vartheta \!=\! (V_\vartheta,E_\vartheta,\Theta)$ is the joint model.  The following hold
\begin{itemize}
\item[] \-\hspace{0.5cm}(i) The transition matrix $\Phi$ of a low-order Markov chain over states $m$ is given by $\Phi \!=\! \Theta \Psi$, where\\ \noindent $\Theta \!=\! U^\top \Pi$.  Here, $[U]_{i,j} \!=\! \gamma_i \psi_{i,j}/\sum_{k=1}^n \gamma_k \psi_{k,j}$ for the probabilistic partition matrix $[\Psi]_{i,j} \!=\! \psi_{i,j}$ found using the updates in proposition 3.1.
\item[] \-\hspace{0.5cm}(ii) Suppose that we have a low-order chain over $m$ states with a transition matrix $\Phi$ and weight matrix $\Theta$ given by (i).  For some $\beta_0$, suppose $\Theta_{\beta_0}$, the matrix $\Theta$ for that value of $\beta_0$, satisfies the following inequality $d^2/d\epsilon^2\, F(\Psi,\alpha;\Pi,\Theta_{\beta_0} \!+\! \epsilon Q,\gamma)|_{\epsilon = 0} > 0$.  Here, $Q \!\in\! \mathbb{R}_+^{m \times n}$ is a matrix such that $\sum_{k=1}^m q_{k,1:n}^\top q_{k,1:n} \!=\! 1$ and\\ \noindent $\sum_{j=1}^n q_{i,j} \!=\! 0$ $\,\forall i$.  A critical value $\beta_c$, $\beta_c \!=\! \textnormal{min}_{\beta > \beta_0}\,(d^2/d\epsilon^2\, F(\Psi,\alpha;\Pi,\Theta_{\beta} \!+\! \epsilon Q,\gamma)|_{\epsilon = 0} \leq 0)$, occurs\\ \noindent whenever the minimum eigenvalue of the matrix
\begin{equation*}
\textnormal{diag}\Bigg(\sum_{i=1}^n \psi_{k,j}\pi_{i,1:n}/\vartheta_{k,1:n}^2\Bigg) - \beta\Bigg(\sum_{i=1}^n \psi_{k,j}(\pi_{i,1:n}/\vartheta_{k,1:n}^2)(\pi_{i,1:n}/\vartheta_{k,1:n}^2)^\top \Bigg)
\end{equation*}
is zero.  The number of rows in $\Theta$ and columns in $\Psi$ needs to be increased for $\beta \!>\! \beta_c$.
\end{itemize}
\vspace{0.05cm}
\end{itemize}


Proposition 3.4 illustrates a major advantage of the value of information cost function for partitioning Markov chains: the number of states in a low-order model does not need to be manually specified.  It is dictated implicitly by the value of the Lagrange multiplier $\beta$ that captures the effects of favoring information retainment over achieving a minimal expected distortion.  This automatic increase in the number of state groups is depicted in figure 3.5.

Choosing a good value for $\beta$ is crucial for practical problems.  There are a variety of ways to do this.  One such approach entails applying perturbation theory to obtain an upper bound on $\beta$.  More specifically, it is known that measurements of Shannon mutual information are always, on average, improperly estimated when considering finite samples \cite{TrevesA-jour1995a}.  That is, for finitely sized state spaces, the probability distributions that comprise the mutual information expression are approximating, thereby leading to errors that propagate into the aggregation process.  Our approach therefore entails modeling this perturbation error and removing it from the value of information.  This leads to a modified criterion for which a value of $\beta$ can be determined that minimizes the estimation error and better fits to the structure of the transition matrix.  Such values typically correspond to the beginning of an asymptotic region of the original value of information expression where favoring a minimum expected distortion over information loss leads to negligible improvements.\vspace{0.05cm}



\begin{itemize}
\item[] \-\hspace{0.0cm}{\small{\sf{\textbf{Proposition 3.5.}}}} Let $R_\pi \!=\! (V_\pi,E_\pi,\Pi)$ and $R_\varphi \!=\! (V_\varphi,E_\varphi,\Phi)$ be transition models of two Markov chains over\\ \noindent $n$ and $m$ states, respectively, where $m \!<\! n$. $R_\vartheta \!=\! (V_\vartheta,E_\vartheta,\Theta)$ is a joint model, with $m \!+\! n$ states.  The systematic underestimation of the information cost of the Shannon mutual information term in definition 3.11 can be second-order minimized by solving the following optimization problem 
\begin{equation*}
\textnormal{min}_{\Psi \in \mathbb{R}_+^{n \times m},\,\Theta \in \mathbb{R}_+^{m \times n}}\Bigg(\!\!\begin{array}{c}\sum_{j=1}^m \alpha_j \sum_{i=1}^n \psi_{i,j}\textnormal{log}(\psi_{i,j}/\gamma_i)\vspace{0.1cm}\\ +\,\sum_{j=1}^m \sum_{i=1}^n \gamma_i\psi_{i,j}^2/2n\textnormal{log}(2)\alpha_j \end{array}\!\!\Bigg|\!\!\!\begin{array}{c} \sum_{i=1}^n\sum_{j=1}^m \gamma_i \psi_{i,j}g(\pi_{i,1:n},\vartheta_{i,1:n})\!\leq\! r\vspace{0.1cm},\\ \; 0 \!\leq\! \vartheta_{i,k},\psi_{i,k} \!\leq\! 1,\; \sum_{k=1}^m \vartheta_{i,k} \!=\! 1,\; \sum_{k=1}^m \psi_{i,k} \!=\! 1\end{array}\!\!\Bigg)
\end{equation*}
where $\beta \!=\! 2^{\sum_{j=1}^m \alpha_j \sum_{i=1}^n \psi_{i,j}\textnormal{log}(\psi_{i,j}/\gamma_i)}\!/2n$.
\end{itemize}

\noindent This corrected version of the value of information has a rescaled slope compared to the original, where a lower bound on the rescaling is given by $\textnormal{log}(2)/\beta \!-\! \textnormal{log}(2)2^{\sum_{j=1}^m \alpha_j \sum_{i=1}^n \psi_{i,j}\textnormal{log}(\psi_{i,j}/\gamma_i)}/2\beta n$.

\subsection*{\small{\sf{\textbf{3.2.3$\;\;\;$Expected Aggregation Performance}}}}

The preceding theory outlines how Markov chains can be aggregated by trading off between expected distortion and expected relative entropy.  We have shown that global-optimal solutions can be uncovered.  However, we have not bounded the aggregation quality of those solutions for arbitrary problems; such bounds are important for understanding how our approach will behave in general. 

Toward this end, we quantify the relationship between stationary distributions of the original and reduced-order stochastic matrix for nearly-completely-decomposable systems.  Many practical examples of Markov chains are typically nearly-completely-decomposable: there are groups of states that possess similar transition dynamics where the chance to jump between states within the group is higher than states outside of the group.

\vspace{0.05cm}\begin{itemize}
\item[] \-\hspace{0.0cm}{\small{\sf{\textbf{Definition 3.12.}}}} The transition model of a first-order, homogeneous, nearly-completely-decomposable Markov chain is a weighted, directed graph $R_\pi$ given by the three-tuple $(V_\pi,E_\pi,\Pi)$ with the following elements
\begin{itemize}
\item[] \-\hspace{0.5cm}(i) A set of $n$ vertices $V_\pi \!=\! v_\pi^1 \cup \ldots \cup v_\pi^n$ representing the states of the Markov chain.\\
\item[] \-\hspace{0.5cm}(ii) A set of $n \!\times\! n$ edge connections $E_\pi \subset V_\pi \!\times\! V_\pi$ between reachable states in the Markov chain.\\
\item[] \-\hspace{0.5cm}(iii) A stochastic transition matrix $\Pi \!\in\! \mathbb{R}_+^{n \times n}$.  Here, $[\Pi]_{i,j} \!=\! \pi_{i,j}$ represents the non-negative transition probability between states $i$ and $j$.  We impose the constraint that the probability of experiencing a state transition is independent of time.  Moreover, for a block-diagonal matrix $\Pi^*$ with zeros along the diagonal, we have that $\Pi \!=\! \Pi^* \!+\! \varepsilon C$.  Here, $\Pi^* \!\in\! \mathbb{R}_+^{n \times n}$ is a completely-decomposable stochastic matrix with $m$ indecomposable sub-matrix blocks $\Pi^*_i$ of order $n_i$.
\begin{equation*}
\Pi^* = \begin{bmatrix} 
\Pi_1^* & 0       & 0       & \ldots & 0\\
0       & \Pi_2^* & 0       & \ldots & 0\\
0       & 0       & \Pi_3^* & \ldots & 0\\
\vdots  & \vdots  & \vdots  & \ddots & \vdots\\
0       & 0       & 0       & \ldots & \Pi_m^*
\end{bmatrix}
\end{equation*}
Since $\Pi$ and $\Pi^*$ are both stochastic, the matrix $C \!\in\! \mathbb{R}_+^{n \times n}$ must satisfy $\sum_{k=1}^{n_i}c_{p_i,k_i} \!=\! -\sum_{j \neq i}\sum_{q=1}^{n_j} c_{p_i,q_j}$\\ \noindent $\forall p_i$, for blocks $\Pi^*_i$ and $\Pi^*_j$.  That is, $\textnormal{max}_{p_i}(\sum_{k=1}^{n_i} |c_{p_i,k_i}|) \!=\! 1$.  Additionally, the maximum degree of coupling between sub-systems $\Pi^*_i$ and $\Pi^*_j$, given by the perturbation factor $\varepsilon$, must obey $\varepsilon \!=\! \textnormal{max}_i(\sum_{i \neq j} \sum_{q=1}^{n_j} \pi_{p_i,q_j})$.
\end{itemize}\vspace{0.05cm}
\end{itemize}

\begin{itemize}
\item[] \-\hspace{0.0cm}{\small{\sf{\textbf{Proposition 3.6.}}}} Let $R_\pi \!=\! (V_\pi,E_\pi,\Pi)$ be a transition model of a Markov chain with $n$ states, where $\Pi \!\in\! \mathbb{R}_+^{n \times n}$ is nearly completely decomposable into $m$ Markov sub-chains.  
\begin{itemize}
\item[] \-\hspace{0.5cm}(i) The associated low-order stochastic matrix $\Phi \!\in\! \mathbb{R}_+^{m \times m}$ found by solving the value of information\\ \noindent is given by $\varphi_{i,j} = \sum_{p_i=1}^{n_i} \sum_{q_j=1}^{n_j} \pi_{p_i,q_i} \gamma_{p_i}/\sum_{q_i}^{n_i} \gamma_{q_i}$, where $p_i,q_i$ represent state indices $p \!=\! 1,\ldots,n_i$\\ \noindent associated with block $i$, while $q_j$ represents a state index $q \!=\! 1,\ldots,n_j$ into block $j$.  The variable $\gamma_{p_i} \!=\! \gamma_{p_i}(\Pi)$ denotes the invariant-distribution probability of state $p$ in block $i$ of $\Pi$.  
\item[] \-\hspace{0.5cm}(ii) Suppose that $\gamma_{p_i}/\sum_{q_i}^{n_i} \gamma_{q_i} \!=\! v_{p_i}^*(1_i)$ is approximated by the entries of the first left-eigenvector $v^*(1_i)$ for block $i$ of $\Pi^*$.  We then have that
\begin{equation*}
\Bigg\|\gamma\Bigg(\sum_{p_i=1}^{n_i} v_{p_i}^*(1_i) \sum_{q_j=1}^{n_j} \pi_{p_i,q_i}  \Bigg) - \gamma(\Pi)\Psi\Bigg\|_1 \sim O(\varepsilon^2)
\end{equation*}
where the first term is the invariant distribution of the low-order matrix $\gamma(\Phi)$, under the simplifying assumption, and $\Psi \!\in\! \mathbb{R}^{n \times m}_+$ is the probabilistic partition matrix found by solving the value of information.
\end{itemize}
\vspace{0.025cm}
\end{itemize}

\noindent The preceding proposition elucidates the behavior of the value-of-information aggregation results: a reduced Markov chain will have similar long-run dynamics as a projected version of the original Markov chain.  This result is made possible by the work of Simon and Ando \cite{SimonHA-jour1961a}.  They proved that, for nearly-completely-decomposable chains, there are two types of dynamics that influence the stationary distribution: short and long term.  In the short term, each completely-decomposable block evolves almost independently towards a local equilibrium, as if the system was completely decomposable.  In the long run, the entire aggregated chain moves toward the steady state defined by the first left-eigenvalue of the original stochastic matrix.  The equilibrium states attained for each block of the original stochastic matrix are approximately the same as those for the short-run dynamics.

More specifically, the local-equilibrium states for the short-term may dynamics be closely approximated by the steady-state vectors of the sub-systems for the completely decomposable stochastic matrix $\Pi^*$.  The macro-transition probability between blocks $\Pi_i$ and $\Pi_j$ of $\Pi$ remain, in the long term, more or less constant in time and are approximately equal to $\Phi$.  Hence, the elements of the steady-state probability vector $\gamma_{1:n}(1_i)$, where $\gamma_{1:n}(1_i)(\Phi \!-\! I_{n \times n}) \!=\! 0$, are so-called macro-variables that yield good approximations to the steady-state probabilities of being in any one state of block $\Pi_i$.  The so-called micro-variables $\gamma_{p_i}(1_i) \!=\! \gamma_i(1)v_{p_i}^*(1_I)$ are good approximations to the steady-state probabilities $v_{p_i}(1_i)$ of being in any particular state $p$ of block $i$.  That is, as we showed in the previous section, both the macro- and micro-variables have an $\ell_1$-norm error that is a square of the perturbation factor compared to those for the original stochastic matrix.  The aggregated chain will thus possess similar long-term dynamics as the original.

\subsection*{\small{\sf{\textbf{4$\;\;\;$Simulations}}}} \addtocounter{section}{1}

In the previous section, we provided an information-theoretic criterion, the value of information, for quantifying the effects of quantizing stochastic matrices associated with Markov chains.  We also provided a first-order approach for optimizing this criterion, which provides a mechanism for simultaneously partitioning and aggregating chain states.  In this section, we assess the empirical performance of this criterion.  The aims of our simulations are multi-fold.  First, we ascertain how well the value of information reduces the complexity of Markov chains when they possess either simple or complex state-transition dynamics.  We also discuss various facets of the criterion within the context of these results.  We then gauge how well the results for the `optimal' free-parameter value, as predicted via perturbation theory, align with the ground truth.  We also illustrate that using Shannon mutual information, versus Shannon entropy, as a constraint for the expected-distortion objective function, avoids returning coincident partitions.


\subsection*{\small{\sf{\textbf{4.1$\;\;\;$Simulation Protocols}}}}

For each of the examples that follow, we adopted the following simulation protocols for value-of-information-based aggregation.  We initialized the aggregation process with a partition matrix of all ones, $\Psi \!=\! [1]_{9 \times 1}$, signifying that each state belongs to a single group.  This is the global optimal solution of Markov chain aggregation for both the binary- and probabilistic-partition cases.  For the latter case, it coincides with a parameter value $\beta$ of zero for the value of information.  We then found the subsequent critical values of $\beta$ and increased the column count of the partition matrix $\Psi$.  We determined which state group would be further split and modified both the new column and an existing column of $\Psi$ to randomly allocate the appropriate states.  This initialization process bootstraps the quantization for the new cluster and typically achieves convergence in only a few iterations.  It also permits the value of information to reliably track the global minimizer for the binary-partition aggregation problem case as $\beta$ increases.

For certain problems, a priori specifying a fixed amount of partition updates may not permit finding a steady-state solution.  We therefore run the alternating updates until no entries of the partition matrix change across two iterations.



\subsection*{\small{\sf{\textbf{4.2$\;\;\;$Simulation Results and Analyses}}}}

\subsection*{\small{\sf{\textbf{4.2.1$\;\;\;$Value-of-Information Aggregation}}}}

\vspace{0.1cm}{\small{\sf{\textbf{Aggregation Performance.}}}} We establish the performance of value-of-information aggregation through two examples.  The first, shown in figure 4.1, corresponds to a Markov chain with nine states and four state groups with strong intra-group interactions and weak inter-group interactions.  This is a relatively simple aggregation problem.  The second example, presented in figure 4.2, is of a nine-state Markov chain with a single dominant state group and six outlying states with near-equal transition probabilities.  This is a more challenging problem than the first, as the outlying states cannot be reliably combined without adversely impacting the mutual dependence.  In both cases, the transition probabilities were randomly generated through knowledge of a limit distribution $\gamma$.

In figures 4.1 and 4.2, we provide partitions and aggregated Markov chains for four critical values of the free parameter $\beta$.  The `optimal' value of $\beta$, as predicted by our perturbation-theory formulation of the value of information, leads to four and seven state groups for the first and second examples, respectively.  The associated partitions align with an inspection of the dynamics of the stochastic matrix: the partitions separate states that are more likely to transition to each other from those that are not.  The `optimal' aggregated stochastic matrix encodes this behavior well.  The remaining aggregated chains do too for their respective partitions, as they all mimic the interaction dynamics of the original chain for the given state groups.  However, those partitions for `non-optimal' $\beta$s either over- or under-quantize the chain states, which is illustrated by the plot of expected distortion $\mathbb{E}[\mathbb{E}[g(\Pi,\Theta)|\Psi]|\gamma]$ versus the critical values of $\beta$; these plots are given in figure 4.3.  That is, for critical $\beta$s before the `optimal' value, there is a steep drop in the distortion, while the remaining $\beta$s only yield modest decreases.  The `optimal' value of $\beta$ for both examples, in contrast, lies at the `knee' of this curve, which is where the expected-distortion minimization, $\textnormal{min}_\Psi\; \mathbb{E}[\mathbb{E}[g(\Pi,\Theta)|\Psi]|\gamma]$, is roughly balanced against the competing objective of state-mutual-dependence maximization with respect to some bound, $\mathbb{E}[D_\textnormal{KL}(\gamma\|\Psi)] \!\leq\! r$.


\setcounter{figure}{0}

\begin{figure*}
\centering
   \includegraphics[width=6.05in]{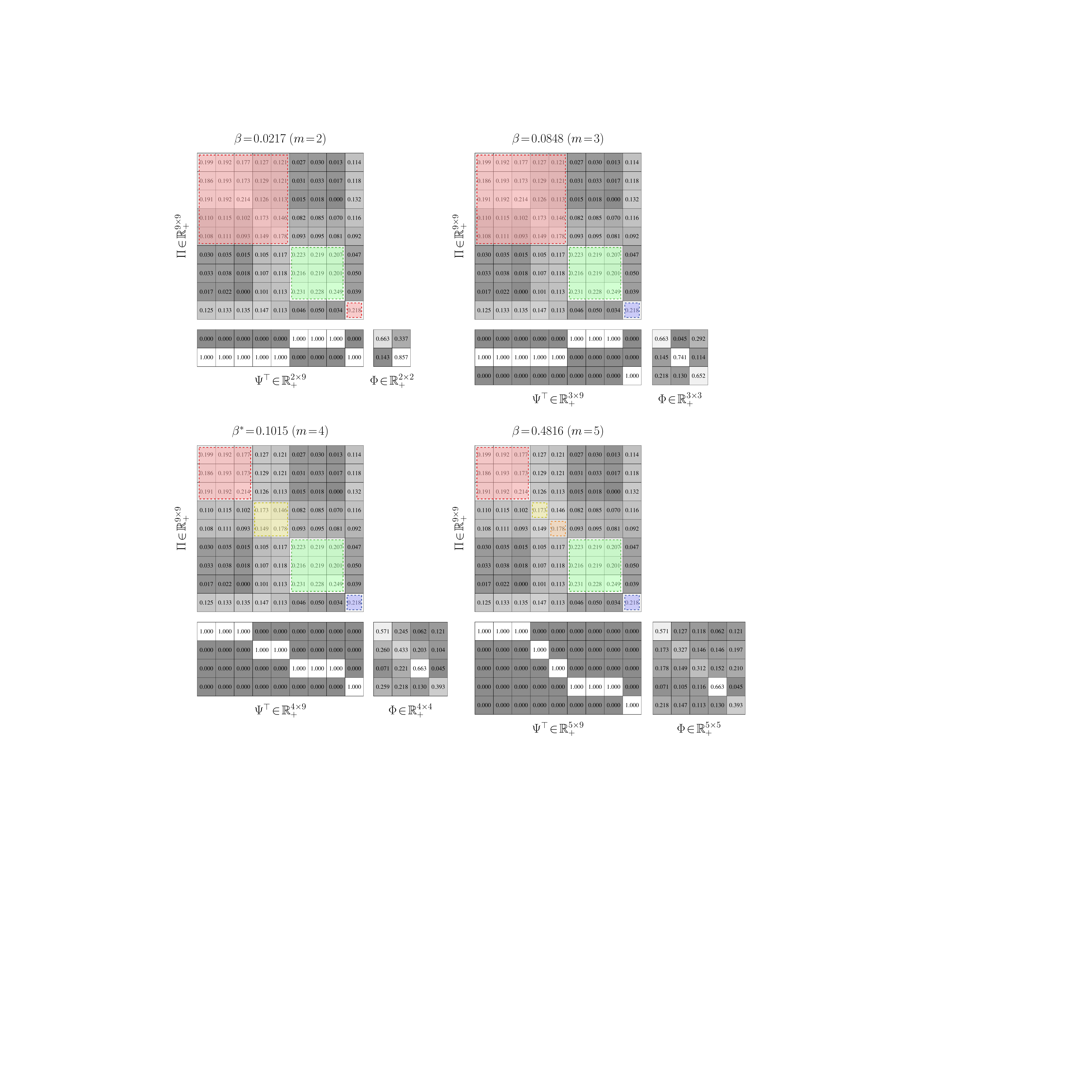}\vspace{-0.075cm}
\caption[]{Value-of-information-based aggregation for a 9-state Markov chain with four discernible state groups.  We show the original stochastic matrix $\Pi \!\in\! \mathbb{R}^{9 \times 9}$ with the partitions $\Psi \!\in\! \mathbb{R}^{m \times 9}$ overlaid for four critical values of $\beta$.  We also show the resulting aggregation $\Theta \!\in\! \mathbb{R}^{m \times m}$, which, in each case, approximately mimics the dynamics of the original stochastic matrix.\vspace{-0.4cm}}
\end{figure*}

\begin{figure*}
\centering
   \includegraphics[width=6.05in]{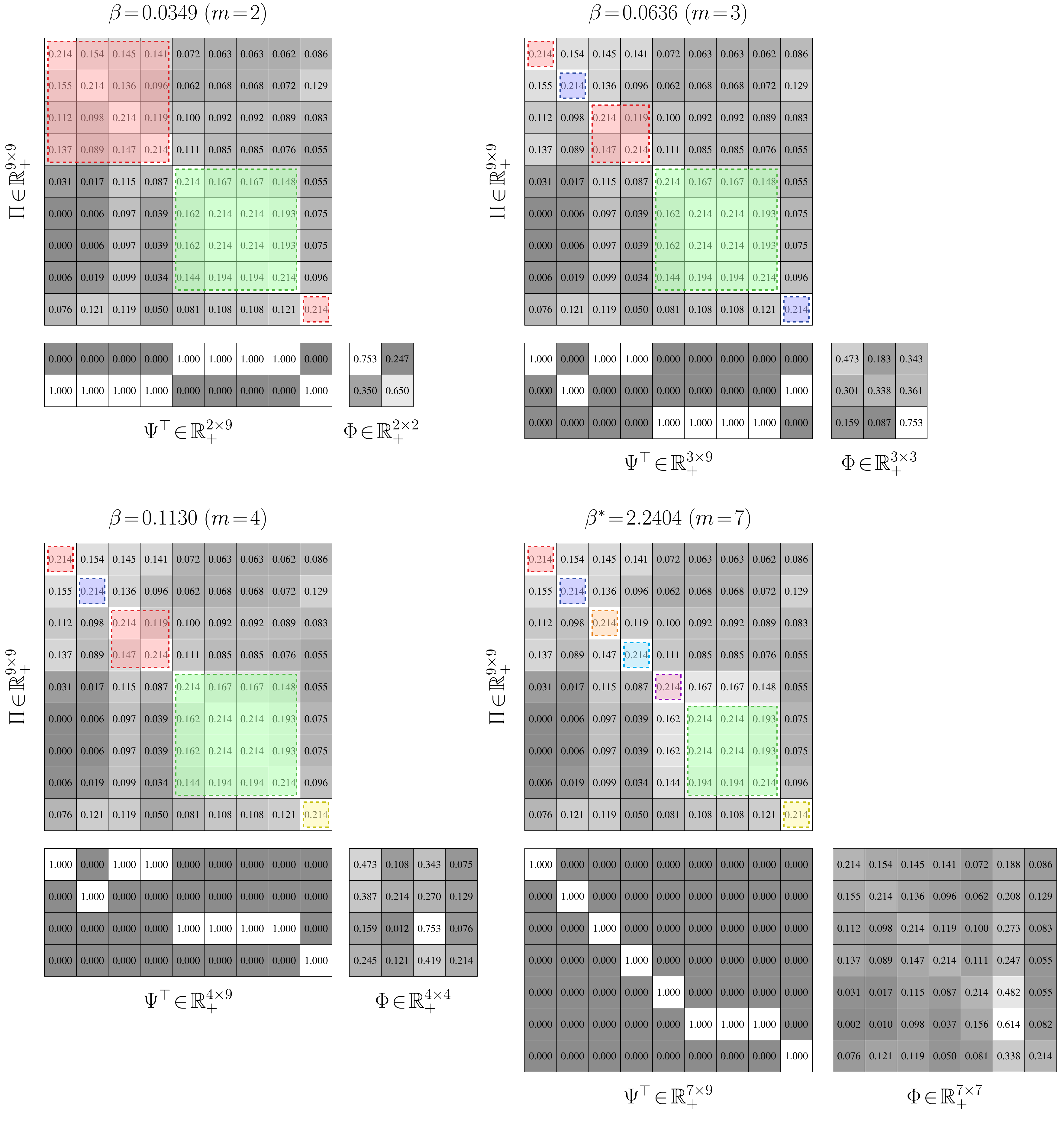}\vspace{-0.275cm}
\caption[]{Value-of-information-based aggregation for a 9-state Markov chain with one discernible state group and 6 outlying states.  We show the original stochastic matrix $\Pi \!\in\! \mathbb{R}^{9 \times 9}$ with the partitions $\Psi \!\in\! \mathbb{R}^{m \times 9}$ overlaid for four critical values of $\beta$.  We also show the resulting aggregation $\Theta \!\in\! \mathbb{R}^{m \times m}$, which, in each case, approximately mimics the dynamics of the original stochastic matrix.\vspace{-0.4cm}}
\end{figure*}

For both examples, we aggregated at critical values of $\beta$ where the number of state groups increases.  We also considered non-critical values of $\beta$ between two phase changes; a thousand Monte Carlo trials were conducted for random $\beta$s.  For each of these trials, the partitions produced between two related critical values were virtually identical, up to a permutation of the rows.  Only minute, arithmetic-error-attributed differences were encountered.  Such results illustrate the validity of our theory: only a finite number of critical values for $\beta$ need to be used for reducing finite-cardinality stochastic matrices.


\vspace{0.1cm}{\small{\sf{\textbf{Convergence.}}}} In figure 4.3, we furnished plots of the decrease in the expected distortion, $\mathbb{E}[\mathbb{E}[g(\Pi,\Theta^{(k)})|\Psi^{(k)}]|\gamma]$ across each iteration $k \!=\! 1,2,\ldots$  This provides a means of gauging the per-iteration solution improvement and hence convergence.  We also provided plots of the partition matrix cross-entropy for consecutive iterations, $\mathbb{E}[-\textnormal{log}(\Psi^{(k-1)})|\Psi^{(k)}]$.  The partition cross-entropy is a bounded measure of change between consecutive partitions and captures how greatly the partition changed across a single update.  Taken together, they offer alternate views of the aggregation improvement during intermediate stages of the dynamics reduction process.  In either example, the average of these quantities across the Monte Carlo trials exhibits a nearly-linear decrease in their respective quantities before plateauing, regardless the critical value of $\beta$.  This finding suggests rapid convergence to the global solution, which was anticipated from our convergence analysis.  That is, due to how we initialize the partitions, we are roughly ensuring that they are in close proximity to the next global optima $\Psi^*$, up to a permutation of the rows.  The $D_\textnormal{KL}(\Psi^*\|\Psi^{(1)})$ term in the approximation error bound dominates over the $k^{-1}$ term in this situation and hence that few changes in $\Psi$ are needed.

To assess the convergence stability of the aggregation process, we performed a thousand Monte Carlo trials on both examples.  In only a very small fraction of the trials did the partitions deviate from those presented in figures 4.1 and 4.2 by more than a simple permutation.  Such occurrences were largely due to a degenerate initialization of a new partition column whereby no states would be associated with that new state group.  Imposing a constraint that a new group must contain at least a single state fixed this issue and led to consistent partitions being produced.  The expectation-maximization-based procedure for solving the value of information was then able to discover global optima well in just a few iterations; the optima often were binary partitions like those presented in figures 4.1 and 4.2.

\begin{figure*}
   \hspace{1.25cm}\begin{tabular}{c c}
      \includegraphics[width=2.4in]{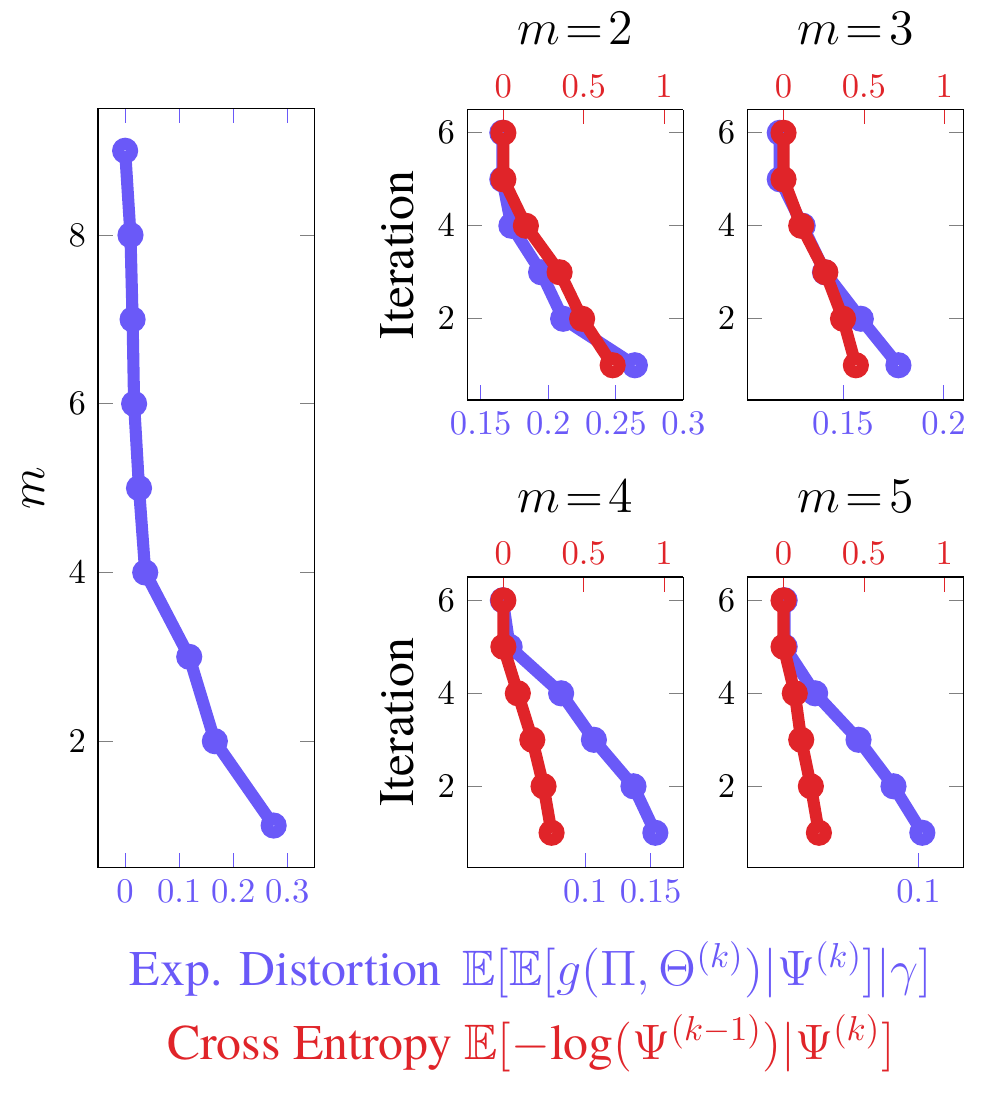} & \;\;\;\includegraphics[width=2.4in]{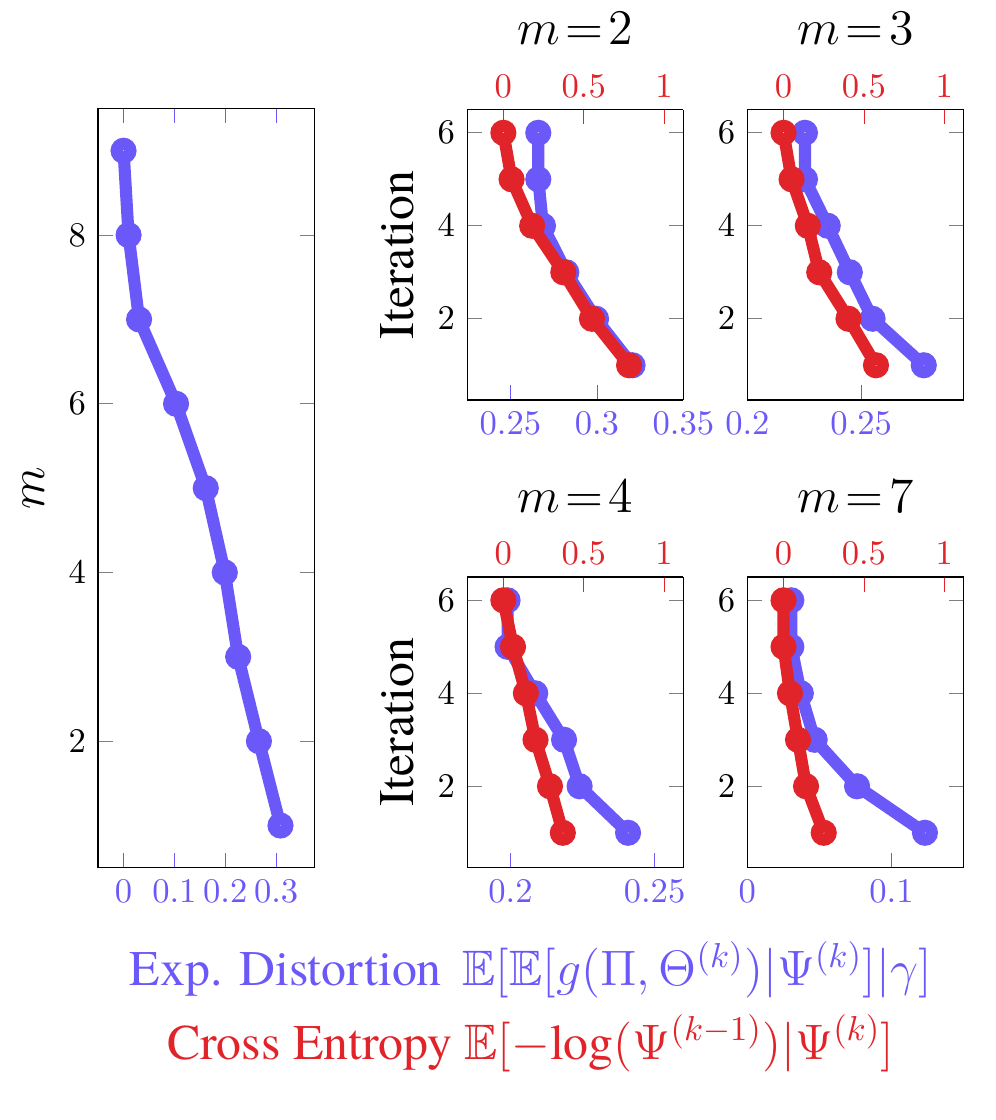}\vspace{-0.275cm}\\
   {\footnotesize (a)} & \;\;\;{\footnotesize (b)}\vspace{-0.175cm}\\
   \end{tabular}
\caption[]{Expected distortion (blue curves) and cross-entropy (red curves) plots for the aggregation results in figure 4.1, shown in (a), and figure 4.2, shown in (b).  For both (a) and (b), the large, left-most plot shows the expected distortion as a function of the number of state groups $m$ after convergence has been achieved.  The `knee' of the plot in (a) is given for $m \!=\! 4$, while for (b) it is at $m \!=\! 7$.  These `knee' regions correspond to the `optimal' number of state groups as returned by our perturbation-theoretic criterion.  They indicate where there are diminishing returns for including more aggregated state groups.  The smaller four plots in (a) and (b) highlight the change in the expected distortion and cross-entropy as a function of the number of alternating-update iterations.  These plots highlight a rapid stabilization of the update process.\vspace{-0.4cm}}
\end{figure*}

\vspace{0.1cm}{\small{\sf{\textbf{Avoiding coincident Partitions.}}}} The results for the preceding examples indicate that the Shannon-information constraint, $\mathbb{E}[D_\textnormal{KL}(\gamma\|\Psi^{(k)})] \!\leq\! r$, has the potential to yield non-coincident partitions.  We now demonstrate using two additional Markov chains that using a Shannon-entropy penalty, $\mathbb{E}[-\textnormal{log}(\Psi^{(k)})] \!\leq\! r$, is more likely to return partitions with duplicated rows.  This over-inflates the state-group cardinality, leading to aggregations with redundant details.

Both of these examples are for Markov chains with nine states.  The first example, shown in the top left-hand corner of figure 4.4, contains three state groups with a high chance to both jump to states in different groups and jump to states within a group.  Moreover, many of the rows in the matrix are the same.  We anticipate that coincident partitions will easily materialize due to these properties.  The second example is given in the bottom left-hand corner of figure 4.4.  It contains two state groups with weakly-interacting intra-group dynamics and strong inter-group dynamics.  Each group has highly distinct transition probabilities.  We hence expect that returning coincident partitions will be more difficult than in the first case.  As before, the transition probabilities for each matrix were randomly generated through knowledge of a limit distribution.

Partitions for to nine state groups are presented in figures 4.4.  The partitions in the middle column of figure 4.4 are the results for the Shannon-information constraint, while those in right column are for the Shannon-entropy constraint.  The Shannon-information case quantizes the data in the manner we would expect for both examples: each state is, more or less, assigned to its own group so that the original stochastic matrix is exactly recovered.  There are hence no degenerate clusters.  For the Shannon-entropy case, three coincident clusters formed for the first problem and this value of $\beta$.  Two states from the first state group were incorrectly viewed as being equivalent.  Two states from the second group and three states from the third group were also improperly treated, leading to further coincident partitions.  Four degenerate clusters thus emerged and the original stochastic matrix could not be recovered; the Kullback-Leibler distortion for this value of $\beta$ and other $\beta$s illustrate this.  For the second problem, every state in the second state group was considered equal.  Seven degenerate groups were thus created, leading to a stochastic matrix with a very different invariant distribution and hence long-term dynamics than the original.

For these examples, we considered the same number of groups as states to highlight the severity of the coincident partition issue when using an uncertainty constraint.  Coincident clusters were also observed when the group count was below the number of states.  




\subsection*{\small{\sf{\textbf{4.2.2$\;\;\;$Value-of-Information Aggregation Discussions}}}}

We have illustrated that the value of information criterion provides an effective mechanism for dynamics reduction of Markov chains for these examples.  Consistently stable partitions of the transition probabilities are produced by optimizing this criterion.  Such partitions induce reduced-order chains that do not have duplicate state groups and are often parsimonious representations.  We have additionally demonstrated that only a finite number of free-parameter values need to be considered for this purpose, the `optimal' value of which can be discerned in a data-driven fashion.

\begin{figure*}
\centering
   \includegraphics[width=5.35in]{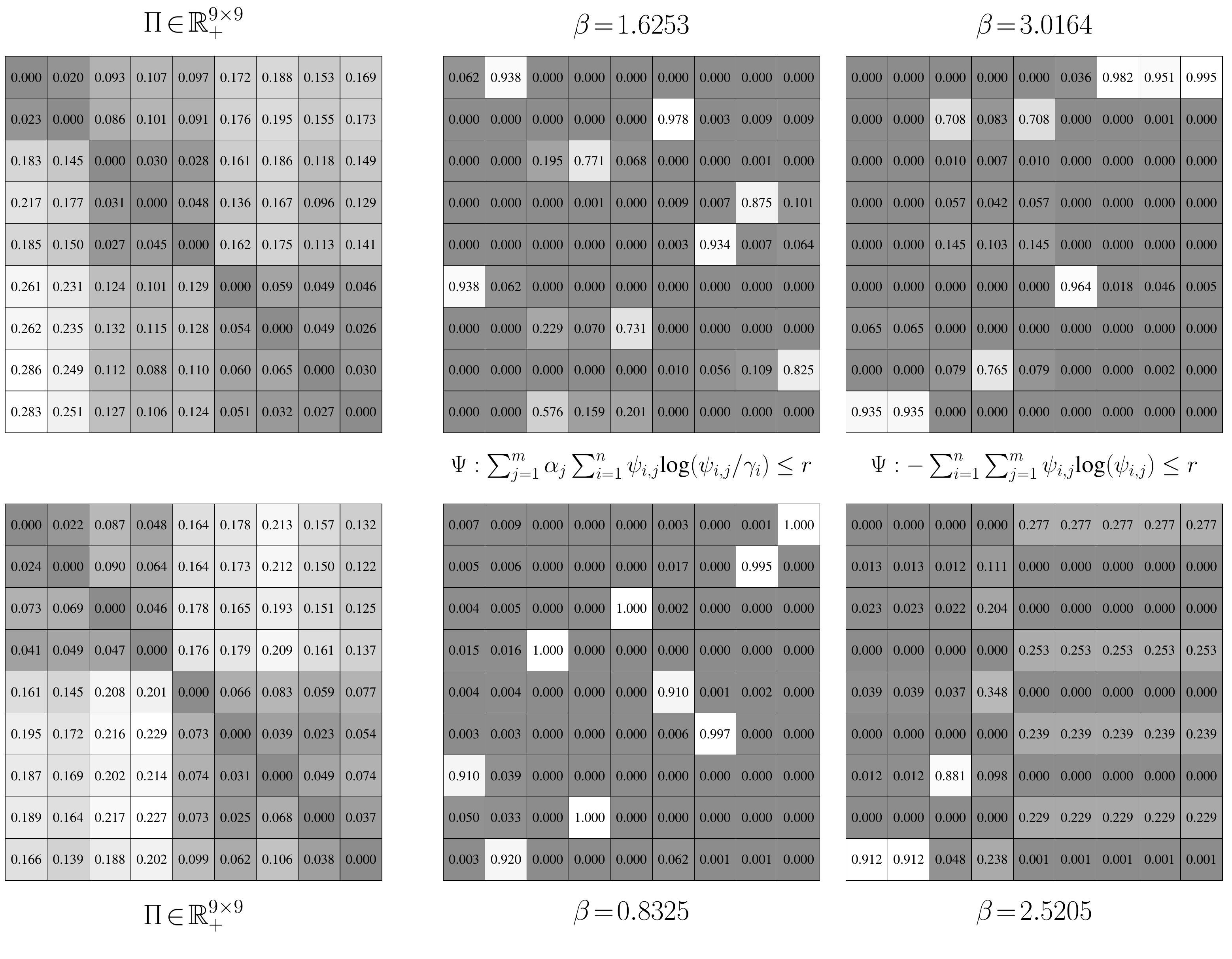}\vspace{-0.275cm}
\caption[]{Value-of-information-based aggregation for a 9-state Markov chain with three discernible state groups (top row) and two discernible state groups (bottom row).  The left-most column shows the original stochastic matrices $\Pi \!\in\! \mathbb{R}^{9 \times 9}$.  The middle column gives the partitions $\Psi \!\in\! \mathbb{R}^{9 \times 9}$ found when using a Shannon mutual information constraint for the expected-distortion objective function.  The right-most column gives the partitions $\Psi \!\in\! \mathbb{R}^{9 \times 9}$ found when using a Shannon entropy constraint for the expected-distortion objective function.  When using Shannon entropy, several columns of the partition matrix are duplicated for high values of $\beta$, leading to an incorrect aggregation of states.\vspace{-0.4cm}}
\end{figure*}

\vspace{0.1cm}{\small{\sf{\textbf{Aggregation Performance: Binary Partitions.}}}} In the previous section, we relaxed the binary-valued constraint on the partition matrices to avoid exactly solving a potentially computationally-intractable problem.  However, our aggregation results for the first two examples indicate that either binary or nearly-binary partition matrices may still be returned when solving the value of information.  The reason for this is the interplay between the expected distortion and the Shannon-mutual-information constraint: while the latter does not explicitly preclude their formation, the former naturally favors binary partitions.

More specifically, non-binary partitions will always have less Shannon mutual information than binary partitions.  This is is because the conditional entropy of the states in the original and aggregated chains increases more quickly than the marginal entropy, which is due to the additional uncertainty in the non-binary partitions.  Hence, for a given upper bound on the Shannon information, if a binary partition can be formed for that bound, then a corresponding non-binary one can also be formed.  The minimization of the distortion term, however, impedes the formation of non-binary partitions.  In the binary case, provided that the partition reflects the underlying structure of the transitions, only related probability vectors will be compared to each other.  Vastly different rows and columns of the stochastic and joint stochastic matrices will not factor into the expected distortion, since the state-assignment probability will be zero if the partition encodes well the underlying transition structure.  Making highly non-binary state-group assignments can raise the expected distortion: the Kullback-Leibler divergence between two, possibly very distinct, probability vectors may be multiplied by a non-zero state-assignment probability.  


This behavior contrasts with the use of a conditional Shannon entropy equality constraint on the entries of the partition matrix.  Such a constraint directly imposes that the partition matrix should have a given amount of uncertainty, potentially at the expense of a worse distortion.  Non-binary partitions hence can be more readily constructed.

\vspace{0.1cm}{\small{\sf{\textbf{Aggregation Performance: `Optimal' State Group Count.}}}} We considered a perturbation-theoretic approach for determining the `optimal' number of state groups.  The approach operates on the assumption that, for finitely-sized stochastic matrices, there is an error in estimating the marginal distribution of the original states.  This poor estimate leads to a systematic error in evaluating the Shannon-information term, which we quantified in a second-order sense.  In the case of binary partitions, a second-order correction of this error introduces a penalty in the value of information for using more aggregated state groups than can be resolved for a particular finitely-sized state space.  Values for the free parameter were returned that, for our examples, aligned well with a balance between the expected distortion of the aggregation and the mutual dependence between states in the original and aggregated chains.

As shown in our experiments, the value of information monotonically decreases for an increasing number of state groups.  The second-order-corrected version shares this trait, as it is a slope-rescaled version of the original value of information.  Ideally, we would like to further transform this slope-scaled value-of-information curve so that it possesses an extremum where both terms of the objective function are balanced.  This would lend further credence to the notion that such a free parameter value, and hence the number of state groups, for any stochastic matrix is `optimal'.  In our upcoming work, we will demonstrate how to perform this transformation.  We will show that the value of information can be written, in some cases, as a variational problem involving two Shannon-mutual-information terms.  Applying the same perturbation-theoretic arguments to this version of the value of information reveals that the corrected criterion is monotonically increasing up to a certain point, after which it is monotonically non-increasing and often is strictly decreasing.  This inflexion point corresponds to the `optimal' parameter value determined here.  This value minimizes the mutual dependence between the original and aggregated states while simultaneously retraining as much information about the original transition dynamics as possible.

\subsection*{\small{\sf{\textbf{5$\;\;\;$Conclusions}}}} 

In this paper, we have provided a novel, two-part information-theoretic approach for aggregating Markov chains.  The first part of our approach is aimed at assessing the distortion between original- and reduced-order chains according to the Kullback-Leibler divergence between rows of their corresponding stochastic matrices.  The discrete nature of the graphs precludes the direct comparison of the transition probabilities according to this divergence, which motivated our construction of a joint transition model.  This joint model encodes all of the information of the reduced-order Markov chain and is of the proper size to compare against the original Markov chain.  The second part of our approach addresses how to combine states in the original chain, according to the Kullback-Leibler divergence, by solving a value-of-information criterion.  This criterion aggregates states together if doing so reduces the total expected distortion between the low- and high-order chains and simultaneously maximizes the bounded mutual dependence between states in the high- and low-order chains.  It thus attempts to find a low-order Markov chain that most resembles the global and local transition structure of the original, high-order Markov chain.

The value of information provides a principled and optimal trade-off between the quality of the aggregation, as measured by the total expected distortion, and the complexity of it, as measured by the state mutual dependence according to Shannon mutual information.  The complexity constraint has dual roles.  The first is that it explicitly dictates the number of states in the low-order chain.  We proved that changing the value of a variable associated with this constraint causes the aggregation process to undergo phase transitions where new groupings emerge by splitting an existing state cluster.  The second role of the constraint is that it relaxes the condition that the partition matrices must be strictly binary.  This relaxation permits the formulation of an efficient procedure for approximately solving the aggregation problem.  While the same effect could be achieved with a Shannon entropy constraint, it has the tendency to yield coincident partitions.  This over-inflates the number of states in the reduced-order model.

We applied our approach to a series of Markov chains.  Our simulation results showed that our value-of-information scheme achieved equal or better performance compared to a range of different aggregation techniques.  A practical advantage of our methodology is that we have derived a data-driven expression for the `optimal' value for a parameter associated with the state mutual dependence constraint.  This expression was based upon correcting the underestimation of the Shannon information term for finitely-sized stochastic matrices.  Employing this expression fits the partitions more to the structure of the data than to the noise, ensuring that it tends not to over-cluster states in the original chain.  It also frees investigators from having to supply the number of state groups.  Many existing aggregation approaches rely on the manual specification of the state-group count, in contrast; a reasonable number of state groups may not be immediately evident for certain problems, which complicates their effective application. 

As we noted at the beginning of the paper, our emphasis is on understanding the effects of the value of information when it is applied to resolve the exploration-exploitation dilemma in reinforcement learning.  In particular, we seek to address the question of if the value of information is implicitly aggregating the Markov chains underlying the Markov decision process during exploration.  Toward this end, our next step will be to show that hidden Markov models can be reduced in a value-of-information-based manner.  Much like our work here, this will entail defining a joint model that allows for comparisons between pairs of hidden Markov models with different state spaces.  We will need to construct the joint model so that it is Markovian, which will ensure that the theory in this paper applies with few modifications.  Following this, for the Markov-decision-process case, we will need to show that a lumpable partition of the state space can be defined by the value of information, where the partition is bounded by the state-transition and cost effects.  States in the same partition will then be viewed as a single state in a reduced-order, aggregated Markov chain.  An aggregated Markov decision process with average cost on the aggregated Markov chain can then be obtained.  As a part of this effort, we will also quantify the local-neighborhood performance difference between this aggregated Markov decision process and the optimal one.

\renewcommand*{\bibfont}{\raggedright}
\renewcommand\bibsection{\subsection*{\small{\sf{\textbf{References}}}}}
{\singlespacing\fontsize{9.75}{10}\selectfont \bibliography{sledgebib} \bibliographystyle{IEEEtran}}

\begin{thebibliography}{10}
\providecommand{\url}[1]{#1}
\csname url@samestyle\endcsname
\providecommand{\newblock}{\relax}
\providecommand{\bibinfo}[2]{#2}
\providecommand{\BIBentrySTDinterwordspacing}{\spaceskip=0pt\relax}
\providecommand{\BIBentryALTinterwordstretchfactor}{4}
\providecommand{\BIBentryALTinterwordspacing}{\spaceskip=\fontdimen2\font plus
\BIBentryALTinterwordstretchfactor\fontdimen3\font minus
  \fontdimen4\font\relax}
\providecommand{\BIBforeignlanguage}[2]{{%
\expandafter\ifx\csname l@#1\endcsname\relax
\typeout{** WARNING: IEEEtran.bst: No hyphenation pattern has been}%
\typeout{** loaded for the language `#1'. Using the pattern for}%
\typeout{** the default language instead.}%
\else
\language=\csname l@#1\endcsname
\fi
#2}}
\providecommand{\BIBdecl}{\relax}
\BIBdecl

\bibitem{ArrudaEF-conf2009a}
\BIBentryALTinterwordspacing
E.~F. Arruda and M.~D. Fragoso, ``Standard dynamic programming applied to time
  aggregated {Markov} decision processes,'' in \emph{Proceedings of the IEEE
  Conference on Decision and Control {(CDC)}}, Shanghai, China, December 15-18
  2009, pp. 2576--2580. Available:
  \url{http://dx.doi.org/10.1109/CDC.2009.5400692}
\BIBentrySTDinterwordspacing

\bibitem{AldhaheriRW-jour1991a}
\BIBentryALTinterwordspacing
R.~W. Aldhaheri and H.~K. Khalil, ``Aggregation of the policy iteration method
  for nearly completely decomposable {Markov} chains,'' \emph{IEEE Transactions
  on Automatic Control}, vol.~36, no.~2, pp. 178--187, 1991. Available:
  \url{http://dx.doi.org/10.1109/9.67293}
\BIBentrySTDinterwordspacing

\bibitem{RenZ-jour2005a}
\BIBentryALTinterwordspacing
Z.~Ren and B.~H. Krogh, ``Markov decision processes with fractional costs,''
  \emph{IEEE Transactions on Automatic Control}, vol.~50, no.~5, pp. 646--650,
  2005. Available: \url{http://dx.doi.org/10.1109/TAC.2005.846520}
\BIBentrySTDinterwordspacing

\bibitem{SunT-jour2007a}
\BIBentryALTinterwordspacing
T.~Sun, Q.~Zhao, and P.~B. Luh, ``Incremental value iteration for
  time-aggregated {Markov} decision processes,'' \emph{IEEE Transactions on
  Automatic Control}, vol.~52, no.~11, pp. 2177--2182, 2007. Available:
  \url{http://dx.doi.org/10.1109/TAC.2007.908359}
\BIBentrySTDinterwordspacing

\bibitem{JiaQS-jour2011a}
\BIBentryALTinterwordspacing
Q.~S. Jia, ``On state aggregation to approximate complex value functions in
  large-scale {Markov} decision processes,'' \emph{IEEE Transactions on
  Automatic Control}, vol.~56, no.~2, pp. 333--334, 2011. Available:
  \url{http://dx.doi.org/10.1109/TAC.2010.2052697}
\BIBentrySTDinterwordspacing

\bibitem{AokiM-jour1978a}
\BIBentryALTinterwordspacing
M.~Aoki, ``Some approximation methods for estimation and control of large scale
  systems,'' \emph{IEEE Transactions on Automatic Control}, vol.~23, no.~2, pp.
  173--182, 1978. Available: \url{http://dx.doi.org/10.1109/TAC.1978.1101705}
\BIBentrySTDinterwordspacing

\bibitem{PrincipeJC-book2010a}
J.~C. {Pr\'{i}ncipe}, \emph{Information Theoretic Learning}.\hskip 1em plus
  0.5em minus 0.4em\relax New York City, NY, USA: Springer-Verlag, 2010.

\bibitem{RachedZ-jour2004a}
\BIBentryALTinterwordspacing
Z.~Rached, F.~Alajaji, and L.~L. Campbell, ``The {Kullback-Leibler} divergence
  rate between {Markov} sources,'' \emph{IEEE Transactions on Information
  Theory}, vol.~50, no.~5, pp. 917--921, 2004. Available:
  \url{http://dx.doi.org/10.1109/TIT.2004.826687}
\BIBentrySTDinterwordspacing

\bibitem{DonskerMD-jour1975a}
\BIBentryALTinterwordspacing
M.~D. Donsker and S.~R.~S. Varadhan, ``Asymptotic evaluation of certain
  {Markov} process expectations for large time {I.}'' \emph{Communications of
  Pure and Applied Mathematics}, vol.~28, no.~1, pp. 1--47, 1975. Available:
  \url{http://dx.doi.org/10.1002/cpa.3160280102}
\BIBentrySTDinterwordspacing

\bibitem{DonskerMD-jour1975b}
\BIBentryALTinterwordspacing
------, ``Asymptotic evaluation of certain {Markov} process expectations for
  large time {II.}'' \emph{Communications of Pure and Applied Mathematics},
  vol.~28, no.~2, pp. 279--301, 1975. Available:
  \url{http://dx.doi.org/10.1002/cpa.3160280206}
\BIBentrySTDinterwordspacing

\bibitem{DengK-jour2011a}
\BIBentryALTinterwordspacing
K.~Deng, P.~G. Mehta, and S.~P. Meyn, ``Optimal {Kullback-Leibler} aggregation
  via spectral theory of {Markov} chains,'' \emph{IEEE Transactions on
  Automatic Control}, vol.~56, no.~12, pp. 2793--2808, 2011. Available:
  \url{http://dx.doi.org/10.1109/TAC.2011.2141350}
\BIBentrySTDinterwordspacing

\bibitem{GeigerBC-jour2015a}
\BIBentryALTinterwordspacing
B.~C. Geiger, T.~Petrov, G.~Kubin, and H.~Koeppl, ``Optimal {Kullback-Leibler}
  aggregation via inforamtion bottleneck,'' \emph{IEEE Transactions on
  Automatic Control}, vol.~60, no.~4, pp. 1010--1022, 2015. Available:
  \url{http://dx.doi.org/10.1109/TAC.2014.2364971}
\BIBentrySTDinterwordspacing

\bibitem{SledgeIJ-jour2017a}
\BIBentryALTinterwordspacing
I.~J. Sledge and J.~C. {Pr\'{i}ncipe}, ``An analysis of the value of
  information when exploring stochastic, discrete multi-armed bandits,''
  \emph{Entropy}, vol.~20, no.~3, pp. 155(1--34), 2018. Available:
  \url{http://dx.doi.org/10.3390/e20030155}
\BIBentrySTDinterwordspacing

\bibitem{SledgeIJ-jour2017b}
\BIBentryALTinterwordspacing
------, ``Analysis of agent expertise in {Ms. Pac-Man} using
  value-of-information-based policies,'' \emph{IEEE Transactions on
  Computational Intelligence and Artificial Intelligence in Games}, 2018,
  (accepted, in press). Available:
  \url{http://dx.doi.org/10.1109/TG.2018.2808201}
\BIBentrySTDinterwordspacing

\bibitem{SledgeIJ-jour2017c}
\BIBentryALTinterwordspacing
I.~J. Sledge, M.~S. Emigh, and J.~C. {Pr\'{i}ncipe}, ``Guided policy
  exploration for {Markov} decision processes using an uncertainty-based
  value-of-information criterion,'' \emph{IEEE Transactions on Neural Networks
  and Learning Systems}, vol.~29, no.~6, pp. 2080--2098, 2018. Available:
  \url{http://dx.doi.org/10.1109/TNNLS.2018.2812709}
\BIBentrySTDinterwordspacing

\bibitem{StratonovichRL-jour1965a}
R.~L. Stratonovich, ``On value of information,'' \emph{Izvestiya of USSR
  Academy of Sciences, Technical Cybernetics}, vol.~5, no.~1, pp. 3--12, 1965.

\bibitem{StratonovichRL-jour1966a}
R.~L. Stratonovich and B.~A. Grishanin, ``Value of information when an
  estimated random variable is hidden,'' \emph{Izvestiya of USSR Academy of
  Sciences, Technical Cybernetics}, vol.~6, no.~1, pp. 3--15, 1966.

\bibitem{CoverTM-book2006a}
T.~M. Cover and J.~A. Thomas, \emph{Elements of Information Theory}.\hskip 1em
  plus 0.5em minus 0.4em\relax New York, NY, USA: John Wiley and Sons, 2006.

\bibitem{vonNeumannJ-book1953a}
J.~{von Neumann} and O.~Morgenstern, \emph{Theory of Games and Economic
  Behavior}.\hskip 1em plus 0.5em minus 0.4em\relax Princeton, NJ, USA:
  Princeton University Press, 1953.

\bibitem{SimonHA-jour1961a}
\BIBentryALTinterwordspacing
H.~A. Simon and A.~Ando, ``Aggregation of variables in dynamic systems,''
  \emph{Econometrica}, vol.~29, no.~2, pp. 111--138, 1961. Available:
  \url{http://dx.doi.org/10.2307/1909285}
\BIBentrySTDinterwordspacing

\bibitem{CourtoisPJ-jour1975a}
\BIBentryALTinterwordspacing
P.~J. Courtois, ``Error analysis in nearly-completely decomposable stochastic
  systems,'' \emph{Econometrica}, vol.~43, no.~4, pp. 691--709, 1975.
  Available: \url{http://dx.doi.org/10.2307/1913078}
\BIBentrySTDinterwordspacing

\bibitem{PervozvanskiiAA-jour1974a}
\BIBentryALTinterwordspacing
A.~A. Pervozvanskii and I.~N. Smirnov, ``Stationary-state evaluation for a
  complex system with slowly varying couplings,'' \emph{Kibernetika}, vol.~3,
  no.~1, pp. 45--51, 1974. Available:
  \url{http://dx.doi.org/10.1007/BF01071538}
\BIBentrySTDinterwordspacing

\bibitem{GaitsgoriVG-jour1975a}
\BIBentryALTinterwordspacing
V.~G. Gaitsgori and A.~A. Pervozvanskii, ``Aggregation of states in a {Markov}
  chain with weak interaction,'' \emph{Kibernetika}, vol.~4, no.~1, pp. 91--98,
  1975. Available: \url{http://dx.doi.org/10.1007/BF01069471}
\BIBentrySTDinterwordspacing

\bibitem{TeneketzisD-conf1980a}
\BIBentryALTinterwordspacing
D.~Teneketzis, S.~H. Javid, and B.~Sridhar, ``Control of weakly-coupled
  {Markov} chains,'' in \emph{Proceedings of the IEEE Conference on Decision
  and Control {(CDC)}}, Albuquerque, NM, USA, December 10-12 1980, pp.
  137--142. Available: \url{http://dx.doi.org/10.1109/CDC.1980.272033}
\BIBentrySTDinterwordspacing

\bibitem{DelebecqueF-jour1981a}
\BIBentryALTinterwordspacing
F.~Delebecque and J.-P. Quadrat, ``Optimal control of {Markov} chains admitting
  strong and weak interactions,'' \emph{Automatica}, vol.~17, no.~2, pp.
  281--296, 1981. Available:
  \url{http://dx.doi.org/10.1016/0005-1098(81)90047-9}
\BIBentrySTDinterwordspacing

\bibitem{ZhangQ-jour1997a}
\BIBentryALTinterwordspacing
Q.~Zhang, G.~Yin, and E.~K. Boukas, ``Controlled {Markov} chains with weak and
  strong interactions,'' \emph{Journal of Optimization Theory and
  Applications}, vol.~94, no.~1, pp. 169--194, 1997. Available:
  \url{http://dx.doi.org/10.1023/A:1022667905086}
\BIBentrySTDinterwordspacing

\bibitem{CourtoisPJ-book1977a}
P.~J. Courtois, \emph{Decomposability, Instabilities, and Saturation in
  Multiprogramming Systems}.\hskip 1em plus 0.5em minus 0.4em\relax New York,
  NY, USA: Academic Press, 1977.

\bibitem{AldhaheriRW-conf1989a}
\BIBentryALTinterwordspacing
R.~W. Aldhaheri and H.~K. Khalil, ``Aggregation and optimal control of nearly
  completely decomposable {Markov} chains,'' in \emph{Proceedings of the IEEE
  Conference on Decision and Control {(CDC)}}, Tampa, FL, USA, December 13-15
  1989, pp. 1277--1282. Available:
  \url{http://dx.doi.org/10.1109/CDC.1989.70343}
\BIBentrySTDinterwordspacing

\bibitem{KotsalisG-conf2003a}
\BIBentryALTinterwordspacing
G.~Kotsalis and M.~Dahleh, ``Model reduction of irreducible {Markov} chains,''
  in \emph{Proceedings of the IEEE Conference on Decision and Control {(CDC)}},
  Maui, HI, USA, December 9-12 2003, pp. 5727--5728. Available:
  \url{http://dx.doi.org/10.1109/CDC.2003.1271917}
\BIBentrySTDinterwordspacing

\bibitem{DeyS-jour2000a}
\BIBentryALTinterwordspacing
S.~Dey, ``Reduced-complexity filtering for partially observed nearly completely
  decomposable {Markov} chains,'' \emph{IEEE Transactions on Signal
  Processing}, vol.~48, no.~12, pp. 3334--3344, 2000. Available:
  \url{http://dx.doi.org/10.1109/78.886997}
\BIBentrySTDinterwordspacing

\bibitem{VantilborghH-jour1985a}
\BIBentryALTinterwordspacing
H.~Vantilborgh, ``Aggregation with an error of {$O(\epsilon^2)$},''
  \emph{Journal of the {ACM}}, vol.~32, no.~1, pp. 162--190, 1985. Available:
  \url{http://dx.doi.org/10.1145/2455.214107}
\BIBentrySTDinterwordspacing

\bibitem{CaoWL-jour1985a}
\BIBentryALTinterwordspacing
W.-L. Cao and W.~J. Stewart, ``Iterative aggregation/disaggregation techniques
  for nearly uncoupled {Markov} chains,'' \emph{Journal of the {ACM}}, vol.~32,
  no.~3, pp. 702--719, 1985. Available:
  \url{http://dx.doi.org/10.1145/3828.214137}
\BIBentrySTDinterwordspacing

\bibitem{KouryJR-jour1984a}
\BIBentryALTinterwordspacing
J.~R. Koury, D.~F. {McAllister}, and W.~J. Stewart, ``Iterative methods for
  computing stationary distributions of nearly completely decomposable {Markov}
  chains,'' \emph{SIAM Journal on Algebraic and Discrete Methods}, vol.~5,
  no.~2, pp. 164--186, 1984. Available: \url{http://dx.doi.org/10.1137/0605019}
\BIBentrySTDinterwordspacing

\bibitem{BarkerGP-jour1986a}
\BIBentryALTinterwordspacing
G.~P. Barker and R.~J. Piemmons, ``Convergent iterations for computing
  stationary distributions of {Markov} chains,'' \emph{SIAM Journal on
  Algebraic and Discrete Methods}, vol.~7, no.~3, pp. 390--398, 1986.
  Available: \url{http://dx.doi.org/10.1137/0607044}
\BIBentrySTDinterwordspacing

\bibitem{DayarT-jour1996a}
\BIBentryALTinterwordspacing
T.~Dayar and W.~J. Stewart, ``On the effects of using the
  {Grassman-Taksar-Heyman} method in iterative aggregation-disaggregation,''
  \emph{SIAM Journal on Scientific Computing}, vol.~17, no.~1, pp. 287--303,
  1996. Available: \url{http://dx.doi.org/10.1137/0917021}
\BIBentrySTDinterwordspacing

\bibitem{PhillipsR-jour1981a}
\BIBentryALTinterwordspacing
R.~Phillips and P.~Kokotovic, ``A singular perturbation approach to modeling
  and control of markov chains,'' \emph{IEEE Transactions on Automatic
  Control}, vol.~26, no.~5, pp. 1087--1094, 1981. Available:
  \url{http://dx.doi.org/10.1109/TAC.1981.1102780}
\BIBentrySTDinterwordspacing

\bibitem{PeponidesG-jour1983a}
\BIBentryALTinterwordspacing
G.~Peponides and P.~Kokotovic, ``Weak connections, time scales, and aggregation
  of nonlinear systems,'' \emph{IEEE Transactions on Automatic Control},
  vol.~28, no.~6, pp. 729--735, 1983. Available:
  \url{http://dx.doi.org/10.1109/TAC.1983.1103300}
\BIBentrySTDinterwordspacing

\bibitem{ChowJ-jour1985a}
\BIBentryALTinterwordspacing
J.~Chow and P.~Kokotovic, ``Time scale modeling of sparse dynamic networks,''
  \emph{IEEE Transactions on Automatic Control}, vol.~30, no.~8, pp. 714--722,
  1985. Available: \url{http://dx.doi.org/10.1109/TAC.1985.1104055}
\BIBentrySTDinterwordspacing

\bibitem{FilarJA-jour2001a}
\BIBentryALTinterwordspacing
J.~A. Filar, V.~Gaitsgory, and A.~B. Haurie, ``Control of singularly perturbed
  hybrid stochastic systems,'' \emph{IEEE Transactions on Automatic Control},
  vol.~46, no.~2, pp. 179--180, 2001. Available:
  \url{http://dx.doi.org/10.1109/9.905686}
\BIBentrySTDinterwordspacing

\bibitem{DengK-conf2009a}
\BIBentryALTinterwordspacing
K.~Deng, Y.~Sun, P.~G. Mehta, and S.~P. Meyn, ``An information-theoretic
  framework to aggregate a {Markov} chain,'' in \emph{Proceedings of the
  American Control Conference {(ACC)}}, St. Louis, MO, USA, June 10-12 2009,
  pp. 731--736. Available: \url{http://dx.doi.org/10.1109/ACC.2009.5160607}
\BIBentrySTDinterwordspacing

\bibitem{DengK-conf2012a}
\BIBentryALTinterwordspacing
K.~Deng and D.~Huang, ``Model reduction of {Markov} chains via low-rank
  approximation,'' in \emph{Proceedings of the American Control Conference
  {(ACC)}}, {Montr\'{e}al}, Canada, June 27-29 2012, pp. 2651--2656. Available:
  \url{http://dx.doi.org/10.1109/ACC.2012.6314781}
\BIBentrySTDinterwordspacing

\bibitem{VidyasagarM-conf2010a}
\BIBentryALTinterwordspacing
M.~Vidyasagar, ``Reduced-order modeling of {Markov} and hidden {Markov}
  processes via aggregation,'' in \emph{Proceedings of the IEEE Conference on
  Decision and Control {(CDC)}}, Atlanta, GA, USA, December 15-17 2010, pp.
  1810--1815. Available: \url{http://dx.doi.org/10.1109/CDC.2010.5717206}
\BIBentrySTDinterwordspacing

\bibitem{VidyasagarM-jour2012a}
\BIBentryALTinterwordspacing
------, ``A metric between probability distributions on finite sets of
  different cardinalities and applications to order reduction,'' \emph{IEEE
  Transactions on Automatic Control}, vol.~57, no.~10, pp. 2464--2477, 2012.
  Available: \url{http://dx.doi.org/10.1109/TAC.2012.2188423}
\BIBentrySTDinterwordspacing

\bibitem{ZangwillWI-book1969a}
W.~I. Zangwill, \emph{Nonlinear Programming: {A} Unified Approach}.\hskip 1em
  plus 0.5em minus 0.4em\relax Upper Saddle River, NJ, USA: Prentice-Hall,
  1969.

\bibitem{TrevesA-jour1995a}
\BIBentryALTinterwordspacing
A.~Treves and S.~Panzeri, ``The upward bias in meausres of information derived
  from limited data samples,'' \emph{Neural Computation}, vol.~7, no.~2, pp.
  399--407, 1995. Available: \url{http://dx.doi.org/10.1162/neco.1995.7.2.399}
\BIBentrySTDinterwordspacing

\bibitem{KatoT-book1966a}
T.~Kato, \emph{Perturbation Theory for Linear Operators}.\hskip 1em plus 0.5em
  minus 0.4em\relax New York, NY, USA: Springer-Verlag, 1966.

\end{thebibliography}

\clearpage\newpage

\subsection*{\small{\sf{\textbf{A$\;\;\;$Appendix}}}}

\begin{itemize}
\item[] \-\hspace{0.0cm}{\small{\sf{\textbf{Proposition 3.1.}}}} For a transition model $R_\pi \!=\! (V_\pi,E_\pi,\Pi)$ over $n$ states and a joint model $R_\vartheta \!=\! (V_\vartheta,E_\vartheta,\Theta)$\\ \noindent and $m \!+\! n$ states, the Lagrangian of the relevant terms for the minimization problem given in definition 3.11 is $F(\Psi,\alpha;\Pi,\Theta,\gamma) = \mathbb{E}[\mathbb{E}[g(\Pi,\Theta)|\Psi]|\gamma] \!-\! \mathbb{E}[D_\textnormal{KL}(\gamma\|\Psi)]/\beta$, or, rather,
\begin{equation*}
F(\Psi,\alpha;\Pi,\Theta,\gamma) = \Bigg(\sum_{i=1}^n\sum_{j=1}^m \gamma_i \psi_{i,j}g(\pi_{i,1:n},\vartheta_{j,1:n})\Bigg) - \frac{1}{\beta}\Bigg(\sum_{j=1}^m \alpha_j \sum_{i=1}^n \psi_{i,j}\textnormal{log}(\psi_{i,j}/\gamma_i)\Bigg).
\end{equation*}
Here, $\beta \!\geq\! 0$ is a Lagrange multiplier that emerges from the Shannon mutual information constraint in the value of information.

Probabilistic partitions $[\Psi]_{i,j} \!=\! \psi_{i,j}$, which are local solutions of $\nabla_{\vartheta_{j,1:n}}F(\Psi,\alpha;\Pi,\Theta,\gamma) \!=\! 0$, can be found by the following expectation-maximization-based alternating updates
\begin{equation*}
\alpha_j \leftarrow \sum_{i=1}^n \gamma_i\psi_{i,j},\;\;\;\;\; \psi_{i,j} \leftarrow \!\Bigg(\alpha_j e^{-\beta g(\pi_{i,1:n},\vartheta_{j,1:n})}\!\Bigg)\!\Bigg/\!\Bigg(\sum_{p=1}^m \alpha_p e^{-\beta g(\pi_{i,1:n},\vartheta_{p,1:n})}\Bigg),
\end{equation*}
which are iterated until convergence.\vspace{0.05cm}
\end{itemize}

\begin{itemize}
\item[] \-\hspace{0.5cm}{\small{\sf{\textbf{Proof:}}}} We can convert the constrained value-of-information into an unconstrained problem using the theory of Lagrange multipliers.  There are five different constraints for which we need to account,
\begin{multline*}
F(\Psi,\alpha;\Pi,\Theta,\gamma) = \Bigg(\sum_{i=1}^n\sum_{j=1}^m \gamma_i \psi_{i,j}g(\pi_{i,1:n},\vartheta_{j,1:n})\Bigg) - \frac{1}{\beta}\Bigg(\sum_{j=1}^m \alpha_j \sum_{i=1}^n \psi_{i,j}\textnormal{log}(\psi_{i,j}/\gamma_i)\Bigg)\;\;\;\;\;\;\;\;\;\;\;\;\;\;\\
+ \Bigg(\sum_{i=1}^n \sum_{j=1}^m \kappa_i (1 \!-\! \psi_{i,j})\Bigg) + \Bigg(\sum_{i=1}^n \sum_{j=1}^m \omega_i(1 \!-\! \vartheta_{j,i})\Bigg) - \Bigg(\sum_{i=1}^n \sum_{j=1}^m \xi_i \psi_{i,j}\Bigg) - \Bigg(\sum_{i=1}^n \sum_{j=1}^m \mu_i \vartheta_{j,i}\Bigg)
\end{multline*}
The first constraint corresponds to the Shannon mutual information term.  The remaining constraints ensure that entries from $\Theta$ and $\Psi$ correspond to valid probabilities.

We can derive the update for the probabilistic partition matrix $\Psi$ by differentiating the Lagrangian and setting it to zero
\begin{equation*}
\frac{\partial}{\partial \psi_{i,j}} F(\Psi,\alpha;\Pi,\Theta,\gamma) = \Bigg(\gamma_i\textnormal{log}\Bigg(\frac{\psi_{i,j}}{\alpha_j}\Bigg) - \frac{\gamma_ig(\pi_{i,1:n},\vartheta_{j,1:n}) - \beta\gamma_i}{\beta}\Bigg) + \kappa_i + \omega_i - \xi_i - \mu_i = 0.
\end{equation*}
Solving for $\psi_{i,j}$ yields $\psi_{i,j} = \sum_{r=1}^n \gamma_r \psi_{r,j}e^{-\beta g(\pi_{i,1:n},\vartheta_{j,1:n})}/\sum_{r=1}^n\sum_{p=1}^m \gamma_r \psi_{r,j}e^{-\beta g(\pi_{i,1:n},\vartheta_{p,1:n})}$, where $\kappa_i$\\ \noindent and $\omega_i$ have been selected such that $\sum_{i=1}^n \psi_{i,j} \!=\! 1$ and $\sum_{i=1}^n \vartheta_{i,j} \!=\! 1$, respectively.  It is apparent that the update for the marginal probabilities $\alpha$ is encoded in this update for $\Psi$.

We now can show that the update for the marginal probabilities is optimal.  For a fixed probabilistic partition matrix $\Psi$, the following inequality
\begin{equation*}
\sum_{i=1}^n\sum_{j=1}^m \gamma_i \psi_{i,j}\textnormal{log}\Bigg(\frac{\psi_{i,j}}{\alpha_j}\Bigg) \geq \sum_{i=1}^n \sum_{j=1}^m \gamma_i \psi_{i,j}\textnormal{log}\Bigg(\frac{\psi_{i,j}}{\sum_{r=1}^n \gamma_r \psi_{j,r}}\Bigg)
\end{equation*}
holds with equality if and only if $\alpha_j \!=\! \sum_{i=1}^n \gamma_i \psi_{i,j}$.  This result is a consequence of applying the divergence\\ \noindent inequality to $\sum_{i=1}^n\sum_{j=1}^m \gamma_i \psi_{i,j}\textnormal{log}(\psi_{i,j}/\alpha_j) - \sum_{i=1}^n \sum_{j=1}^m \gamma_i \psi_{i,j}\textnormal{log}(\psi_{i,j}/\sum_{r=1}^n \gamma_r \psi_{j,r})$.  That is,
\begin{align*}
\sum_{i=1}^n\sum_{j=1}^m \gamma_i \psi_{i,j}\textnormal{log}\Bigg(\frac{\psi_{i,j}}{\alpha_j}\Bigg) - \sum_{i=1}^n \sum_{j=1}^m \gamma_i \psi_{i,j}\textnormal{log}\Bigg(\frac{\psi_{i,j}}{\sum_{r=1}^n \gamma_r \psi_{j,r}}\Bigg) &= \sum_{i=1}^n \sum_{j=1}^m \gamma_i \psi_{i,j}\textnormal{log}\Bigg(\frac{\sum_{r=1}^n \gamma_r \psi_{j,r} }{\alpha_j}\Bigg)\\
 &\geq \Bigg(\sum_{i=1}^n\sum_{j=1}^m \gamma_i\psi_{i,j}\Bigg) - \Bigg(\sum_{j=1}^m \alpha_j\Bigg)\\
 &\geq 0,
\end{align*}
where equality to zero is only obtained if and only if $\alpha_j \!=\! \sum_{i=1}^n \gamma_i \psi_{i,j}$.  This implies that, for a fixed $\Psi$, the\\ \noindent update for $\alpha$ globally solves the problem $\textnormal{min}_\alpha\; F(\Psi,\alpha;\Pi,\Theta,\gamma)$, establishing its optimality.

We now demonstrate that the probabilistic partition update is optimal.  For a fixed marginal probability vector $\alpha$, the following inequality
\begin{align*}
-\sum_{i=1}^n \gamma_i \textnormal{log}\Bigg(\sum_{j=1}^m \alpha_j e^{-\beta g(\pi_{i,1:n},\vartheta_{j,1:n})} \Bigg) &\leq \sum_{i=1}^n\sum_{j=1}^m \gamma_i \psi_{i,j}\textnormal{log}\Bigg(\frac{\psi_{i,j}}{\alpha_j}\Bigg) - \frac{1}{\beta}\Bigg(\sum_{i=1}^n\sum_{j=1}^m \psi_{i,j} g(\pi_{i,1:n},\vartheta_{j,1:n})\Bigg)\\
&\leq \sum_{i=1}^n\sum_{j=1}^m \gamma_i \psi_{i,j}\textnormal{log}\Bigg(\frac{\psi_{i,j} e^{-\beta g(\pi_{i,1:n},\vartheta_{j,1:n})}}{\alpha_j}\Bigg)
\end{align*}
holds with equality if and only if $\psi_{i,j} \!=\! \alpha_j e^{-\beta g(\pi_{i,1:n},\vartheta_{j,1:n})}/\sum_{p=1}^m \alpha_p e^{-\beta g(\pi_{i,1:n},\vartheta_{p,1:n})}$.  This follows from showing that
\begin{equation*}
\!\!\!\!\!\!\!\!\!\!\!\!\!\!\!\!\!\!\!\!\!\!\!\!\!\!\!\!\!\!\!\!\!\!\!\!\!\!\!\!\!\!\!\!\!\!\!\!\!\!\!\!\!\!\!\!\!\!\!\!\!\!\!\!\!\!\!\!\!\!\!\!\!\!\!\!\!\!\!\!\!\!\!\!\!\!\!\!\!\!\!\!\!\!\!\!\!\!\!\!\!\!\!\!\!\!\!\!\!\!\!\!\!\!\!\!\!\!\!\!\!\!\!\!\!\!\!\!\!\!\!\!\!\!\!\!\!\!\!\!\!\!\!\!\!\!\!\!\!\!\!\!\!\!\!\!\!\!\!\!\!\!\!\begin{array}{l}
\displaystyle\sum_{i=1}^n\sum_{j=1}^m \gamma_i \psi_{i,j}\textnormal{log}\Bigg(\frac{\psi_{i,j} e^{-\beta g(\pi_{i,1:n},\vartheta_{j,1:n})}}{\alpha_j}\Bigg) \hfill
\end{array}\hfill
\end{equation*}
\begin{equation*}
\begin{array}{l}
\displaystyle = \sum_{i=1}^n\sum_{j=1}^m \gamma_i \psi_{i,j}\textnormal{log}\Bigg(\frac{\psi_{i,j}\sum_{p=1}^m \alpha_p e^{-\beta g(\pi_{i,1:n},\vartheta_{p,1:n})}}{ \alpha_j e^{-\beta g(\pi_{i,1:n},\vartheta_{j,1:n})} \sum_{p=1}^m \alpha_p e^{-\beta g(\pi_{i,1:n},\vartheta_{p,1:n})} }\Bigg)\\
\displaystyle = \sum_{i=1}^n\sum_{j=1}^m \gamma_i \psi_{i,j}\textnormal{log}\Bigg(\frac{\psi_{i,j}}{ \alpha_j e^{-\beta g(\pi_{i,1:n},\vartheta_{j,1:n})}/\sum_{p=1}^m \alpha_p e^{-\beta g(\pi_{i,1:n},\vartheta_{p,1:n})}}\Bigg)\\
\;\;\;\;\;\;\;\;\;\;\;\;\;\;\;\;\;\displaystyle +\; \sum_{i=1}^n\sum_{j=1}^m \gamma_i\psi_{i,j}\textnormal{log}\Bigg(\frac{1}{\sum_{p=1}^m \alpha_p e^{-\beta g(\pi_{i,1:n},\vartheta_{p,1:n})}}\Bigg)\\
\displaystyle \geq 0 + \sum_{i=1}^n\sum_{j=1}^m \gamma_i\psi_{i,j}\textnormal{log}\Bigg(\frac{1}{\sum_{p=1}^m \alpha_p e^{-\beta g(\pi_{i,1:n},\vartheta_{p,1:n})}}\Bigg)\\
\displaystyle \geq \sum_{i=1}^n \gamma_i \textnormal{log}\Bigg(\frac{1}{\sum_{p=1}^m \alpha_p e^{-\beta g(\pi_{i,1:n},\vartheta_{p,1:n})}}\Bigg)
\end{array}
\end{equation*}
Here, we used the divergence inequality in the second-to-last step.  As well, we have that the last step can be\\ \noindent written as $\sum_{i=1}^n \gamma_i \textnormal{log}(1/\sum_{p=1}^m \alpha_p e^{-\beta g(\pi_{i,1:n},\vartheta_{p,1:n})}) = -\sum_{i=1}^n \gamma_i \textnormal{log}(\sum_{j=1}^n \alpha_j e^{-\beta g(\pi_{i,1:n},\vartheta_{j,1:n})})$, where\\ \noindent the right-hand side is the desired expression.  Substituting the update for $\Psi$ in the original inequality leads to equivalency.  Hence, for a fixed $\alpha$, the update for $\Psi$ globally solves $\textnormal{min}_\Psi\; F(\Psi,\alpha;\Pi,\Theta,\gamma)$. \hfill $\blacksquare$
\end{itemize}\vspace{0.15cm}

\begin{itemize}
\item[] \-\hspace{0.0cm}{\small{\sf{\textbf{Proposition 3.2.}}}} Let $R_\pi \!=\! (V_\pi,E_\pi,\Pi)$ and $R_\varphi \!=\! (V_\varphi,E_\varphi,\Phi)$ be transition models of two Markov chains\\ \noindent over $n$ and $m$ states, respectively, where $m \!<\! n$.  If $[\Psi^*]_{i,j} \!=\! \psi_{i,j}^*$ is an optimal probabilistic partition and\\ \noindent $[\alpha^*]_j \!=\! \alpha^*_j$ an optimal marginal probability vector, then, for the updates in proposition 3.1, we have that: 
\begin{itemize}
\item[] \-\hspace{0.5cm}(i) The approximation error is non-negative
\begin{equation*}
\Bigg(F(\Psi^{(k)},\alpha^{(k)};\Pi,\Theta,\gamma) - F(\Psi^*,\alpha^*;\Pi,\Theta,\gamma)\Bigg) = \sum_{i=1}^n \gamma_i \textnormal{log}\Bigg(\frac{\sum_{j=1}^m \alpha^*_je^{-\beta g(\pi_{i,1:n},\vartheta_{j,1:n})}}{\sum_{j=1}^m \alpha^{(k)}_je^{-\beta g(\pi_{i,1:n},\vartheta_{j,1:n})}}\Bigg) \geq 0.
\end{equation*}
\item[] \-\hspace{0.5cm}(ii) The modified free energy monotonically decreases $F(\Psi^{(k)},\alpha^{(k)};\Pi,\Theta,\gamma) \!\geq\! F(\Psi^{(k+1)},\alpha^{(k+1)};\Pi,\Theta,\gamma)$ across all iterations $k$.
\item[] \-\hspace{0.5cm}(iii) For any $K \!\geq\! 1$, we have the following bound for the sum of approximation errors
\begin{equation*}
\Bigg(\sum_{k=1}^K F(\Psi^{(k)},\alpha^{(k)};\Pi,\Theta,\gamma) - F(\Psi^*,\alpha^*;\Pi,\Theta,\gamma)\Bigg) \leq \sum_{i=1}^n \sum_{j=1}^m \gamma_i \psi_{i,j}^* \textnormal{log}\Bigg(\frac{\psi_{i,j}^*}{\psi_{i,j}^{(1)}}\Bigg).
\end{equation*}
\end{itemize}
Here, $F(\Psi,\alpha;\Pi,\Theta,\gamma) \!=\! \mathbb{E}[\mathbb{E}[g(\Pi,\Theta)|\Psi]|\gamma] \!-\! \mathbb{E}[D_\textnormal{KL}(\gamma\|\Psi)]/\beta$ is the Lagrangian.\vspace{0.05cm}
\end{itemize}

\begin{itemize}
\item[] \-\hspace{0.5cm}{\small{\sf{\textbf{Proof:}}}} Parts (i) and (ii) follow immediately from proposition 3.1.  For part (iii), we have the following equality expressions for iterations $k$ and $k \!+\! 1$ of the expectation-maximization updates
\begin{align*}
\sum_{i=1}^n \sum_{j=1}^m \gamma_i \psi^*_{j,i}\textnormal{log}\Bigg(\frac{\psi_{i,j}^{(k+1)}}{\psi_{i,j}^{(k)}}\Bigg) &= \sum_{i=1}^n \sum_{j=1}^m \gamma_i \psi^*_{j,i} \textnormal{log}\Bigg( \frac{\alpha_j^{(k)}e^{-g(\pi_{i,1:n},\vartheta_{j,1:n})/\beta}}{ \psi_{i,j}^{(k)} \sum_{p=1}^m\alpha_p^{(k)}e^{-g(\pi_{i,1:n},\vartheta_{p,1:n})/\beta} } \Bigg)\\
  &= \Bigg(F(\Psi^{(k)},\alpha^{(k)};\Pi,\Theta,\gamma) - F(\Psi^*,\alpha^*;\Pi,\Theta,\gamma)\Bigg)\\
  &\;\;\;\;\;\;\;\;\;\;\;\;\;\;\;\;\; + \sum_{i=1}^n \sum_{j=1}^m \gamma_i \psi_{i,j}^* \textnormal{log}\Bigg(\frac{1}{\psi_{i,j}^{(k)}}\frac{\alpha_j^{(k)}}{\alpha_j^*}\frac{\alpha_j^*e^{-g(\pi_{i,1:n},\vartheta_{j,1:n})/\beta}}{\sum_{p=1}^m\alpha_p^*e^{-g(\pi_{i,1:n},\vartheta_{p,1:n})/\beta}} \Bigg).
\end{align*}
We obtain that the last term in the last inequality expression is non-negative, since, from part (i), we have that $\psi_{i,j}^* = \alpha_j^* e^{-g(\pi_{i,1:n},\vartheta_{j,1:n})/\beta}/\sum_{p=1}^m \alpha_p^* e^{-g(\pi_{i,1:n},\vartheta_{p,1:n})/\beta}$.  We thus have that.
\begin{align*}
\sum_{i=1}^n \sum_{j=1}^m \gamma_i \psi^*_{j,i}\textnormal{log}\Bigg(\frac{\psi_{i,j}^{(k+1)}}{\psi_{i,j}^{(k)}}\Bigg) &= \Bigg(F(\Psi^{(k)},\alpha^{(k)};\Pi,\Theta,\gamma) - F(\Psi^*,\alpha^*;\Pi,\Theta,\gamma)\Bigg) + \sum_{i=1}^n \sum_{j=1}^m \gamma_i \psi_{i,j}^* \textnormal{log}\Bigg(\frac{\psi_{i,j}^*}{\psi_{i,j}^{(k)}}\frac{\alpha_j^{(k)}}{\alpha_j^*}\Bigg)\\
  &\geq \Bigg(F(\Psi^{(k)},\alpha^{(k)};\Pi,\Theta,\gamma) - F(\Psi^*,\alpha^*;\Pi,\Theta,\gamma)\Bigg) + \sum_{i=1}^n \sum_{j=1}^m \gamma_i\psi_{i,j}^*\Bigg(1 - \frac{\alpha_j^*\psi_{i,j}^{(k)}}{\alpha_j^{(k)}\psi_{i,j}^*}\Bigg).
\end{align*}
Since $\alpha_j^*\psi_{i,j}^{(k)}/\alpha_j^{(k)}\psi_{i,j}^* = 1$, it can be seen that $\sum_{i=1}^n \sum_{j=1}^m \gamma_i\psi_{i,j}^*(1 - \alpha_j^*\psi_{i,j}^{(k)}/\alpha_j^{(k)}\psi_{i,j}^*) = 0$.  We hence recover the following inequality
\begin{equation*}
\sum_{i=1}^n \sum_{j=1}^m \gamma_i \psi^*_{j,i}\textnormal{log}\Bigg(\frac{\psi_{i,j}^{(k+1)}}{\psi_{i,j}^{(k)}}\Bigg) \geq \Bigg(F(\Psi^{(k)},\alpha^{(k)};\Pi,\Theta,\gamma) - F(\Psi^*,\alpha^*;\Pi,\Theta,\gamma)\Bigg),
\end{equation*}
which holds for any $k \!=\! 1,2,\ldots$ Summing up this inequality up to $K$, we have that
\begin{align*}
\Bigg(\sum_{k=1}^K F(\Psi^{(k)},\alpha^{(k)};\Pi,\Theta,\gamma) - F(\Psi^*,\alpha^*;\Pi,\Theta,\gamma)\Bigg) &\leq \sum_{i=1}^n \sum_{j=1}^m \gamma_i \psi_{i,j}^*\textnormal{log}\Bigg(\frac{\psi_{i,j}^{(K+1)}}{\psi_{i,j}^{(1)}}\Bigg)\\
     &\leq \sum_{i=1}^n \sum_{j=1}^m \gamma_i \psi_{i,j}^*\Bigg(\frac{\psi_{i,j}^{(K+1)}}{\psi_{i,j}^{(1)}} - 1\Bigg)
\end{align*}
from which we get the desired inequality, since $\textnormal{log}(\psi_{i,j}^{(K+1)}/\psi_{i,j}^{(1)}) = \textnormal{log}(\psi_{i,j}^{(K+1)}/\psi_{i,j}^*) + \textnormal{log}(\psi_{i,j}^*/\psi_{i,j}^{(1)})$. \hfill $\blacksquare$

\end{itemize}\vspace{0.15cm}

\begin{itemize}
\item[] \-\hspace{0.0cm}{\small{\sf{\textbf{Proposition 3.3.}}}} Let $R_\pi \!=\! (V_\pi,E_\pi,\Pi)$ and $R_\varphi \!=\! (V_\varphi,E_\varphi,\Phi)$ be transition models of two Markov chains\\ \noindent over $n$ and $m$ states, respectively, where $m \!<\! n$.  If $[\Psi^*]_{i,j} \!=\! \psi_{i,j}^*$ is an optimal probabilistic partition and\\ \noindent $[\alpha^*]_j \!=\! \alpha^*_j$ an optimal marginal probability vector, then, the approximation error
\begin{equation*}
\Bigg(F(\Psi^*,\alpha^*;\Pi,\Theta,\gamma) - F(\Psi^{(k)},\alpha^{(k)};\Pi,\Theta,\gamma)\Bigg) \leq \frac{1}{k}\sum_{i=1}^n \sum_{j=1}^m \gamma_i \psi_{i,j}^* \textnormal{log}\Bigg(\frac{\psi_{i,j}^*}{\psi_{i,j}^{(1)}}\Bigg).
\end{equation*}
falls off as a function of the inverse of the iteration count $k$.  Here, the constant factor of the error bound is a Kullback-Leibler divergence between the initial partition matrix $\Psi^{(1)}$ and the global-best partition matrix $\Psi^*$.
\end{itemize}
\vspace{0.05cm}
\begin{itemize}
\item[] \-\hspace{0.5cm}{\small{\sf{\textbf{Proof:}}}} From propositions 3.2(ii) and 3.2(iii), we get that
\begin{equation*}
k\Bigg(F(\Psi^*,\alpha^*;\Pi,\Theta,\gamma) - F(\Psi^{(k)},\alpha^{(k)};\Pi,\Theta,\gamma)\Bigg) \leq \Bigg(\sum_{k=1}^K F(\Psi^{(k)},\alpha^{(k)};\Pi,\Theta,\gamma) - F(\Psi^*,\alpha^*;\Pi,\Theta,\gamma)\Bigg).
\end{equation*}
The desired inequality follows after dividing both sides by $k$ and substituting the bound obtained in proposition 3.2(iii) on the right-hand side of the inequality. \hfill $\blacksquare$
\end{itemize}\vspace{0.15cm}

\begin{itemize}
\item[] \-\hspace{0.0cm}{\small{\sf{\textbf{Proposition 3.4.}}}} Let $R_\pi \!=\! (V_\pi,E_\pi,\Pi)$ and $R_\varphi \!=\! (V_\varphi,E_\varphi,\Phi)$ be transition models of two Markov chains\\ \noindent over $n$ and $m$ states, respectively, where $m \!<\! n$.  Let $g(\pi_{i,1:n},\vartheta_{i,1:n}) \!=\! \sum_{j=1}^n \gamma_i \pi_{i,j} \textnormal{log}(\pi_{i,j}/\vartheta_{i,j})$, where\\ \noindent $R_\vartheta \!=\! (V_\vartheta,E_\vartheta,\Theta)$ is the joint model.  The following hold:
\begin{itemize}
\item[] \-\hspace{0.5cm}(i) The transition matrix $\Phi$ of a low-order Markov chain over states $m$ is given by $\Phi \!=\! \Theta \Psi$, where\\ \noindent $\Theta \!=\! U^\top \Pi$.  Here, $[U]_{i,j} \!=\! \gamma_i \psi_{i,j}/\sum_{k=1}^n \gamma_k \psi_{k,j}$ for the probabilistic partition matrix $[\Psi]_{i,j} \!=\! \psi_{i,j}$ found using the updates in proposition 3.1.
\item[] \-\hspace{0.5cm}(ii) Suppose that we have a low-order chain over $m$ states with a transition matrix $\Phi$ and weight matrix $\Theta$ given by (i).  For some $\beta_0$, suppose $\Theta_{\beta_0}$, the matrix $\Theta$ for that value of $\beta_0$, satisfies the following inequality $d^2/d\epsilon^2\, F(\Psi,\alpha;\Pi,\Theta_{\beta_0} \!+\! \epsilon Q,\gamma)|_{\epsilon = 0} > 0$.  Here, $Q \!\in\! \mathbb{R}_+^{m \times n}$ is a matrix such that $\sum_{k=1}^m q_{k,1:n}^\top q_{k,1:n} \!=\! 1$ and\\ \noindent $\sum_{j=1}^n q_{i,j} \!=\! 0$ $\,\forall i$.  A critical value $\beta_c$, $\beta_c \!=\! \textnormal{min}_{\beta > \beta_0}\,(d^2/d\epsilon^2\, F(\Psi,\alpha;\Pi,\Theta_{\beta} \!+\! \epsilon Q,\gamma)|_{\epsilon = 0} \leq 0)$, occurs\\ \noindent whenever the minimum eigenvalue of the matrix
\begin{equation*}
\textnormal{diag}\Bigg(\sum_{i=1}^n \psi_{k,j}\pi_{i,1:n}/\vartheta_{k,1:n}^2\Bigg) - \beta\Bigg(\sum_{i=1}^n \psi_{k,j}(\pi_{i,1:n}/\vartheta_{k,1:n}^2)(\pi_{i,1:n}/\vartheta_{k,1:n}^2)^\top \Bigg)
\end{equation*}
is zero.  The number of rows in $\Theta$ and columns in $\Psi$ needs to be increased once $\beta \!>\! \beta_c$.
\end{itemize}
\vspace{0.05cm}
\end{itemize}
\begin{itemize}
\item[] \-\hspace{0.5cm}{\small{\sf{\textbf{Proof:}}}} For part (i), we first substitute the partition matrix update into the Lagrangian $F(\Psi,\alpha;\Pi,\Theta,\gamma)$.  After some simplification and when ignoring irrelevant terms, we find that
\begin{equation*}
\frac{\partial}{\partial \vartheta_{j,i}} F(\Psi,\alpha;\Pi,\Theta,\gamma) = \frac{\partial}{\partial \vartheta_{j,i}}\Bigg(\sum_{i=1}^n \gamma_i \textnormal{log}\Bigg(\sum_{j=1}^m \alpha_j e^{-\beta g(\pi_{i,1:n},\vartheta_{j,1:n})}\!\Bigg)\!\Bigg).
\end{equation*}
Setting this expression to zero and solving, we arrive at $\vartheta_{j,i} \!=\! \sum_{r=1}^n \gamma_r \psi_{r,j}\pi_{r,i}$.  Due to the conditions that\\ \noindent $\sum_{i=1}^n \vartheta_{j,i} \!=\! \sum_{i=1}^n \pi_{i,j} \!=\! 1$ $\forall j$, we thus arrive at $\vartheta_{j,i} \!=\! \sum_{r=1}^n \gamma_r u_{r,j}$, where $u_{r,j} \!=\! \psi_{r,j}/\sum_{p=1}^n \gamma_p \psi_{p,j}$.  It is\\ \noindent apparent that the entries of $\Theta$ are non-negative.  As well, every entry in $\Phi$ is a convex combination of a column in $\Theta$.  Therefore, $\Phi$ is a probabilistic transition matrix.

For part (ii), the second variation of $\sum_{i=1}^n \gamma_i \textnormal{log}(\sum_{j=1}^m \alpha_j e^{-\beta g(\pi_{i,1:n},\vartheta_{j,1:n})})$ at the optimal weighting matrix $\Theta^*$ is given by

\begin{align*}
\Delta^2F(\Psi,\alpha;\Pi,\Theta^*,\gamma) &= d^2/d\epsilon^2\, F(\Psi,\alpha;\Pi,\Theta^*_{\beta_0} \!+\! \epsilon Q,\gamma)|_{\epsilon = 0}\\
&= \sum_{j=1}^m \sum_{i=1}^n \gamma_i \psi_{i,j} q_{j,1:n}^\top \Bigg(\textnormal{diag}\Bigg(\sum_{i=1}^n \psi_{k,j}\pi_{i,1:n}/(\vartheta_{k,1:n}^*)^2\Bigg)\\
&\;\;\;\;\;\;\;\;\;\;\;\;\;\;\;\;\;- \beta\Bigg(\sum_{i=1}^n \psi_{k,j}(\pi_{i,1:n}/(\vartheta_{k,1:n}^*)^2)(\pi_{i,1:n}/(\vartheta_{k,1:n}^*)^2)^\top \Bigg)\!\Bigg)q_{j,1:n}\\
&\;\;\;\;\;\;\;\;\;\;\;\;\;\;\;\;\;+ \beta\Bigg(\sum_{i=1}^n \Bigg(\sum_{q=1}^m \frac{\gamma_i \psi_{i,q}}{\sum_{r=1}^n \gamma_r\psi_{r,q}} (\pi_{i,1:n}/\vartheta^*_{k,1:n})^\top q_{q,1:n} \Bigg)^{\!\!2}\,\Bigg)\\
&\geq \textnormal{min}_j\Bigg(\textnormal{min-eig}\Bigg(\sum_{j=1}^m \sum_{i=1}^n \gamma_i \psi_{i,j} q_{j,1:n}^\top \Bigg(\textnormal{diag}\Bigg(\sum_{i=1}^n \psi_{k,j}\pi_{i,1:n}/(\vartheta_{k,1:n}^*)^2\Bigg)\\
&\;\;\;\;\;\;\;\;\;\;\;\;\;\;\;\;\;- \beta\Bigg(\sum_{i=1}^n \psi_{k,j}(\pi_{i,1:n}/(\vartheta_{k,1:n}^*)^2)(\pi_{i,1:n}/(\vartheta_{k,1:n}^*)^2)^\top \Bigg)\!\Bigg)q_{j,1:n} \Bigg)\!\Bigg)\\
&\geq 0.
\end{align*}

\noindent Here, we have used the fact that $\Delta^2F(\Psi,\alpha;\Pi,\Theta^*,\gamma)$ is continuous and strictly positive for $\beta_0 \!<\! \beta$.  We have\\ \noindent also used the spectral theorem \cite{KatoT-book1966a} to obtain a lower bound on the lowest eigenvalue for a self-adjoint operator.  Equality to zero, $\Delta^2F(\Psi,\alpha;\Pi,\Theta^*,\gamma) \!=\! 0$, can thus only be obtained for $\beta \!<\! \beta_c$ if and only if 
\begin{equation*}
\textnormal{min-eig}\Bigg(\textnormal{diag}\Bigg(\sum_{i=1}^n \psi_{k,j}\pi_{i,1:n}/\vartheta_{k,1:n}^2\Bigg) - \beta\Bigg(\sum_{i=1}^n \psi_{k,j}(\pi_{i,1:n}/\vartheta_{k,1:n}^2)(\pi_{i,1:n}/\vartheta_{k,1:n}^2)^\top \Bigg)\!\Bigg) = 0.
\end{equation*}
At such a point $\beta$, a bifurcation in $d^2/d\epsilon^2\, F(\Psi,\alpha;\Pi,\Theta_{\beta_0}^* \!+\! \epsilon Q,\gamma)|_{\epsilon = 0}$ occurs, for a finite perturbation term $Q$, and the minimum is no longer stable.  That is, there is a bifurcation on a solution branch that is fixed by the algebraic group of all permutations on $m \!<\! n$ symbols, $\textnormal{Sym}(m)$, which follows from the equivariant branching lemma and\\ \noindent the Smoller-Wasserman theorem; this bifurcation is symmetry breaking.  The equivariant branching lemma gives explicit bifurcating directions of the $m$ branching solutions, each of which has symmetry $\textnormal{Sym}(m\!-\! 1)$.  The branches are hence associated with a $\Theta$ and $\Psi$ of different cardinalities $m$. \hfill $\blacksquare$

\end{itemize}\vspace{0.15cm}

\begin{itemize}
\item[] \-\hspace{0.0cm}{\small{\sf{\textbf{Proposition 3.5.}}}} Let $R_\pi \!=\! (V_\pi,E_\pi,\Pi)$ and $R_\varphi \!=\! (V_\varphi,E_\varphi,\Phi)$ be transition models of two Markov chains over\\ \noindent $n$ and $m$ states, respectively, where $m \!<\! n$. $R_\vartheta \!=\! (V_\vartheta,E_\vartheta,\Theta)$ is a joint model, with $m \!+\! n$ states.  The systematic underestimation of the information cost of the Shannon mutual information term in definition 3.11 can be second-order minimized by solving the following optimization problem 
\begin{equation*}
\textnormal{min}_{\Psi \in \mathbb{R}_+^{n \times m},\,\Theta \in \mathbb{R}_+^{m \times n}}\Bigg(\!\!\begin{array}{c}\sum_{j=1}^m \alpha_j \sum_{i=1}^n \psi_{i,j}\textnormal{log}(\psi_{i,j}/\gamma_i)\vspace{0.1cm}\\ +\,\sum_{j=1}^m \sum_{i=1}^n \gamma_i\psi_{i,j}^2/2n\textnormal{log}(2)\alpha_j \end{array}\!\!\Bigg|\!\!\!\begin{array}{c} \sum_{i=1}^n\sum_{j=1}^m \gamma_i \psi_{i,j}g(\pi_{i,1:n},\vartheta_{i,1:n})\!\leq\! r\vspace{0.1cm},\\ \; 0 \!\leq\! \vartheta_{i,k},\psi_{i,k} \!\leq\! 1,\; \sum_{k=1}^m \vartheta_{i,k} \!=\! 1,\; \sum_{k=1}^m \psi_{i,k} \!=\! 1\end{array}\!\!\Bigg)
\end{equation*}
where $\beta \!=\! 2^{\sum_{j=1}^m \alpha_j \sum_{i=1}^n \psi_{i,j}\textnormal{log}(\psi_{i,j}/\gamma_i)}\!/2n$.
\end{itemize}\vspace{0.05cm}

\begin{itemize}
\item[] \-\hspace{0.5cm}{\small{\sf{\textbf{Proof:}}}} Let us assume that we approximate the marginal distribution $\gamma_i'$ by $\gamma_i' \!=\! \gamma_i \!+\! \delta\gamma_i$, where $\gamma_i$ is the true\\ \noindent marginal, with zero average over all possible realizations.  There is hence an underestimation of the Shannon mutual information $\sum_{j=1}^m \alpha_j \sum_{i=1}^n \psi_{i,j}\textnormal{log}(\psi_{i,j}/\gamma_i)$, which can be found by taking the multi-order Taylor expansion about $\gamma_i$,
\begin{align*}
\sum_{j=1}^m \alpha_j \sum_{i=1}^n \psi_{i,j}\textnormal{log}\Bigg(\frac{\psi_{i,j}}{\gamma_i}\Bigg)\Bigg|_{\gamma_i + \delta\gamma_i}\\
 &\!\!\!\!\!\!\!\!\!\!\!\!\!\!\!\!\!\!\!\!\!\!\!\!\!\!\!\!\!\!\!\!\!\!\!\!\!\!\!\!\!\!\!\!\!\!\!\!\!\!\!\!\!\!\!\!=\sum_{j=1}^m \alpha_j \sum_{i=1}^n \psi_{i,j}\textnormal{log}\Bigg(\frac{\psi_{i,j}}{\gamma_i}\Bigg) + \sum_{p=2}^\infty \frac{(-1)^p}{p(p \!-\! 1)}\frac{1}{\textnormal{log}(2)\alpha_j^{p-1}}\Bigg(\sum_{j=1}^m \sum_{w=1}^n\sum_{s=1}^m \gamma_w\psi_{w,s}\Bigg(\sum_{r=1}^n \psi_{r,j}\delta\gamma_r\Bigg)^{\!\!p}\,\Bigg)\\
 &\!\!\!\!\!\!\!\!\!\!\!\!\!\!\!\!\!\!\!\!\!\!\!\!\!\!\!\!\!\!\!\!\!\!\!\!\!\!\!\!\!\!\!\!\!\!\!\!\!\!\!\!\!\!\!\!\geq \sum_{j=1}^m \alpha_j \sum_{i=1}^n \psi_{i,j}\textnormal{log}\Bigg(\frac{\psi_{i,j}}{\gamma_i}\Bigg) + \frac{1}{2p\textnormal{log}(2)}\Bigg(\sum_{j=1}^m \sum_{w=1}^n\sum_{s=1}^m \frac{\gamma_w\psi_{w,s}}{\alpha_j}\Bigg(\sum_{r=1}^n \psi_{r,j}^2\gamma_r\Bigg)\!\Bigg)\\
 &\!\!\!\!\!\!\!\!\!\!\!\!\!\!\!\!\!\!\!\!\!\!\!\!\!\!\!\!\!\!\!\!\!\!\!\!\!\!\!\!\!\!\!\!\!\!\!\!\!\!\!\!\!\!\!\!\geq \sum_{j=1}^m \alpha_j \sum_{i=1}^n \psi_{i,j}\textnormal{log}\Bigg(\frac{\psi_{i,j}}{\gamma_i}\Bigg) + \frac{1}{2p\textnormal{log}(2)}2^{\sum_{j=1}^m \alpha_j \sum_{i=1}^n \psi_{i,j}\textnormal{log}(\psi_{i,j}/\gamma_i)}.
\end{align*}
For the last and second-to-last steps, we considered only the second-order Taylor series expansion term.  For the last step, we used the following equality $\sum_{i=1}^n \sum_{j=1}^m \rho_{i,j} \psi_{i,j}/\alpha_j = \sum_{i=1}^n \sum_{j=1}^m \rho_{i,j} 2^{\textnormal{log}(\psi_{i,j}/\alpha_j)}$, which is\\ \noindent bounded below by $2^{\sum_{j=1}^m \alpha_j \sum_{i=1}^n \psi_{i,j}\textnormal{log}(\psi_{i,j}/\gamma_i)}$.  Here, $\rho_{i,j}$ is the joint distribution of the random variables.  It can be seen that the Shannon mutual information term is underestimated by $2^{\sum_{j=1}^m \alpha_j \sum_{i=1}^n \psi_{i,j}\textnormal{log}(\psi_{i,j}/\gamma_i)}/2p\textnormal{log}(2)$ bits.  The bound on the second-order Taylor expansion hence has a rescaled slope.

Plugging the augmented Shannon information term into the value of information Lagrangian, in place of the original Shannon mutual information constraint, and solving yields following update for $\Psi$, 
\begin{multline*}
\psi_{i,j} = \frac{\alpha_j}{z} \textnormal{exp}\Bigg(\!\!-\beta\textnormal{log}(2)g(\pi_{i,1:n},\vartheta_{j,1:n}) +\sum_{j,s=1}^m \sum_{p=2}^\infty \frac{(-1)^p}{p\alpha_j^p} \gamma_i\psi_{i,s}\Bigg(\sum_{r=1}^n \psi_{r,j}\delta\gamma_r\Bigg)^{\!\!p}\\
- \frac{1}{(p \!-\! 1)\gamma_i\alpha_j^{p-1}}\sum_{w=1}^n \sum_{s=1}^m \gamma_i \delta \gamma_w\psi_{w,s} \Bigg(\sum_{r=1}^n \psi_{r,j}\delta\gamma_r\Bigg)^{\!\!p-1}
\Bigg)
\end{multline*}
where $z$ is a normalization factor that ensures the entries of $\Psi$ are probabilities.  Considering only the second-order terms from the Taylor expansion and using $\beta \!=\! 2^{\sum_{j=1}^m \alpha_j \sum_{i=1}^n \psi_{i,j}\textnormal{log}(\psi_{i,j}/\gamma_i)}\!/2n$ leads to a second-order minimization of the underestimation of information. \hfill $\blacksquare$

\end{itemize}\vspace{0.15cm}

\begin{itemize}
\item[] \-\hspace{0.0cm}{\small{\sf{\textbf{Proposition 3.6.}}}} Let $R_\pi \!=\! (V_\pi,E_\pi,\Pi)$ be a transition model of a Markov chain with $n$ states, where $\Pi \!\in\! \mathbb{R}_+^{n \times n}$ is nearly completely decomposable into $m$ Markov sub-chains.  
\begin{itemize}
\item[] \-\hspace{0.5cm}(i) The associated low-order stochastic matrix $\Phi \!\in\! \mathbb{R}_+^{m \times m}$ found by solving the value of information\\ \noindent is given by $\varphi_{i,j} = \sum_{p_i=1}^{n_i} \sum_{q_j=1}^{n_j} \pi_{p_i,q_i} \gamma_{p_i}/\sum_{q_i}^{n_i} \gamma_{q_i}$, where $p_i,q_i$ represent state indices $p \!=\! 1,\ldots,n_i$\\ \noindent associated with block $i$, while $q_j$ represents a state index $q \!=\! 1,\ldots,n_j$ into block $j$.  The variable $\gamma_{p_i} \!=\! \gamma_{p_i}(\Pi)$ denotes the invariant-distribution probability of state $p$ in block $i$ of $\Pi$.  
\item[] \-\hspace{0.5cm}(ii) Suppose that $\gamma_{p_i}/\sum_{q_i}^{n_i} \gamma_{q_i} \!=\! v_{p_i}^*(1_i)$ is approximated by the entries of the first left-eigenvector $v^*(1_i)$ for block $i$ of $\Pi^*$.  We then have that
\begin{equation*}
\Bigg\|\gamma\Bigg(\sum_{p_i=1}^{n_i} v_{p_i}^*(1_i) \sum_{q_j=1}^{n_j} \pi_{p_i,q_i}  \Bigg) - \gamma(\Pi)\Psi\Bigg\|_1 \sim O(\varepsilon^2)
\end{equation*}
where the first term is the invariant distribution of the low-order matrix $\gamma(\Phi)$, under the simplifying assumption, and $\Psi \!\in\! \mathbb{R}^{n \times m}_+$ is the probabilistic partition matrix found by solving the value of information.
\end{itemize}
\vspace{0.05cm}
\end{itemize}

\begin{itemize}
\item[] \-\hspace{0.5cm}{\small{\sf{\textbf{Proof:}}}} For part (i), we note that the aggregated stochastic matrix for nearly-completely decomposable chains, obtained via $\Theta \!=\! U^\top \Pi$ and $\Phi \!=\! \Theta \Psi$, satisfies the implicit equation
\begin{equation*}
\begin{array}{c}
\displaystyle \varphi_{1:n,j} = \sum_{k=1}^{m} \sum_{p_i=1}^{n_k} \vartheta_{j,p_i} \pi_{p_i,1:n}\vspace{0.15cm}\\
\displaystyle \textnormal{where}\; u_{j,p_i} = \frac{\psi_{p_i,j}\gamma_{p_i}}{\sum_{q=1}^m \sum_{s_q=1}^{n_q}\psi_{s_q,q}\gamma_{s_q}} \;\textnormal{and}\; \psi_{p_i,j} = \frac{e^{-\beta g(\pi_{p_i,1:n},\vartheta_{j,1:n})}}{\sum_{q=1}^m \sum_{s_q=1}^{n_q}e^{-\beta g(\pi_{s_q,1:n},\vartheta_{q,1:n})}}.
\end{array}
\end{equation*}
In what follows, we want to assess the form of the aggregated stochastic matrix $\Phi$ when it contains $m$ state groups.  However, $\beta$ dictates the number of state groups in a manner that is dependent on the original stochastic matrix $\Pi$.  We therefore simply consider what happens when $\beta$ is infinite and note that the same expression for $\Phi$ can be obtained for finite $\beta$s.  In the former case, we have that
\begin{equation*}
\textnormal{lim}_{\beta \to \infty}\, u_{j,p_i} = \frac{\mathbb{I}_{g(\pi_{p_i,1:n},\vartheta_{j,1:n}) \,=\, \textnormal{min}{j'}\, g(\pi_{p_i,1:n},\vartheta_{j',1:n})}\gamma_{p_i}}{\sum_{q=1}^m \sum_{s_q=1}^{n_q} \mathbb{I}_{g(\pi_{s_q,1:n},\vartheta_{q,1:n}) \,=\, \textnormal{min}{q'}\, g(\pi_{s_q,1:n},\vartheta_{q',1:n})}\gamma_{s_q} }\vspace{0.05cm}
\end{equation*}
\begin{equation*}
\textnormal{lim}_{\beta \to \infty}\, \psi_{p_i,j} = \frac{e^{-\beta g(\pi_{p_i,1:n},\vartheta_{j,1:n}) - \textnormal{min}{j'}\, g(\pi_{p_i,1:n},\vartheta_{j',1:n})}}{\sum_{q=1}^m e^{-\beta g(\pi_{p_i,1:n},\vartheta_{q,1:n}) - \textnormal{min}{j'}\, g(\pi_{p_i,1:n},\vartheta_{j',1:n})}}
\end{equation*}
where $\mathbb{I}$ is the indicator function.  Due to the nearly-completely decomposable structure of the Markov chain, $\mathbb{I}_{g(\pi_{p_i,1:n},\vartheta_{j,1:n}) = \textnormal{min}{j'}\, g(\pi_{p_i,1:n},\vartheta_{j',1:n})}\gamma_{p_i} \!=\! \gamma_{p_i}$ $\forall j$.  Hence,
\begin{equation*}
\textnormal{lim}_{\beta \to \infty}\, \varphi_{i,j} \!=\! (\textnormal{lim}_{\beta \to \infty}\, u_{j,p_i})\Pi(\textnormal{lim}_{\beta \to \infty}\, \psi_{p_i,j}) = \sum_{p_i=1}^{n_i} \frac{\gamma_{p_i}}{\sum_{q_i}^{n_i} \gamma_{q_i}} \sum_{q_j=1}^{n_j} \pi_{p_i,q_i}.
\end{equation*}
Part (ii) follows from the work of Courtois \cite{CourtoisPJ-jour1975a}.

\end{itemize}

\end{document}